\documentclass[oneside,11pt]{article}

\usepackage[utf8]{inputenc}
\usepackage[T1]{fontenc}

\usepackage{amsthm}
\usepackage{amsmath}
\usepackage{amsfonts}
\usepackage{amssymb}

\usepackage{graphicx}
\usepackage{graphbox}
\usepackage{float}
\usepackage{booktabs}
\usepackage{multirow}
\usepackage{rotating}
\usepackage{array}
\usepackage{xcolor}
\usepackage{subcaption}
\usepackage[ruled]{algorithm2e}

\newcolumntype{C}[1]{>{\centering\arraybackslash}p{#1}}
\newcolumntype{R}[1]{>{\raggedleft\let\newline\\\arraybackslash\hspace{0pt}}m{#1}}
\newcolumntype{L}[1]{>{\raggedright\let\newline\\\arraybackslash\hspace{0pt}}m{#1}}

\usepackage{breakcites}

\usepackage{nowidow}

\usepackage{enumitem}

\usepackage[top=1.5cm,bottom=2cm,right=2.5cm,left=2.5cm]{geometry}
\usepackage{titling}

\usepackage{natbib}

\usepackage{ifplatform}

\usepackage{textcomp}
\usepackage{bbm}

\usepackage{comment}

\ifwindows
\fi

\allowdisplaybreaks

\newcommand{\lc}{\left[}
\newcommand{\rc}{\right]}
\newcommand{\lb}{\left\{}
\newcommand{\rb}{\right\}}

\newcommand{\R}{\mathbb{R}}

\newcommand{\rd}{\mathrm{d}}

\newcommand{\bx}{\boldsymbol{x}}
\newcommand{\bw}{\boldsymbol{w}}
\newcommand{\be}{\boldsymbol{e}}

\newcommand{\by}{\boldsymbol{y}}
\newcommand{\bX}{\boldsymbol{X}}

\newcommand{\bU}{\boldsymbol{U}}

\newcommand{\bmu}{\boldsymbol\mu}
\newcommand{\bu}{\boldsymbol{u}}
\newcommand{\bv}{\boldsymbol{v}}

\newcommand{\bb}{\boldsymbol{b}}
\newcommand{\bba}{\boldsymbol{a}}

\newcommand{\bt}{\boldsymbol{t}}

\newcommand{\zero}{\boldsymbol{0}}

\newcommand{\bxi}{\boldsymbol\xi}

\newcommand{\bga}{\boldsymbol\gamma}

\newcommand{\bIcal}{\boldsymbol{\mathcal{I}}}

\newcommand{\bSigma}{{\boldsymbol\Sigma}}

\newcommand{\bO}{\boldsymbol{O}}

\newcommand{\bB}{\boldsymbol B}
\newcommand{\bC}{\boldsymbol C}

\newcommand{\bA}{\boldsymbol{A}}

\newcommand{\bI}{\boldsymbol I}
\newcommand{\bS}{\boldsymbol{S}}

\newcommand{\bXi}{\boldsymbol\Xi}

\newcommand{\inlaw}{\rightsquigarrow}
\DeclareMathAlphabet\mathbfcal{OMS}{cmsy}{b}{n}

\newcommand{\Sym}{\boldsymbol{S}}

\newcommand{\bD}{\boldsymbol{D}}

\newcommand{\Sd}{\mathbb{S}^{d}}
\newcommand{\Sdm}{\mathbb{S}^{d-1}}

\newcommand{\defin}{:=}
\newcommand{\indef}{=:}

\newcommand{\lrp}[1]{\left(#1\right)}
\newcommand{\lrc}[1]{\left[#1\right]}
\newcommand{\lrb}[1]{\left\{#1\right\}}
\newcommand{\lrpbig}[1]{\big(#1\big)}
\newcommand{\lrcbig}[1]{\big[#1\big]}
\newcommand{\lrbbig}[1]{\big\{#1\big\}}

\newcommand{\lrbBig}[1]{\Big\{#1\Big\}}
\newcommand{\lrpbigg}[1]{\bigg(#1\bigg)}

\newcommand{\lrbbigg}[1]{\bigg\{#1\bigg\}}

\newcommand{\Prob}[1]{\mathrm{P}\lb #1\rb}
\newcommand{\E}[1]{\mathrm{E}\lc #1\rc}
\newcommand{\V}[1]{\mathrm{V}\mathrm{ar}\lc #1\rc}
\newcommand{\Ebig}[1]{\mathrm{E}\big[#1\big]}
\newcommand{\Vbig}[1]{\mathrm{V}\mathrm{ar}\big[#1\big]}

\newcommand{\Covbig}[2]{\mathrm{C}\mathrm{ov}\big[#1,#2\big]}

\newcommand{\om}[1]{\omega_{#1}}

\DeclareFontFamily{OT1}{pzc}{}
\DeclareFontShape{OT1}{pzc}{m}{it}{<-> s * [1.10] pzcmi7t}{}
\DeclareMathAlphabet{\mathpzc}{OT1}{pzc}{m}{it}

\newtheorem{definition}{Definition}[section]
\newtheorem{theorem}{Theorem}[section]
\newtheorem{corollary}{Corollary}[section]

\newtheorem{proposition}{Proposition}[section]
\newtheorem{lemma}{Lemma}[section]

\newcommand{\pct}{\%}

\usepackage[pdftex,final,bookmarksnumbered,bookmarksopen=false,breaklinks,colorlinks]{hyperref}
\hypersetup{
	final,
	bookmarksnumbered=true,
	bookmarksopen=true,
	bookmarksopenlevel=0,
	unicode=false,
	pdftoolbar=true,
	pdfmenubar=true,
	pdffitwindow=false,
	pdftitle={On the spherical cardioid distribution and its goodness-of-fit},
	pdfdisplaydoctitle=true,
	pdftoolbar=true,
	pdfmenubar=true,
	pdflang={English},
	pdfauthor={Eduardo Garcia-Portugues},
	pdfsubject={arXiv paper},
	pdfcreator={Eduardo Garcia-Portugues},
	pdfproducer={Eduardo Garcia-Portugues},
	pdfkeywords={Directional Statistics}{Gegenbauer Polynomials}{Chebyshev Polynomials}{Projections},
	pdfnewwindow=true,
	breaklinks=true,
	hidelinks,
	linkcolor=black,
	citecolor=black,
	filecolor=magenta,
	urlcolor=cyan
}

\newif\ifmain
\maintrue
\newif\ifsupplement
\supplementtrue

\begin{document}

\ifmain

\title{On the spherical cardioid distribution and its goodness-of-fit}
\setlength{\droptitle}{-1cm}
\predate{}%
\postdate{}%
\date{}

\author{Eduardo Garc\'ia-Portugu\'es$^{1,2}$}
\footnotetext[1]{Department of Statistics, Universidad Carlos III de Madrid (Spain).}
\footnotetext[2]{Corresponding author. e-mail: \href{mailto:edgarcia@est-econ.uc3m.es}{edgarcia@est-econ.uc3m.es}.}
\maketitle

\begin{abstract}
	In this paper, we study the spherical cardioid distribution, a higher-dimensional and arbitrary-order generalization of the circular cardioid distribution. This distribution is rotationally symmetric and generates unimodal, multimodal, axial, and girdle-like densities. We identify various properties of the spherical cardioid that make it highly tractable: simple density evaluation, closedness under convolution, explicit expressions for vectorized moments, and efficient simulation. The moments of the spherical cardioid of order $k$ up to order $k-1$ coincide with those of the uniform distribution on the sphere, highlighting its closeness to the latter. We derive estimators by the method of moments and maximum likelihood, their asymptotic distributions, and their asymptotic relative efficiencies. We give the machinery for bootstrap goodness-of-fit tests based on the projected empirical cumulative distribution function approach, including the projected distribution and closed-form expressions for test statistics. An application to modeling the orbits of long-period comets shows the usefulness of the spherical cardioid distribution in real data analyses.
\end{abstract}
\begin{flushleft}
\end{flushleft}

\section{Introduction}
\label{sec:intro}

%
The \emph{(circular) cardioid distribution} was introduced by Harold Jeffreys in 1948 \citep[Eq. (1) in p. 328 of][reprint]{Jeffreys2003} as that with density
\begin{align}
    f_{\mathrm{C}}(\theta;\mu,\rho)\defin \frac{1}{2\pi}\{1+\rho\cos(\theta-\mu)\},\quad \theta\in[0,2\pi), \label{eq:circard}
\end{align}
for a concentration $\rho\in[-1,1]$ and a location $\mu\in[0,2\pi)$. The name ``cardioid'' stems from the fact that the polar curve $r(\theta)=f_{\mathrm{C}}(\theta;\mu,\rho)$ traces a cardioid when $|\rho|=1$ and a limaçon when $|\rho|<1$. Originally proposed as a ``continuous departure from a uniform distribution of chance'', the cardioid distribution is arguably the simplest non-uniform distribution on the circle $\mathbb{S}^1\equiv [0,2\pi)$, with $0\equiv2\pi$ identified. It reduces to the uniform density on the circle for $\rho=0$ and gives a unimodal density for $\rho>0$ with mode at $\mu$. In the literature, the cardioid density is often parametrized as $\theta\mapsto(2\pi)^{-1}(1+2\rho\cos(\theta-\mu))$ and $|\rho|\in[0, 1/2)$ (\citealp[Sec.~3.5.5]{Mardia1999a}; \citealp[Sec.~2.2.2]{Jammalamadaka2001}) or $\rho\in[0,1/2]$ \citep[Sec.~4.3.4]{Pewsey2013}, but we adopt the original parametrization in \eqref{eq:circard} for later convenience.

The cardioid distribution has received considerable attention in the literature. The main reasons are threefold. First, the model is highly tractable, which makes it particularly easy to understand and rich in terms of available results that are scarcer for other circular distributions: closedness under convolution, closed-form cumulative distribution function, straightforward trigonometric moments, and direct estimation and inference. \cite{Pewsey2026} gave a recent and critical review of these basic properties. Second, the density \eqref{eq:circard} arises as a submodel of various circular models that have been proposed in the literature, like the $M$-truncated nonnegative Fourier series models of \cite{Fernandez-Duran2004} (\eqref{eq:circard} arises with the choice $M=1$), the symmetric and unimodal family of circular distributions of \cite{Jones2005} (choice $\psi=1$), and the unimodal family of \citet{Kato2015} (choice $\rho=0$). Third, due to its simplicity, the cardioid density has been used as the building block for more flexible models, through generators and transformations. We refer to \cite{Pewsey2026} for references on the latter use.

Generalizations of \eqref{eq:circard} to higher-dimensional supports have been mainly concentrated on the two-dimensional torus $(\mathbb{S}^1)^2$. Extensions of \eqref{eq:circard} to the sphere $\mathbb{S}^d\defin\{\bx\in\mathbb{R}^{d+1}:\|\bx\|=1\}$ have been scarcer and closely related. \cite{Saw1984} and \cite{Baringhaus2024} used Gegenbauer (or ultraspherical) polynomials to represent rotationally symmetric distributions on $\Sd$, in particular generalizing \eqref{eq:circard} to $\Sd$. A particular case of this construction was used in \cite{Borodavka2026} as an auxiliary device to study the power of maximal projection-based uniformity tests on $\Sd$. On $\mathbb{S}^2$, \cite{Fernandez-Duran2014a} considered a bivariate nonnegative Fourier series on the angles $(\theta,\phi)\in[0,2\pi)\times[0,\pi]$. To the best of the author's knowledge, no systematic statistical analysis (including, e.g., arbitrary moments, estimation, inference, simulation, and goodness-of-fit testing) of a distribution generalizing \eqref{eq:circard} to $\mathbb{S}^d$, $d\geq 1$, had been conducted before the present work.

This paper studies in depth the \emph{spherical cardioid distribution} on $\Sd$, $d\geq1$. Its density is a $k$-order generalization of the circular cardioid density \eqref{eq:circard} based on Chebyshev and Gegenbauer polynomials. Our main methodological contributions are threefold. First, we identify various properties of the spherical cardioid that make it an appealing model: ability to produce unimodal/multimodal patterns, connections with the von Mises--Fisher and Watson distributions, explicit vectorized moments of arbitrary order, closedness under convolution, a simple characteristic function, and efficient simulation. The $m$-moments of the cardioid of order $k$, with $k>m$, are shown to coincide with those of the uniform distribution. This makes a large-$k$ spherical cardioid distribution a formidably challenging alternative for uniformity tests. Second, we derive estimation and inference by the method of moments, for $k=1,2$, and a moment estimator for the concentration parameter when the location is known. We also study maximum likelihood estimation and inference for $k\geq 1$. We obtain the asymptotic distributions of these estimators and their asymptotic relative efficiencies, showing that the efficiency of moment estimators decreases for large absolute values of the concentration. Third, we give the machinery for bootstrap goodness-of-fit tests based on a statistic that integrates, across projecting directions, discrepancies between the projected empirical cumulative distribution function and the null projected cumulative distribution function. To that aim, we provide closed forms for the projected distribution and several test statistics. Numerical experiments corroborate the adequate empirical behavior of the estimators and the goodness-of-fit tests. In an application to modeling the orbits of long-period comets, we show that the spherical cardioid distribution of order two on $\mathbb{S}^2$ is a practically relevant model.

The rest of this paper is organized as follows. Section~\ref{sec:gegen} provides general notation and background on orthogonal polynomials used throughout the paper. Section~\ref{sec:cardioid} introduces the spherical cardioid distribution (Sec.~\ref{sec:genesis}) and its basic properties: closedness under convolution (Sec.~\ref{sec:convolution}), moments (Sec.~\ref{sec:moments}), characteristic function (Sec.~\ref{sec:characteristic}), and simulation (Sec.~\ref{sec:simulation}). Section~\ref{sec:estimation} deals with estimation and inference by the method of moments (Sec.~\ref{sec:mom}) and maximum likelihood (Sec.~\ref{sec:mle}). Goodness-of-fit tests for spherical cardioidness are provided in Sec.~\ref{sec:gof}, which are based on the projected cumulative distribution function (Sec.~\ref{sec:projections}), closed forms of the test statistics (Sec.~\ref{sec:teststat}), and a parametric bootstrap procedure (Sec.~\ref{sec:boot}). Section~\ref{sec:experiments} reports numerical experiments on the finite-sample performance of the estimators and bootstrap tests. An application to modeling the orbits of comets is presented in Sec.~\ref{sec:application}. Some concluding remarks are given in Sec.~\ref{sec:discussion}. Proofs and additional results are deferred to the Supplementary Materials (SM).

\section{Background on orthogonal polynomials}
\label{sec:gegen}

%
We denote the surface area of $\Sd=\{\bx\in\R^{d+1} : \|\bx\|=1\}$, $d\geq1$, as $\om{d}\defin2\pi^{(d+1)/2}/\Gamma((d+1)/2)$. The surface area measure on $\Sd$ is denoted by $\sigma_d$. We set $\mathbb{S}^0\defin\{-1,1\}$, $\sigma_0$ as the counting measure on $\mathbb{S}^0$, and $\om{0}\defin2$. The uniform distribution on $\Sd$, denoted $\mathrm{Unif}(\Sd)$, has constant density $1/\om{d}$ with respect to $\sigma_d$. In the paper, densities on $\Sd$ are always with respect to~$\sigma_d$.

Gegenbauer and Chebyshev polynomials are central to the spherical cardioid construction. Both are orthogonal polynomial families on the space of square-integrable functions on $[-1,1]$ with respect to the weight functions $x\mapsto(1-x^2)^{d/2-1}$, $d\geq2$, and $x\mapsto(1-x^2)^{-1/2}$, respectively. We denote by $L_d^2([-1,1])$ this weighted $L^2$ space. \citet[Ch.~18]{NIST} offers a compendium of results related to Gegenbauer and Chebyshev polynomials. Gegenbauer polynomials are also known as ultraspherical polynomials.

The Gegenbauer polynomials $\{C_k^{(d-1)/2}\}_{k=0}^\infty$ of index $\lambda=(d-1)/2$, $d\geq2$, form an orthogonal basis on $L_d^2([-1,1])$. They satisfy the orthogonality relation
\begin{align*}
	\int_{-1}^1 C_k^{(d-1)/2}(x)C_\ell^{(d-1)/2}(x)(1-x^2)^{d/2-1}\,\rd x=\delta_{k,\ell}c_{k,d},%
\end{align*}
where $\delta_{k,\ell}$ denotes the Kronecker delta and
\begin{align}
	c_{k,d}&\defin \frac{2^{3-d}\pi\Gamma(d+k-1)}{(d+2k-1)k!\Gamma((d-1)/2)^2}=\frac{\om{d}}{\om{d-1}}\lrp{1+\frac{2k}{d-1}}^{-2}d_{k,d},\label{eq:ckd}\\
  d_{k,d}&\defin %
  \lrp{1+\frac{2k}{d-1}}\frac{\Gamma(d-1+k)}{\Gamma(d-1)k!}=\lrp{1+\frac{2k}{d-1}}C_k^{(d-1)/2}(1).\label{eq:dkd}
\end{align}
Above, $d_{k,d}$ is the dimension of the vector space of spherical harmonics of degree $k$, i.e., the harmonic homogeneous polynomials of degree $k$ defined on $\Sd$. The first Gegenbauer polynomials are $C_0^{(d-1)/2}(x)=1$, $C_1^{(d-1)/2}(x)=(d-1)x$, and $C_2^{(d-1)/2}(x)=[(d-1)/2][(d+1)x^2 -1]$, while the next follow from the recurrence relation
\begin{align*}
    C_{k+1}^{(d-1)/2}(x)=\frac{2k+d-1}{k+1}\,x\,C_k^{(d-1)/2}(x)-\frac{k+d-2}{k+1}C_{k-1}^{(d-1)/2}(x),\quad k\geq 1.
\end{align*}

The Chebyshev polynomials (of the first kind) $\{T_k\}_{k=0}^\infty$ are expressible as $T_k(x)=\cos(k \cos^{-1}(x))$ for $x\in[-1,1]$. They form an orthogonal basis on $L_1^2([-1,1])$, satisfying
\begin{align*}
    \int_{-1}^1 T_k(x)T_\ell(x)(1-x^2)^{-1/2}\,\rd x=\delta_{k,\ell}c_{k,1},
\end{align*}
where
\begin{align}
    c_{k,1}\defin\frac{1+\delta_{k,0}}{2}\pi=\frac{\om{1}}{\om{0}}\lrp{2-\delta_{k,0}}^{-2}d_{k,1},\quad d_{k,1}\defin 2-\delta_{k,0}=(2-\delta_{k,0})T_k(1). \label{eq:ck1}
\end{align}
The first Chebyshev polynomials are $T_0(x)=1$, $T_1(x)=x$, and $T_2(x)=2x^2 -1$, with the next following from
\begin{align*}
    T_{k+1}(x)=2x\,T_k(x)-T_{k-1}(x),\quad k\geq 1.
\end{align*}

Chebyshev polynomials can be seen as the scaled limit of Gegenbauer polynomials when their index shrinks to zero: $\lim_{\lambda\to0^+}\lambda^{-1}C_k^{\lambda}(x)=(2/k)T_k(x)$ for $k\geq1$. To unify notation, we adopt the convention $C_k^{0}(x)\defin T_k(x)$ for $k\geq0$. By defining
\begin{align*}
\tau_{k,d}\defin \begin{cases}
  2-\delta_{k,0}, & d=1,\\
  1+2k/(d-1), & d\geq2,
\end{cases}
\end{align*}
for $k\geq 0$, it follows that \eqref{eq:ckd}--\eqref{eq:ck1} are expressible, for $d\geq 1$, as
\begin{align}
    c_{k,d}=\frac{\om{d}}{\om{d-1}}\tau_{k,d}^{-2}\,d_{k,d},\quad C_k^{(d-1)/2}(1)=\tau_{k,d}^{-1} \, d_{k,d}.\label{eq:ckd2}
\end{align}
Using this notation, we define the normalized polynomials
\begin{align*}
  \tilde{C}_k^{(d-1)/2}(x)\defin \frac{C_k^{(d-1)/2}(x)}{C_k^{(d-1)/2}(1)}=\frac{C_k^{(d-1)/2}(x)}{\tau_{k,d}^{-1}\, d_{k,d}}%
  =\frac{\om{d}}{\om{d-1}}\frac{C_k^{(d-1)/2}(x)}{c_{k,d}\,\tau_{k,d}}
\end{align*}
for $k\geq0$ and $d\geq1$. In particular, $\tilde{C}_0^{(d-1)/2}(x)=1$, $\tilde{C}_1^{(d-1)/2}(x)=x$, and $\tilde{C}_2^{(d-1)/2}(x)=d^{-1}[(d+1)x^2 -1]$ for $d\geq1$. The normalized polynomials satisfy $|\tilde{C}_k^{(d-1)/2}(x)|\leq 1$ for $x\in[-1,1]$, $k\geq0$, and $d\geq1$. The parity of $C_k^{(d-1)/2}$ and $\tilde{C}_k^{(d-1)/2}$ is that of $k$, i.e., they are even (odd) functions when $k$ is even (odd).

Since $\{C_k^{(d-1)/2}\}_{k=0}^\infty$ form an orthogonal basis of $L_d^2([-1,1])$, any function $g\in L_d^2([-1,1])$, $d\geq 1$, admits the expansion
\begin{align}
    g(x)=&\;\sum_{k=0}^{\infty} b_{k,d}(g) \,C_{k}^{(d-1)/2}(x),\label{eq:gexp}\\
    b_{k,d}(g)\defin&\; \frac{1}{c_{k,d}}\int_{-1}^1 g(x)C_k^{(d-1)/2}(x)(1-x^2)^{d/2-1}\,\rd x,\nonumber
\end{align}
with the series \eqref{eq:gexp} converging in $L_d^2([-1,1])$. As a consequence of \eqref{eq:gexp}, a function $f\in L^2(\Sd)$ rotationally symmetric about $\bmu\in\Sd$, defined as $f(\bx)\defin g(\bx^\top\bmu)$, admits the zonal spherical harmonics expansion
\begin{align}
    f(\bx)=&\;\sum_{k=0}^{\infty} b_{k,d}(g) \,C_{k}^{(d-1)/2}(\bx^\top\bmu),\label{eq:sphharmexp}
\end{align}
with the series converging in $L^2(\Sd)$. \citet[Theorem~2]{Kalf1995} can be applied to guarantee uniform convergence of \eqref{eq:sphharmexp} if $f$ is sufficiently smooth. Indeed, if $f\in C^\infty(\Sd)$, then \eqref{eq:sphharmexp} is uniformly absolutely convergent.

A useful result involving Gegenbauer and Chebyshev polynomials is the formula
\begin{align}
  \frac{1}{\om{d}}\int_{\Sd} C_k^{(d-1)/2}(\bga^\top\bu) C_m^{(d-1)/2}(\bga^\top\bv) \,\sigma_d(\rd\bga)&=\tau_{k,d}^{-1}\,C_k^{(d-1)/2}(\bu^\top\bv)\delta_{k,m} \label{eq:addgegen}
\end{align}
for $k,m\geq0$ and $d\geq1$. A proof of \eqref{eq:addgegen} can be found in \citet[Lemma B.7]{Garcia-Portugues2020b}.

\section{Spherical cardioid distribution}
\label{sec:cardioid}

\subsection{Genesis}
\label{sec:genesis}

%
The generalization of \eqref{eq:circard} to the sphere is motivated by the fact that $\cos(\theta-\mu)$ can be expressed as the Chebyshev polynomial $\tilde{C}_1^{0}(\cos(\theta-\mu))=\cos(\theta-\mu)$. Switching to Cartesian coordinates on $\mathbb{S}^1$, $\bx=(\cos\theta,\sin\theta)^\top$ and $\bmu=(\cos\mu,\sin\mu)^\top$, it follows that $\cos(\theta-\mu)=\cos\theta\cos\mu+\sin\theta\sin\mu=\bx^\top\bmu$, and hence $\cos(\theta-\mu)=\tilde{C}_1^{0}(\bx^\top\bmu)$. The generalization to the sphere $\Sd $, $d\geq1$, and higher orders $k\geq1$ is then evident.

\begin{definition}[Spherical cardioid distribution] \label{def:sphcard}
The \emph{spherical cardioid distribution} on $\Sd$, $d\geq1$, with location $\bmu\in\Sd$, concentration $\rho\in[-1,1]$, and (integer) order $k\geq1$, has density
\begin{align}
  f_{\mathrm{C}_k}(\bx;\bmu,\rho)\defin\frac{1}{\om{d}}\lrb{1+\rho\, \tilde{C}_k^{(d-1)/2}(\bx^\top\bmu)}\label{eq:sphcard}
\end{align}
with respect to the surface area measure $\sigma_d$ on $\Sd$. The distribution is denoted $\mathrm{C}_k(\bmu,\rho)$.
\end{definition}

Just as the circular cardioid is the simplest Fourier-based density, the spherical cardioid is the simplest spherical harmonic-based density (see \eqref{eq:sphharmexp}). The connection between Gegenbauer polynomials and distributions on $\Sd$ was already noted by \cite{Saw1984}, who considered the role of these polynomials as an orthogonal basis for statistical analysis on spheres. \citet[Eq. (27)]{Baringhaus2024} and \citet[Eq. (5.3)]{Borodavka2026} had considered \eqref{eq:sphcard} before as auxiliary means for infinite mixture representation and power investigation in uniformity tests on $\Sd$, respectively. As a particular spherical harmonics-based density, more general instances of \eqref{eq:sphcard} can be traced back to \citet[Theorem 5.3, Proposition 6.5]{Gine1975} in the context of Sobolev tests of uniformity on $\Sd$. On $\mathbb{S}^2$, $\mathrm{C}_k((0,0,\pm 1)^\top,\rho)$ is a radially symmetric ($M_1=0$, $M_2=k$) member of the (non-rotationally invariant) Fourier-based family of \citet{Fernandez-Duran2014a}.

The density \eqref{eq:sphcard} is well-defined since $|\tilde{C}_k^{(d-1)/2}(\bx^\top \bmu)|\leq 1$ and $\int_{\Sd} C_k^{(d-1)/2}(\bx^\top \bmu)\,\sigma_d(\rd\bx)=0$ for $k\geq1$ (e.g., by \eqref{eq:addgegen}). For $d=1$, using polar coordinates, \eqref{eq:sphcard} reduces to
\begin{align*}
    f_{\mathrm{C}_k}(\theta;\mu,\rho)=\frac{1}{2\pi}\lrb{1+\rho\cos(k(\theta-\mu))},
\end{align*}
which yields the circular cardioid density \eqref{eq:circard} when $k=1$. When $d=2$ and $k=1,2$, \eqref{eq:sphcard} becomes
\begin{align*}
    f_{\mathrm{C}_k}(\bx;\bmu,\rho)=\begin{cases}
    \displaystyle\frac{1}{4\pi}\lrb{1+\rho\, \bx^\top\bmu}, & k=1,\\
    \displaystyle\frac{1}{4\pi}\lrb{1+\rho\, \frac{3(\bx^\top\bmu)^2 -1}{2}}, & k=2.
    \end{cases}
\end{align*}
For large $d$, the density \eqref{eq:sphcard} converges to $(1/\om{d})\lrb{1+\rho\, (\bx^\top\bmu)^k}$ since $\lim_{d\to\infty}\tilde{C}_k^{(d-1)/2}(x)=x^k$ for $x\in[-1,1]$ and $k\geq1$ \citep[Eq. 18.6.4]{NIST}.

If $\bX\sim \mathrm{C}_k(\bmu,\rho)$, then for any $(d+1)\times(d+1)$ orthogonal matrix $\bO$, $\bO\bX\sim \mathrm{C}_k(\bO\bmu,\rho)$. Therefore, $\mathrm{C}_k(\bmu,\rho)$ is rotationally symmetric about $\bmu\in \Sd$, since $\bO\bX$ and $\bX$ are equal in distribution for any orthogonal transformation $\bO$ such that $\bO\bmu=\bmu$. Thus, in particular, $f_{\mathrm{C}_k}(\bx;\bmu,\rho)=f_{\mathrm{C}_k}(\bmu;\bx,\rho)=f_{\mathrm{C}_k}(-\bx;-\bmu,\rho)$. When $\rho=0$, $\mathrm{C}_k(\bmu,0)=\mathrm{Unif}(\Sd)$ for any $k\geq 1$.

Because the parity of $\tilde{C}_k^{(d-1)/2}$ is that of $k$, the density \eqref{eq:sphcard} is non-identifiable on $(\bmu,\rho)\in\Sd\times[-1,1]$ for odd $k$, as $f_{\mathrm{C}_k}(\bx;\bmu,\rho)=f_{\mathrm{C}_k}(\bx;-\bmu,-\rho)$. Therefore, we can restrict the parameter space to $\rho\in[0,1]$ for odd $k$ without loss of generality. In contrast, for even $k$, both signs of $\rho$ yield different densities for $d\geq 2$ (see Figure~\ref{fig:cardioid_d2}), but $\pm\bmu$ give the same density. In particular, the density is axially symmetric, i.e., $f_{\mathrm{C}_k}(\bx;\bmu,\rho)=f_{\mathrm{C}_k}(-\bx;\bmu,\rho)=f_{\mathrm{C}_k}(\bx;-\bmu,\rho)$. The parameter space of $\bmu$ can be thus restricted to $\Sd_{+}\defin\{\bmu\in\Sd:\mu_1>0\}$ for $k$ even without loss of generality. This symmetry also makes the density non-identifiable in $\rho$ for $d=1$, as $f_{\mathrm{C}_k}(\theta;\mu,\rho)=f_{\mathrm{C}_k}(\theta;\mu+\pi/k,-\rho)$ since $\cos(k(\theta-(\mu+\pi/k)))=-\cos(k(\theta-\mu))$ (see Figure~\ref{fig:cardioid_d1}); a restriction to $\rho\in[0,1]$ prevents it. Actually, for $d=1$ and any $k\geq 1$, the density is $k$-fold symmetric, so only values of $\bmu$ in $\mathbb{S}_{k}^1\defin\{(\cos(\theta),\sin(\theta))^\top:\theta\in[0,2\pi/k)\}$ give different densities.

To remove all these non-identifiabilities from the full parameter space $\Sd\times [-1,1]$, we reduce the parameter space of $(\bmu,\rho)$ in \eqref{eq:sphcard} to
\begin{align*}
  \Theta_{k,d}\defin\begin{cases}
  \mathbb{S}^1_k\times[0,1], & d=1,\,k\geq 1,\\
  \mathbb{S}^d \times[0,1], & d\geq 2,\,k\text{ odd},\\
  \Sd_+\times[-1,1], & d\geq 2,\,k\text{ even}.
  \end{cases} %
\end{align*}
We may further restrict to $\rho\neq 0$ (excludes uniform density; $\bmu$ is not identifiable) and $|\rho|<1$ (ensures the density is always strictly positive). We denote this reduced parameter space as $\Theta_{k,d}^\circ$.

\begin{figure}[h!]
  \centering
  \includegraphics[width=0.8\textwidth]{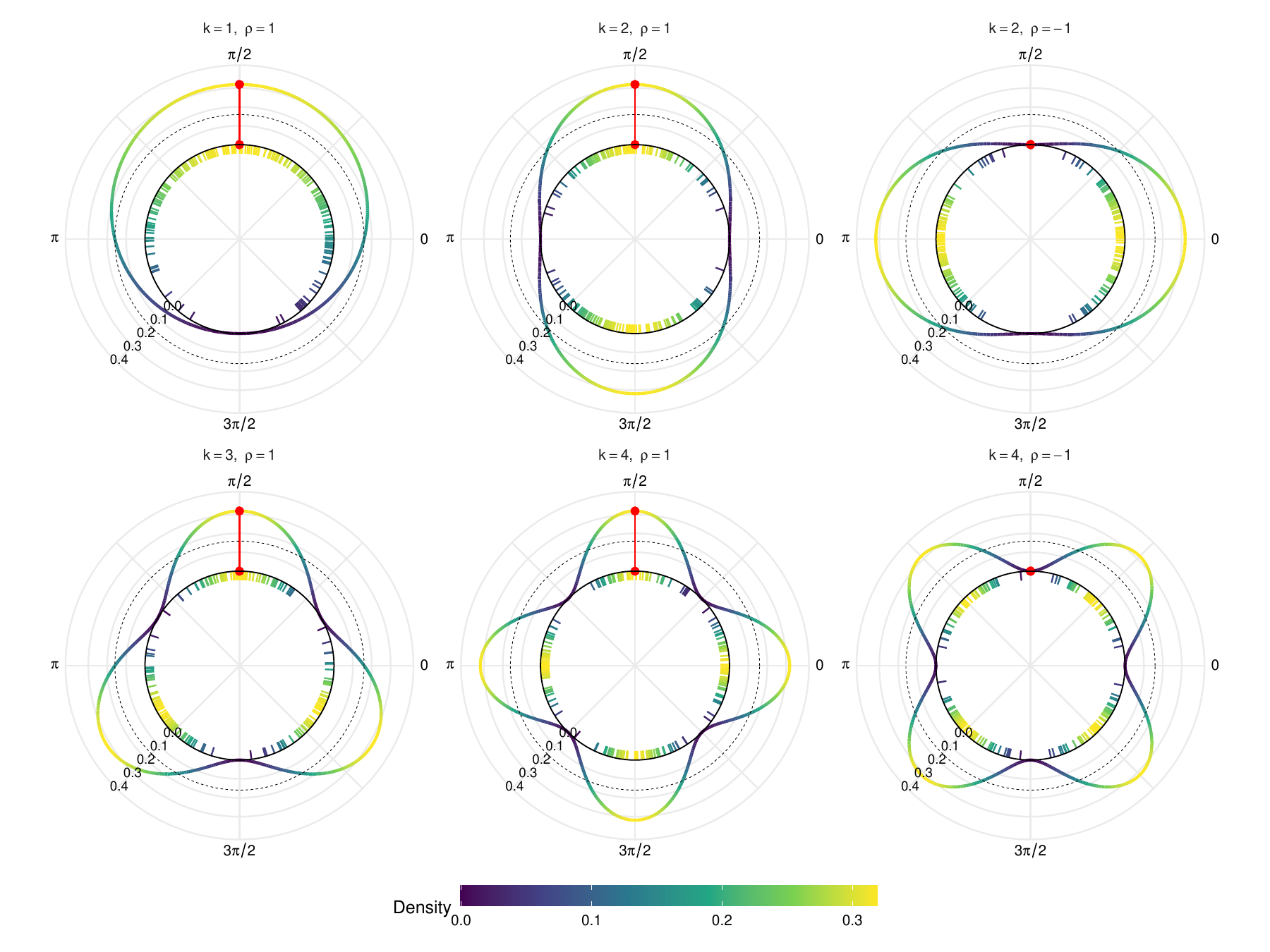}
  \caption{\small Samples and density of the spherical cardioid on $\mathbb{S}^1$ with $\bmu=\be_2$ (red point, north) and $(k,\rho)\in\{(1,1), (2, 1), (2, -1), (3, 1), (4, 1), (4, -1)\}$. For each panel, a random sample of $n=200$ observations is shown. The dashed curve gives the uniform density $1/(2\pi)$ as reference. Negative-$\rho$ panels illustrate overparametrization. \label{fig:cardioid_d1}}
\end{figure}

Figure~\ref{fig:cardioid_d1} shows the shapes of the density \eqref{eq:sphcard} on $\mathbb{S}^1$ when $\bmu=\be_{2}$, $k=1,2,3,4$ and $\rho\in\{-1,1\}$. For $k\geq 1$ and $\rho>0$, there are $k$ modes located at $\mu+2\pi j/k$ and $k$ antimodes (local minima) located at $\mu+2\pi (j+1/2)/k$, $j=0,1,\ldots,k-1$. The density vanishes at the antimodes only if $\rho=1$. The figure illustrates how negative values of $\rho$, also for even $k$, result in non-identifiability.

\begin{figure}[h!]
  \centering
  \includegraphics[width=0.8\textwidth]{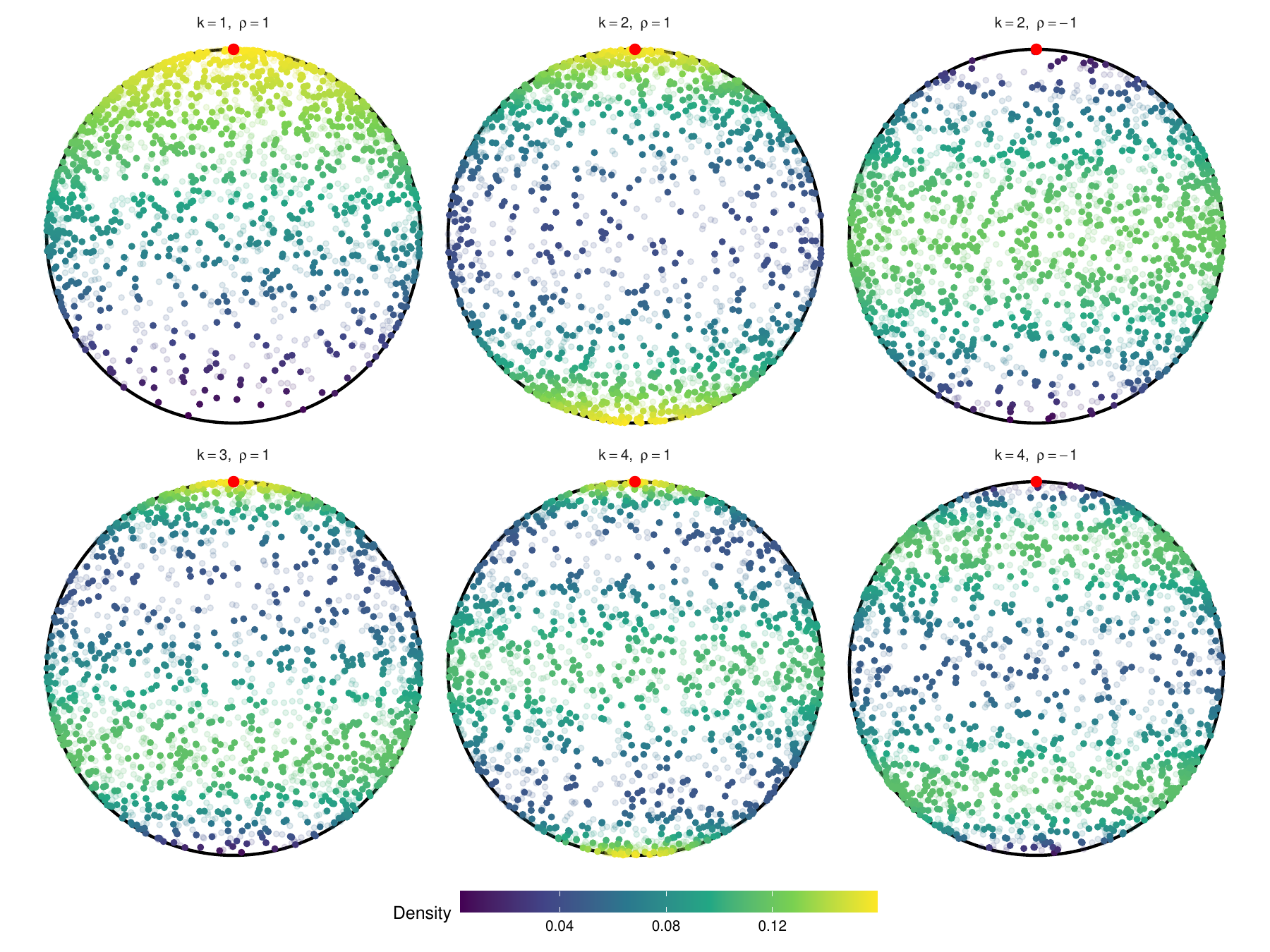}
  \caption{\small Samples and density of the spherical cardioid on $\mathbb{S}^2$ with $\bmu=\be_3$ (red point, north) and $(k,\rho)\in\{(1,1), (2, 1), (2, -1), (3, 1), (4, 1), (4, -1)\}$. The plots show the front hemisphere of $\mathbb{S}^2$, with shading applied to the points in the back hemisphere. The sample, with $n=2000$ observations, is colored according to the value of the density at the observations. \label{fig:cardioid_d2}}
\end{figure}

For $k>1$, the shapes of the spherical cardioid density are multimodal and exhibit multiple girdle-like structures. Figure~\ref{fig:cardioid_d2} shows them on $\mathbb{S}^2$ for $\bmu=\be_{3}$, $k=1,2,3,4$ and $\rho\in\{-1,1\}$. Because $f_{\mathrm{C}_k}(\bx;\bmu,\rho)$ depends on $\bx$ only through $t=\bx^\top\bmu$, its critical sets are either the poles $\pm\bmu$ (corresponding to $t=\pm1$) or latitude subspheres $\{\bx\in\Sd:\bx^\top\bmu=t\}$, creating density ridges or valleys. The $t$-latitudes, $t\in(-1,1)$, of the density ridges/valleys correspond to the zeros of $[C_k^{(d-1)/2}(t)]'=(d-1)C_{k-1}^{(d+1)/2}(t)$, which can be obtained numerically or localized through inequalities \citep[Sec.~18.16(ii)]{NIST}. In general, the location and type of the extrema depend on the parity of $k$ and the sign of $\rho$: (\textit{i}) for $k=1$ and $\rho>0$, there is a single mode at $\bmu$ and a single antimode at $-\bmu$; (\textit{ii}) for $k=2$ and $\rho>0$, there are two antipodal modes at $\pm\bmu$ and one density valley at the equator; (\textit{iii}) for $k\geq 3$ and $\rho>0$, there is a mode at $\bmu$, $\lfloor (k-1)/2\rfloor$ density ridges, and a mode (antimode) at $-\bmu$ if $k$ is even (odd); (\textit{iv}) for even $k$ and $\rho<0$, the poles $\pm\bmu$ are antimodes and there are $k/2$ density ridges, distributed on latitude subspheres symmetric with respect to the equator (with the central ridge at the equator when $k/2$ is odd). These statements hold for all $d\geq 2$.

Just as the von Mises distribution with low concentration is approximately a circular cardioid \citep[see, e.g.,][Eq. (3.5.20)]{Mardia1999a}, the von Mises--Fisher distribution $\mathrm{vMF}(\bmu,\kappa)$ on $\Sd$, $d\geq1$, with density proportional to $\bx\mapsto\exp\{\kappa\bx^\top\bmu\}$, with location $\bmu\in\Sd$ and concentration $\kappa\geq0$, is connected with the spherical cardioid. Precisely,
\begin{align}
  \mathrm{vMF}(\bmu,\kappa)\approx \mathrm{C}_1(\bmu,\kappa)\quad \text{when }\kappa\approx 0, \label{eq:approxM}
\end{align}
as $\exp\{\kappa \bx^\top\bmu\}\approx 1+\kappa \bx^\top\bmu=1+\kappa\,\tilde{C}^{(d-1)/2}_1(\bx^\top\bmu)$. Similarly, the Watson distribution $\mathrm{W}(\bmu,\kappa)$, with density proportional to $\bx\mapsto\exp\{\kappa(\bx^\top\bmu)^2\}$ for $\kappa\in\R$, satisfies
\begin{align}
  \mathrm{W}(\bmu,\kappa)\approx\mathrm{C}_2\lrp{\bmu,\frac{d\kappa}{d+1+\kappa}}\quad \text{when }\kappa\approx 0, \label{eq:approxW}
\end{align}
as $\exp\{\kappa(\bx^\top\bmu)^2\}\approx 1+\kappa(\bx^\top\bmu)^2=
[1+\kappa/(d+1)]\lrbbig{1+[(d\kappa)/(d+1+\kappa)]\tilde{C}^{(d-1)/2}_2(\bx^\top\bmu)}$. Approximations \eqref{eq:approxM}--\eqref{eq:approxW} are useful because it is easier to evaluate the density of $\mathrm{C}_k(\bmu,\kappa)$ and to simulate from it (Section~\ref{sec:simulation}) than from the von Mises--Fisher or Watson distributions. The distribution $\mathrm{C}_1(\bmu,\rho)$ is a special case (when $\psi=1$) of the extension to $\Sd$ of the unimodal family of circular distributions of \cite{Jones2005}. \citet[Sec.~3]{Baringhaus2024} show how rotationally symmetric distributions on $\Sd$ about $\bmu$, particularly the von Mises--Fisher and Cauchy-like distributions, can be expressed in terms of infinite mixtures of $\{\mathrm{C}_k(\bmu,\rho)\}_{k=1}^\infty$, a result related to \eqref{eq:sphharmexp}.

\subsection{Closedness under convolution}
\label{sec:convolution}

%
The circular cardioid family is closed under convolution, in the sense that if $\Theta_1\sim \mathrm{C}_1(\mu_1,\rho_1)$ and $\Theta_2\sim \mathrm{C}_1(\mu_2,\rho_2)$ independently, then $(\Theta_1+\Theta_2)\mod 2\pi \sim \mathrm{C}_1(\mu_1+\mu_2,\rho_1\rho_2/2)$ \citep[see, e.g.,][p. 2, under a slightly different parametrization]{Pewsey2026}. For brevity, we write this result as $\mathrm{C}_1(\mu_1,\rho_1)+\mathrm{C}_1(\mu_2,\rho_2)\stackrel{d}{=}\mathrm{C}_1(\mu_1+\mu_2,\rho_1\rho_2/2)$.

The spherical cardioid family defined by \eqref{eq:sphcard} is also closed under convolution in the following general sense.

\begin{proposition}[Closedness under convolution] \label{prp:conv}
Let $\bmu_1,\bmu_2\in\Sd$, $\rho_1,\rho_2\in[-1,1]$, and $k_1,k_2\geq1$. Then,
\begin{align*}
  \int_{\Sd} f_{\mathrm{C}_{k_1}}(\bx;\bmu_1,\rho_1)f_{\mathrm{C}_{k_2}}(\bx;\bmu_2,\rho_2)\,\sigma_d(\rd\bx)=f_{\mathrm{C}_{k_1}}\lrp{\bmu_1;\bmu_2,\delta_{k_1,k_2}\frac{\rho_1\rho_2}{d_{k_1,d}}}
\end{align*}
and, if $\bX\mid \bXi\sim \mathrm{C}_{k_1}(\bXi,\rho_1)$ and $\bXi\sim \mathrm{C}_{k_2}(\bmu_2,\rho_2)$, then $\bX\sim \mathrm{C}_{k_1}(\bmu_2,\delta_{k_1,k_2} \rho_1\rho_2/d_{k_1,d})$.
\end{proposition}

For $d=1$, using polar coordinates, the previous result readily implies that
\begin{align*}
  (f_{\mathrm{C}_{k_1}}(\cdot;\mu_1,\rho_1)*f_{\mathrm{C}_{k_2}}(\cdot;\mu_2,\rho_2))(\theta)&=\int_{0}^{2\pi} f_{\mathrm{C}_{k_1}}(\theta-\varphi;\mu_1,\rho_1)f_{\mathrm{C}_{k_2}}(\varphi;\mu_2,\rho_2)\,\rd\varphi\\
  &=f_{\mathrm{C}_{k_1}}\lrp{\theta;\mu_1+\mu_2,\delta_{k_1,k_2}\frac{\rho_1\rho_2}{2}}.
\end{align*}
Hence, $\mathrm{C}_{k_1}(\mu_1,\rho_1)+\mathrm{C}_{k_2}(\mu_2,\rho_2)\stackrel{d}{=}\mathrm{C}_{k_1}(\mu_1+\mu_2,\delta_{k_1,k_2}\rho_1\rho_2/2)$, which in particular shows that $\mathrm{Unif}(\mathbb{S}^1)$ can arise as the convolution of two non-uniform cardioid distributions of different orders.

Alternatively, for $d\geq 1$, Proposition \ref{prp:conv} gives the closedness under the Cartesian-coordinates ``convolution'' on $\Sd$
\begin{align*}
    (f_{\mathrm{C}_{k_1}}(\cdot;\cdot,\rho_1)*f_{\mathrm{C}_{k_2}}(\cdot;\bmu,\rho_2))(\bx)\defin&\;\int_{\Sd} f_{\mathrm{C}_{k_1}}(\bx;\by,\rho_1)f_{\mathrm{C}_{k_2}}(\by;\bmu,\rho_2) \,\sigma_d(\rd \by)\\
    =&\; f_{\mathrm{C}_{k_1}}\lrp{\bx;\bmu,\delta_{k_1,k_2}\frac{\rho_1\rho_2}{d_{k_1,d}}}
\end{align*}
considered in, e.g., \cite{Klemela2000} in the context of kernel smoothing on the sphere.

\subsection{Moments}
\label{sec:moments}

%
We compute in this section the vectorized moments of the $\mathrm{C}_k(\bmu,\rho)$ distribution. These moments quantify the similarity of $\mathrm{C}_k(\bmu,\rho)$ to $\mathrm{Unif}(\Sd)$, and are key to deriving the inference approach in Section~\ref{sec:mom}.

We require some matrix notation. We use $\operatorname{vec}:\mathcal{M}_{p,q}\to\R^{pq}$ to denote the vectorization map that stacks the columns of a $p\times q$ matrix into a $pq$-dimensional vector, and $\operatorname{vec}^{-1}_{p,q}$ to denote the inverse map. We denote $\Sym_{d+1,r}$ as the \emph{symmetrizer} $(d+1)^r\times (d+1)^r$ matrix such that
\begin{align*}
    \Sym_{d+1,r}(\bv_{1}\otimes\cdots\otimes\bv_{r})=\frac{1}{r!}\sum_{\sigma\in S_r} \bv_{\sigma(1)}\otimes\cdots\otimes\bv_{\sigma(r)}
\end{align*}
for any collection $\{\bv_{1},\ldots,\bv_{r}\}\subset\R^{d+1}$ and $S_r$ being the set of permutations of $\{1,\ldots,r\}$ \citep[see, e.g.,][p. 95]{Chacon2018}. The symmetrizer matrix satisfies $\Sym_{d+1,r}^2=\Sym_{d+1,r}$ and $\Sym_{d+1,r}^\top=\Sym_{d+1,r}$. The standard Kronecker product of matrices is denoted with $\otimes$, and $\bx^{\otimes r}=\otimes_{j=1}^r \bx$ represents the $r$-fold Kronecker product of the column vector $\bx$.

Our first result shows, in particular, that all the moments of order $m<k$ of $\mathrm{C}_k(\bmu,\rho)$ coincide with those of $\mathrm{Unif}(\Sd)$, and so do the moments of order $m>k$ with $m-k$ odd. The result uses the vectorized uniform moments \eqref{eq:unifmom} given in \citet[Lemma 5]{Chacon2026}.

\begin{theorem}[Vectorized moments] \label{thm:mom}
Let $\bX\sim \mathrm{C}_k(\bmu,\rho)$ for $(\bmu,\rho)\in\Sd\times [-1,1]$, $k\geq 1$, and $d\geq 1$. Let $m\geq 1$. The $m$th vectorized moment of $\bX$, $\E{\bX^{\otimes m}}$, is given by:
\begin{enumerate}[label=(\roman{*})., ref=(\roman{*})]
  \item For $m<k$, $\E{\bX^{\otimes m}}=\bmu_{d+1,m}$, where \label{thm:mom:1}
  \begin{align}
    \bmu_{d+1,m}\defin\Ebig{\bU^{\otimes m}}=\frac{(m-1)!!}{\prod_{r=0}^{m/2-1} (d+1+2r)}\,\Sym_{d+1,m}(\operatorname{vec}\bI_{d+1})^{\otimes m/2}1_{\{m\;\text{even}\}}\label{eq:unifmom}
  \end{align}
  for $\bU\sim\mathrm{Unif}(\Sd)$.
  \item For $m=k$,\label{thm:mom:2}
  \begin{align*}
    \Ebig{\bX^{\otimes k}}=\bmu_{d+1,k}+\frac{\rho}{d_{k,d}}\, \Sym_{d+1,k} \sum_{j=0}^{\lfloor k/2\rfloor} a_{k,j} (\operatorname{vec} \bI_{d+1})^{\otimes j}\otimes \bmu^{\otimes k-2j},
  \end{align*}
  where
  \begin{align*}
    a_{k,j}&=(-1)^j\frac{k!}{2^j(k-2j)!j!}\frac{1}{\prod_{r=1}^{j} (2(k-r)+d-1)},\quad a_{k,0}=1.
  \end{align*}
  \item For $m>k$ and $m-k$ odd, $\Ebig{\bX^{\otimes m}}=\bmu_{d+1,m}$. \label{thm:mom:3}
  \item For general $m\geq 1$, \label{thm:mom:4}
  \begin{align*}
    &\hspace*{-0.5cm}\Ebig{\bX^{\otimes m}}=\bmu_{d+1,m}\\
    &+ \frac{\rho}{C_k^{(d-1)/2}(1)}\frac{\om{d-1}}{\om{d}}\,\Sym_{d+1,m} \sum_{j=0}^{\lfloor m/2\rfloor} e_{j,k,m}\, \bmu^{\otimes (m-2j)}\otimes \lrb{\lrpbig{\operatorname{vec}\bI_{d+1}-\bmu^{\otimes 2}}^{\otimes j}},
  \end{align*}
  where
  \begin{align*}
    e_{j,k,m}\defin&\;\binom{m}{2j} \frac{(2j-1)!!}{\prod_{r=0}^{j-1}(d+2r)} f_{j,m,k} 1_{\{m+k\;\text{even}\}},\\
    f_{j,m,k}\defin&\;\int_{-1}^1 t^{m-2j} (1-t^2)^{d/2-1+j} C_k^{(d-1)/2}(t)\,\rd t\\
    =&\;\frac{\Gamma(d/2+j)}{g_{k,d}}\\
    &\times \sum_{s=0}^{\lfloor k/2\rfloor} (-1)^s\frac{2^{k-2s}\Gamma((d-1)/2+k-s)}{s!(k-2s)!} \frac{\Gamma((m+k+1)/2-j-s)}{\Gamma((d+k+m+1)/2-s)},\\
    g_{k,d}\defin&\;\begin{cases}
      k/2,&d=1,\\
      [\Gamma((d-1)/2)]^{-1}, & d\geq 2.
    \end{cases}
  \end{align*}
\end{enumerate}
\end{theorem}

The next two corollaries exploit Theorem \ref{thm:mom} to provide explicit expressions for particular moments and their covariance matrices, useful for Section~\ref{sec:mom}.

\begin{corollary}[Specific moments]\label{cor:momspecial}
Let $\bX_k\sim \mathrm{C}_k(\bmu,\rho)$ for $(\bmu,\rho)\in\Sd\times [-1,1]$, $k\geq 1$, and $d\geq 1$. Then, the $k$th moments for $k=1,\ldots,4$ are:
\begin{align*}
  \E{\bX_1}=&\;\frac{\rho}{d+1}\bmu,\\
  \Ebig{\bX_2^{\otimes 2}}=&\;\frac{1}{d+1}\operatorname{vec}\bI_{d+1}+\frac{2\rho}{d(d+3)}\lrbBig{\bmu^{\otimes 2}-\frac{1}{d+1}\operatorname{vec}\bI_{d+1}},\\
  \Ebig{\bX_3^{\otimes 3}}=&\;\frac{6\rho}{d(d+1)(d+5)}\Sym_{d+1,3}\lrbBig{\bmu^{\otimes 3}-\frac{3}{d+3}\,\operatorname{vec}\bI_{d+1}\otimes \bmu},\\
  \Ebig{\bX_4^{\otimes 4}}=&\;\frac{3}{(d+1)(d+3)} \Sym_{d+1,4}(\operatorname{vec}\bI_{d+1})^{\otimes 2}\\
  & + \frac{24\rho}{d(d+1)(d+2)(d+7)}\Sym_{d+1,4}\bigg\{\bmu^{\otimes 4}-\frac{6}{d+5}\,\operatorname{vec}\bI_{d+1}\otimes \bmu^{\otimes 2}\\
  &+\frac{3}{(d+3)(d+5)}(\operatorname{vec}\bI_{d+1})^{\otimes 2}\bigg\}.
\end{align*}
\end{corollary}

\begin{corollary}[Covariance matrices of vectorized moments] \label{cor:scatter}
Let $\bX\sim \mathrm{C}_k(\bmu,\rho)$ for $(\bmu,\rho)\in\Sd\times [-1,1]$, $k\geq 1$, and $d\geq 1$. Let $m\geq 1$. The covariance matrix of $\bX^{\otimes m}$ is
\begin{align*}
  \Vbig{\bX^{\otimes m}}=&\;\operatorname{vec}_{(d+1)^{m},(d+1)^{m}}^{-1}\lrp{\Ebig{\bX^{\otimes 2m}}-\Ebig{\bX^{\otimes m}}^{\otimes 2}}.
\end{align*}

If $k$ is odd, then $\Ebig{\bX^{\otimes 2m}}=\bmu_{d+1,2m}$. In particular, for $k=1$,
\begin{align*}
  \Vbig{\bX}=\frac{1}{d+1}\bI_{d+1}-\frac{\rho^2}{(d+1)^2}\bmu\bmu^\top.
\end{align*}
\end{corollary}

As a consequence of the previous result, we have the following Central Limit Theorem (CLT) for the sample $m$th vectorized moment:
\begin{align*}
  \sqrt{n}\lrpbig{\overline{\bX^{\otimes m}} - \Ebig{\bX^{\otimes m}}} \inlaw \mathcal{N}\lrpbig{\mathbf{0}, \Vbig{\bX^{\otimes m}}},
\end{align*}
with relatively explicit forms for $k=m=1$ and $k=m=2$.

\subsection{Characteristic function}
\label{sec:characteristic}

%
The characteristic and moment generating functions, $\varphi_{\bX}(\bt)\defin\Ebig{e^{\mathrm{i}\bt^\top \bX}}$ and $M_{\bX}(\bt)\defin\Ebig{e^{\bt^\top \bX}}$, admit explicit expressions for $\bX\sim \mathrm{C}_k(\bmu,\rho)$, as shown next.

\begin{proposition}[Characteristic and moment generating functions] \label{prp:mgf}
Let $\bX\sim \mathrm{C}_k(\bmu,\rho)$ for $(\bmu,\rho)\in\Sd\times [-1,1]$, $k\geq 1$, and $d\geq 1$. Then, for $\bt\in\R^{d+1}\setminus\{\zero\}$ ($\bt=\zero$ follows by continuity),
\begin{align*}
    M_{\bX}(\bt)=&\;\lrp{\frac{2}{\|\bt\|}}^{(d-1)/2} \!\bigg\{e_{0,d} \,\mathcal{I}_{(d-1)/2}(\|\bt\|) + \frac{\rho}{d_{k,d} } e_{k,d} \,\mathcal{I}_{(2k+d-1)/2}(\|\bt\|)\, C_k^{(d-1)/2}\!\bigg(\frac{\bmu^\top\bt}{\|\bt\|}\bigg)\! \bigg\},
\end{align*}
where $\mathcal{I}_\nu$ is the modified Bessel function of the first kind and order $\nu$, and
\begin{align*}
e_{\ell,d}\defin\begin{cases}
  \Gamma((d-1)/2) (\ell+(d-1)/2), & d\geq2,\\
  2-\delta_{\ell,0}, & d=1.
\end{cases}
\end{align*}
$\varphi_{\bX}(\bt)$ is obtained by replacing $\mathcal{I}_{(2\ell+d-1)/2}(\|\bt\|)$ with $\mathrm{i}^\ell\mathcal{J}_{(2\ell+d-1)/2}(\|\bt\|)$, $\ell=0,k$, in the expression of $M_{\bX}(\bt)$, where $\mathcal{J}_\nu$ is the Bessel function of the first kind and order $\nu$.
\end{proposition}

The explicit form of the characteristic function can be used to derive goodness-of-fit tests based on the squared distance between the population and empirical characteristic functions, as done in \cite{Ebner2024}. We do not pursue such an approach here, and instead focus on a different goodness-of-fit test strategy in Sec.~\ref{sec:gof}.

\subsection{Projected distributions}
\label{sec:projections}

%
When $\bX\sim \mathrm{Unif}(\Sd)$, the distribution of $\bga^\top\bX$ is independent of $\bga$. Its density and cdf are, respectively,
\begin{align}
	f_d(x)&\defin\frac{\om{d-1}}{\om{d}}(1-x^2)^{d/2-1}\label{eq:fd}
\end{align}
and
\begin{align}
	F_d(x)&\defin \frac{\om{d-1}}{\om{d}}\int_{-1}^x(1-t^2)^{d/2-1}\,\mathrm{d}t
	=\frac{1}{2}\left\{1+\mathrm{sign}(x)\mathrm{I}_{x^2}(1/2,d/2)\right\},\label{eq:Fd}
\end{align}
where $x\in[-1,1]$ and $\mathrm{I}_x(a,b):=\mathrm{B}(a,b)^{-1}\int_{0}^xt^{a-1}(1-t)^{b-1}\,\mathrm{d}t$, $a,b>0$, is the regularized incomplete beta function. For $x\in[-1,1]$, $F_{1}(x)=1-\cos^{-1}(x)/\pi$ and $F_{2}(x)=(x+1)/2$. For $d\geq3$ we have that
\begin{align*}
	F_{d}(x)&=F_{d-2}(x) + \frac{x(1-x^2)^{d/2-1}}{(d-2)\mathrm{B}(1/2,(d-2)/2)},
\end{align*}
reducing the evaluation of $F_d(x)$ to polynomials in $x$ and $\{\mathrm{B}(1/2,(d-2k)/2)\}_{k=0}^{\lfloor d/2\rfloor}$.

The following result gives closed-form expressions for the density and cdf of $\bga^\top\bX$ when $\bX\sim \mathrm{C}_k(\bmu,\rho)$ as perturbations of their projected uniform counterparts.

\begin{theorem}[Projected distribution of $\mathrm{C}_k(\bmu,\rho)$] \label{thm:projgamma}
Let $\bX\sim \mathrm{C}_k(\bmu,\rho)$ for $(\bmu,\rho)\in\Sd\times [-1,1]$, $k\geq 1$, and $d\geq 1$. Let $\bga\in\Sd$. Then, the density and cdf of $\bga^\top\bX$ are, respectively,
\begin{align}
    f_{\bga}(x)=f_d(x)\lrb{1+\rho\, \tilde{C}_{k}^{(d-1)/2}(\bga^\top\bmu)\tilde{C}_{k}^{(d-1)/2}(x)}, \label{eq:projpdf}
\end{align}
and
\begin{align}
    F_{\bga}(x)=F_d(x)-\rho\,\eta_k(\bga^\top\bmu) G_k(x),\label{eq:projcdf}
\end{align}
where $x\in[-1,1]$,
\begin{align*}
    \eta_k(\bga^\top\bmu)\defin\frac{\om{d-1}}{\om{d}}\frac{C_{k}^{(d-1)/2}(\bga^\top\bmu)}{\lrcbig{C_k^{(d-1)/2}(1)}^2},
\end{align*}
and
\begin{align*}
    G_k(x)\defin\begin{cases}
    \displaystyle\frac{1}{k}\sin(k\cos^{-1}(x)),&d=1,\\
    \displaystyle\frac{d-1}{k(k+d-1)}C_{k-1}^{(d+1)/2}(x)(1-x^2)^{d/2},&d\geq 2.
    \end{cases}
\end{align*}
\end{theorem}

When $\bga=\bmu$, the density simplifies to $f_{\bmu}(x)=f_d(x)\lrbbig{1+\rho\, \tilde{C}_{k}^{(d-1)/2}(x)}$ and \eqref{eq:projcdf} can be used for simulation, as shown next.

\subsection{Simulation}
\label{sec:simulation}

%
Simulation from $\mathrm{C}_k(\bmu,\rho)$ is immediate by acceptance--rejection sampling from $\mathrm{Unif}(\Sd)$, since
\begin{align*}
  \frac{f_{\mathrm{C}_k}(\bx;\bmu,\rho)}{1/\om{d}}=1+\rho\, \tilde{C}_k^{(d-1)/2}(\bx^\top\bmu)\leq 1+|\rho|\indef M.
\end{align*}
The acceptance probability is $1/M\in[1/2,1]$. Algorithm \ref{algo:sim1} sets out the steps.

\begin{algorithm}
\caption{\small Acceptance--rejection sampling from $\mathrm{C}_k(\bmu,\rho)$\label{algo:sim1}}
\begin{enumerate}
  \item Simulate $\bU\sim\mathrm{Unif}(\Sd)$. \label{step1}
  \item Simulate $V\sim\mathrm{Unif}(0,1)$.
  \item If $V\leq (1+\rho\, \tilde{C}_k^{(d-1)/2}(\bU^\top\bmu))/M$, accept $\bU$, else return to the first step.
\end{enumerate}
\end{algorithm}

Rejection-free simulation can be derived using the rotational symmetry of the model and either the parity of $k$ or Theorem \ref{thm:projgamma}. Because of rotational symmetry, if $\bX\defin T\bmu + \sqrt{1-T^2}\bB_{\bmu}\boldsymbol{\Xi}$, where $\boldsymbol{\Xi}\sim\mathrm{Unif}(\mathbb{S}^{d-1})$ is independent of the random variable $T \sim F_{\bmu}$ on $[-1,1]$ (see Theorem \ref{thm:projgamma}) and $\bB_{\bmu}$ is a semi-orthogonal $(d+1)\times d$ matrix such that $\bmu^\top\bB_{\bmu}=\zero^\top$, then $\bX\sim \mathrm{C}_k(\bmu,\rho)$. For example, for $d=2$ and $k=2$, Theorem \ref{thm:projgamma} gives
\begin{align*}
    F_{\bmu}(x)%
    =\frac{\rho x^3+(2-\rho) x+2}{4},
\end{align*}
which enables inverse transform sampling by solving $T=F_{\bmu}^{-1}(U)$ for $U\sim\mathrm{Unif}(0,1)$. In particular, for $\rho>0$, the cubic equation has an explicit valid root given by $T=\sqrt[3]{-q/2+\sqrt{\Delta}}+\sqrt[3]{-q/2-\sqrt{\Delta}}$, where $q=2(1-2u)/\rho$, $p=(2-\rho)/\rho$, and $\Delta=(q/2)^2+(p/3)^3$.

Due to the cost of inverting the distribution function, the inverse transformation method is rarely computationally more efficient than acceptance--rejection sampling except for $k=1,2$. However, for odd $k\geq 1$, we can borrow from the stochastic representation of skew symmetric distributions \citep[e.g.,][Proposition 1.3]{Azzalini2014} to simulate $\mathrm{C}_k(\bmu,\rho)$ exactly without rejections due to the parity of $f_d$ (see \eqref{eq:fd}) and oddness of $\tilde{C}_k^{(d-1)/2}$. Algorithm \ref{algo:sim2} and the proof of Proposition \ref{prp:algo2} give the details.

\begin{algorithm}
\caption{\small Rejection-free sampling from $\mathrm{C}_k(\bmu,\rho)$ for odd $k$\label{algo:sim2}}
\begin{enumerate}
  \item Simulate $\bU\sim\mathrm{Unif}(\Sd)$.
  \item Set $X=\bU^\top \be_1$ and $R=|X|$.
  \item Simulate $S\in\{-1,1\}$ with $\Prob{S=1\mid R}=\!\lrbbig{1+\rho\, \tilde{C}_k^{(d-1)/2}(R)}/2$ and set $T=SR$.\!\!
  \item Simulate $\boldsymbol{\Xi}\sim\mathrm{Unif}(\mathbb{S}^{d-1})$ independent of $(R,S)$.
  \item Return $\bX=T \bmu + \sqrt{1-T^2}\bB_{\bmu}\boldsymbol{\Xi}$.
\end{enumerate}
\end{algorithm}

\begin{proposition} \label{prp:algo2}
The random vector $\bX$ generated in Algorithm \ref{algo:sim2} is distributed as $\mathrm{C}_k(\bmu,\rho)$.
\end{proposition}

\section{Estimation and inference}
\label{sec:estimation}

%
Let $\bX_1,\ldots,\bX_n$ be an independent and identically distributed (iid) sample from $\mathrm{C}_k(\bmu,\rho)$. We address in this section the estimation of $(\bmu,\rho)$ and asymptotic inferential results.

\subsection{Method of moments}
\label{sec:mom}

%
The moments obtained in Theorem~\ref{thm:mom} can be used for estimation of $(\bmu,\rho)$ through the method of moments. Its applicability, as anticipated by the expressions of the moments in Corollary \ref{cor:momspecial}, depends on the order $k$. We first provide moment estimators of $(\bmu,\rho)$ for the specific orders $k=1,2$. We then follow a different approach to obtain an estimator of $\rho$ for arbitrary $k\geq1$, when $\bmu$ is known, using Gegenbauer moments.

Our first result concerns the case $k=1$, for which $\E{\bX}=[\rho/(d+1)]\bmu,$
and hence $\hat{\bmu}_{\mathrm{MM},1}\defin\bar{\bX}/\|\bar{\bX}\|$ and $\hat{\rho}_{\mathrm{MM},1}\defin (d+1)\|\bar{\bX}\|$ %
are natural moment estimators. The estimators $(\hat{\bmu}_{\mathrm{MM},1},\hat{\rho}_{\mathrm{MM},1})$, computed from the sample $\bX_1,\ldots,\bX_n$, are rotationally equivariant ($\hat{\bmu}_{\mathrm{MM},1}$) and rotationally invariant ($\hat{\rho}_{\mathrm{MM},1}$): if they are computed from the rotated sample $\bO\bX_1,\ldots,\bO\bX_n$, their resulting values are $(\bO\hat{\bmu}_{\mathrm{MM},1},\hat{\rho}_{\mathrm{MM},1})$.

\begin{theorem}[Moment estimators when $k=1$] \label{thm:mm1}
Let $\bX_1,\ldots,\bX_n\sim \mathrm{C}_1(\bmu,\rho)$, for $(\bmu,\rho)\in\Theta_{1,d}^\circ$ and $d\geq 1$. Let
\begin{align*}
  \hat{\bmu}_{\mathrm{MM},1}=\frac{\bar{\bX}}{\|\bar{\bX}\|} \quad\text{and}\quad \hat{\rho}_{\mathrm{MM},1}=(d+1)\|\bar{\bX}\|.
\end{align*}
Then:
\begin{enumerate}[label=(\roman{*})., ref=(\roman{*})]
  \item The estimators are strongly consistent: $(\hat{\bmu}_{\mathrm{MM},1}, \hat{\rho}_{\mathrm{MM},1}) \to (\bmu, \rho)$ a.s. as $n\to\infty$. \label{thm:mm1:1}
  \item The estimators are asymptotically normal: \label{thm:mm1:2}
  \begin{align*}
  \sqrt{n}
  \begin{pmatrix}
    \hat{\bmu}_{\mathrm{MM},1}-\bmu\\
    \hat{\rho}_{\mathrm{MM},1}-\rho
  \end{pmatrix}
  \inlaw \mathcal{N}_{d+2}\lrp{\mathbf{0},
  \begin{pmatrix}
    \sigma_{\mathrm{MM},1}^2(\bmu) (\bI_{d+1} - \bmu\bmu^\top) & \zero\\
    \zero^\top & \sigma^2_{\mathrm{MM},1}(\rho)
  \end{pmatrix}}
  \end{align*}
  as $n\to\infty$, where $\sigma_{\mathrm{MM},1}^2(\bmu)=(d+1)\rho^{-2}$ and $\sigma^2_{\mathrm{MM},1}(\rho)=d+1-\rho^2$.
\end{enumerate}
\end{theorem}

The asymptotic covariance matrix is block-diagonal and singular. The former gives the asymptotic independence of $(\hat{\bmu}_{\mathrm{MM},1},\hat{\rho}_{\mathrm{MM},1})$, as in any rotationally symmetric model, while the latter is a consequence of the constraint $\|\bmu\|=1$, and is ubiquitous in models in directional statistics \citep[see, e.g.,][p. 199]{Mardia1999a}.

Specializing Theorem \ref{thm:mm1}\ref{thm:mm1:2} to $d=1$ and taking polar coordinates, the result in \citet[Eq. (10)]{Pewsey2026} follows after adjusting for the different parametrization of $\rho$.

In the case $k=2$, the first moment of $\mathrm{C}_2(\bmu,\rho)$ is null and the second is required to identify both parameters. By Corollary \ref{cor:momspecial}, we know that
\begin{align*}
  \Ebig{\bX\bX^\top}&=\frac{1}{d+1}\bI_{d+1}+\frac{2\rho}{d(d+3)}\lrbBig{\bmu\bmu^\top-\frac{1}{d+1}\bI_{d+1}}\\
  &= \frac{d+3+2\rho}{(d+1)(d+3)} \bmu\bmu^\top+\frac{(d+3)-2\rho/d}{(d+1)(d+3)}(\bI_{d+1}-\bmu\bmu^\top)\\
  &=: a(\rho) \bmu\bmu^\top+b(\rho)(\bI_{d+1}-\bmu\bmu^\top).
\end{align*}
Note that $a(\rho)>b(\rho)$ for any $\rho>0$, and $a(\rho)<b(\rho)$ for any $\rho<0$. Hence,
\begin{align}
  (\lambda_{\pm},\bu_{\pm})=\lrp{a(\rho), \bmu} \label{eq:eig}
\end{align}
is the first ($+$) or last ($-$) eigenpair of $\bSigma$, depending on the sign of $\rho$.

The eigenpair \eqref{eq:eig} suggests the following estimators based on the sample scatter matrix $\bS=(1/n)\sum_{i=1}^n \bX_i\bX_i^\top$:
\begin{align*}
  \hat{\bmu}_{\mathrm{MM},2}\defin\bu_{\pm}(\bS),\quad \hat{\rho}_{\mathrm{MM},2}\defin\frac{d+3}{2}\left((d+1)\lambda_{\pm}(\bS)-1\right),
\end{align*}
where $\lambda_{+}(\bS)$ is the largest (resp., $\lambda_{-}(\bS)$ is the smallest) eigenvalue of $\bS$ for $\mathrm{C}_2(\bmu,\rho)$ with $\rho>0$ (resp., $\rho<0$) and $\bu_{\pm}(\bS)$ is its associated unit eigenvector. Since the scatter matrix of the rotated sample $\bO\bX_1,\ldots,\bO\bX_n$ is $\bO\bS\bO^\top$, the estimators for the rotated sample are $(\bO\hat{\bmu}_{\mathrm{MM},2},\hat{\rho}_{\mathrm{MM},2})$, showing their rotational equivariance. To avoid testing for the eigenpair with single multiplicity in $\bS$, we assume the sign of $\rho$ is fixed in advance.

\begin{theorem}[Moment estimators when $k=2$]\label{thm:mm2}
Let $\bX_1,\ldots,\bX_n\sim \mathrm{C}_2(\bmu,\rho)$, for $(\bmu,\rho)\in\Theta^\circ_{2,d}$ and $d\geq 1$. Let
\begin{align*}
  \hat{\bmu}_{\mathrm{MM},2}=\bu_{\pm}(\bS)\quad\text{and}\quad\hat{\rho}_{\mathrm{MM},2}=\frac{d+3}{2}\left((d+1)\lambda_{\pm}(\bS)-1\right),
\end{align*}
with $(\hat{\bmu}_{\mathrm{MM},2},\hat{\rho}_{\mathrm{MM},2})\in\Theta_{2,d}^\circ$, where the sign in $\bu_{\pm}(\bS)$ and $\lambda_{\pm}(\bS)$ is chosen according to the sign of $\rho$.
Then:
\begin{enumerate}[label=(\roman{*})., ref=(\roman{*})]
  \item The estimators are strongly consistent: $(\hat{\bmu}_{\mathrm{MM},2}, \hat{\rho}_{\mathrm{MM},2}) \to (\bmu, \rho)$ a.s. as $n\to\infty$. \label{thm:mm2:1}
  \item The estimators are asymptotically normal: \label{thm:mm2:2}
\begin{align*}
\sqrt{n}
\begin{pmatrix}
  \hat{\bmu}_{\mathrm{MM},2}-\bmu \\
  \hat{\rho}_{\mathrm{MM},2}-\rho
\end{pmatrix}
\inlaw \mathcal{N}_{d+2}\lrp{\mathbf{0},
\begin{pmatrix}
    \sigma_{\mathrm{MM},2}^2(\bmu) (\bI_{d+1} - \bmu\bmu^\top) & \zero\\
    \zero^\top & \sigma_{\mathrm{MM},2}^2(\rho)
  \end{pmatrix}}
\end{align*}
as $n\to\infty$, where
\begin{align*}
  \sigma_{\mathrm{MM},2}^2(\bmu)&=\frac{d(d+3)[d(d+5)+2(d-1)\rho]}{4\rho^2(d+1)(d+5)},\\
  \sigma_{\mathrm{MM},2}^2(\rho)&=\frac{d(d+3)}{2}+ \frac{2(d-1)(d+3)}{d+5}\rho -\rho^2.
\end{align*}
For $d=1$, $\sigma_{\mathrm{MM},2}^2(\bmu)=\sigma_{\mathrm{MM},1}^2(\bmu)$ and $\sigma_{\mathrm{MM},2}^2(\rho)=\sigma_{\mathrm{MM},1}^2(\rho)$.
\end{enumerate}
\end{theorem}

When $\bmu$ is known, a natural estimator for $\rho$ is obtained by equating the theoretical and sample Gegenbauer moments. The estimator has the advantage of being straightforward for any integer $k\geq1$ and seamlessly handling positive and negative $\rho$. Its asymptotic variance for $k=1,2$ coincides with that of the method of moments estimators in Theorems \ref{thm:mm1}--\ref{thm:mm2}, but this estimator is unbiased.

\begin{theorem}[Estimation of $\rho$ via Gegenbauer moments]
\label{thm:gegen-moments}
Let $\bX_1,\dots,\bX_n\sim \mathrm{C}_k(\bmu,\rho)$ for $\bmu\in\Sd$ (known), $\rho\in[-1,1]$, and $d\geq 1$. Define the estimator
\begin{align*}
  \hat\rho_{\mathrm{GM}}\defin \frac{\tau_{k,d}}{n} \sum_{i=1}^n C_k^{(d-1)/2}(\bmu^\top\bX_i).
\end{align*}
Then:
\begin{enumerate}[label=(\roman{*})., ref=(\roman{*})]
  \item The estimator is unbiased. \label{thm:gegen-moments:i}
  \item The estimator is strongly consistent: $\hat{\rho}_{\mathrm{GM}} \to \rho$ a.s. as $n\to\infty$. \label{thm:gegen-moments:ii}
  \item The estimator is asymptotically normal: \label{thm:gegen-moments:iii}
  \begin{align*}
    \sqrt{n}\lrp{\hat\rho_{\mathrm{GM}}-\rho}\inlaw \mathcal{N}\lrpbig{0,\sigma^2_{\mathrm{GM},k}(\rho)}
  \end{align*}
  as $n\to\infty$, where
  \begin{align*}
    \sigma^2_{\mathrm{GM},k}(\rho)&= d_{k,d} + \rho\, \eta_{k,d}\, 1_{\{k \text{ even}\}} -\rho^2,
  \end{align*}
  with
  \begin{align*}
    \eta_{k,d}&=\frac{(2k+d-1)^2}{(3k+d-1)(d-1)} \frac{k!}{((k/2)!)^3} \frac{\Gamma((d+k-1)/2)^3 \Gamma(d+3k/2-1)}{\Gamma(d+k-1) \Gamma((d-1)/2)^2 \Gamma((d+3k-1)/2)}
  \end{align*}
  for $d\geq 2$ and $\eta_{k,1}=0$ for $d=1$. For $k=1,2$, $\sigma^2_{\mathrm{GM},k}(\rho)=\sigma^2_{\mathrm{MM},k}(\rho)$.
\end{enumerate}
\end{theorem}

The moment estimators $\hat{\rho}_{\mathrm{MM},k}$, $k=1,2$, and $\hat{\rho}_{\mathrm{GM}}$ may fall outside $[-1,1]$ or $[0,1]$ for finite sample sizes. Hence, in practice they should be truncated to ensure valid parameter estimates. Such truncation preserves consistency; asymptotic normality is preserved too by the delta method if $\rho$ is an interior point.

\subsection{Maximum likelihood}
\label{sec:mle}

%
It is operationally convenient to reparametrize the model in terms of a less constrained parameter $\bxi\defin\rho\bmu\in \{\bx \in \R^{d+1} : \|\bx\| \leq 1\}$, for $\rho>0$. With this parametrization, \eqref{eq:sphcard} becomes
\begin{align}
  f_{\mathrm{C}_k}(\bx;\bxi)\defin\frac{1}{\om{d}}\lrb{1+\|\bxi\| \tilde{C}_k^{(d-1)/2}(\bx^\top\bxi/\|\bxi\|)}\label{eq:cardxi}
\end{align}
for $\bxi\neq \zero$, with the convention that $f_{\mathrm{C}_k}(\bx;\zero)=1/\om{d}$. For $\rho<0$ and even $k$, an analogous reparametrization follows with $f_{\mathrm{C}_k}(\bx;-\bxi)$.

The Maximum Likelihood (ML) estimator of $\bxi$ is
\begin{align}
  \hat{\bxi}_{\mathrm{ML}} \defin \arg\max_{\|\bxi\|\leq 1}\ell_n(\bxi),\quad \ell_n(\bxi)\defin \sum_{i=1}^n \log f_{\mathrm{C}_k}(\bX_i;\bxi), \label{eq:llxi}
\end{align}
with $\ell_n(\bxi)$ denoting the log-likelihood in terms of $\bxi$. It readily follows that $\hat{\bmu}_{\mathrm{ML}}=\hat{\bxi}_{\mathrm{ML}}/\|\hat{\bxi}_{\mathrm{ML}}\|$ and $\hat{\rho}_{\mathrm{ML}}=\|\hat{\bxi}_{\mathrm{ML}}\|$. Maximization of $\bxi\mapsto\ell_n(\bxi)$ can be done by inspecting its local maxima, related with the roots $\hat{\bxi}_n$ of the score equations:
\begin{align*}
  \frac{\partial}{\partial\bxi}\ell_n(\hat{\bxi}_n) = \zero.
\end{align*}
Both $\hat{\bxi}_{\mathrm{ML}}$ and the roots $\hat{\bxi}_n$ are rotationally equivariant, as implied by the rotational equivariance of the density function: $f_{\mathrm{C}_k}(\bO\bx;\bO\bxi)=f_{\mathrm{C}_k}(\bx;\bxi)$ for any orthogonal matrix $\bO$. Hence, the corresponding estimators computed from the rotated sample $\bO\bX_1,\ldots,\bO\bX_n$ are $\bO\hat{\bxi}_{\mathrm{ML}}$ and $\bO\hat{\bxi}_n$.

The following result establishes the existence of a consistent sequence of roots $\hat{\bxi}_n$ that are local maxima of $\bxi\mapsto\ell_n(\bxi)$ and the asymptotic normality of that sequence. The result is stated in terms of $(\bmu,\rho)$ rather than $\bxi$ to ease interpretation.

\begin{theorem}[Maximum likelihood estimation]\label{thm:mle}
Let $\bX_1,\dots,\bX_n\sim \mathrm{C}_k(\bmu,\rho)$ for
\begin{align}
  (\bmu,\rho)\in\Theta_{k,d}^\star\defin\begin{cases}
  \{(\cos(\theta),\sin(\theta))^\top: \theta\in(0,2\pi/k)\}\times(0,\rho_\star), & d=1,\,k\geq 1,\\
  \mathbb{S}^d \times(0,\rho_\star), & d\geq 2,\,k\text{ odd},\\
  \Sd_+\times(0,\rho_\star), & d\geq 2,\,k\text{ even},
  \end{cases} \label{eq:paramspace}
\end{align}
with $0<\rho_\star<1$. Then:
\begin{enumerate}[label=(\roman{*})., ref=(\roman{*})]
  \item The probability that $(\bmu,\rho)\mapsto\ell_n(\bmu,\rho)$ has at least one local maximum tends to one as $n\to\infty$. \label{thm:mle:1}
  \item There exists a sequence of local maxima $(\hat{\bmu}_n,\hat{\rho}_n)$ such that $(\hat{\bmu}_n,\hat{\rho}_n)\to(\bmu,\rho)$ in probability as $n\to\infty$. \label{thm:mle:2}
  \item The sequence $(\hat{\bmu}_n,\hat{\rho}_n)$ is asymptotically normal: \label{thm:mle:3}
  \begin{align*}
  \sqrt{n}
  \begin{pmatrix}
    \hat{\bmu}_n - \bmu \\
    \hat{\rho}_n - \rho
  \end{pmatrix}
  \inlaw \mathcal{N}_{d+2}\lrp{\mathbf{0},\begin{pmatrix}
      \sigma^2_{\mathrm{ML}}(\bmu) (\bI_{d+1} - \bmu\bmu^\top) & \zero\\
      \zero^\top & \sigma^2_{\mathrm{ML}}(\rho)\\
    \end{pmatrix}}
  \end{align*}
  as $n\to\infty$, where, for $d=1$,
  \begin{align*}
    \sigma^2_{\mathrm{ML}}(\rho)&=\frac{\rho^2\sqrt{1-\rho^2}}{1-\sqrt{1-\rho^2}},\quad
    \sigma^2_{\mathrm{ML}}(\bmu)=\frac{1}{k^2(1-\sqrt{1-\rho^2})}
  \end{align*}
  and, for $d\geq 2$,
  \begin{align*}
    [\sigma^2_{\mathrm{ML}}(\rho)]^{-1}&= \frac{\om{d-1}}{\om{d}C_k^{(d-1)/2}(1)}\int_{-1}^1 \frac{C_k^{(d-1)/2}(t)^2(1-t^2)^{d/2-1}}{C_k^{(d-1)/2}(1)+\rho\, C_k^{(d-1)/2}(t)}\,\rd t,\\
    [\sigma^2_{\mathrm{ML}}(\bmu)]^{-1}&= \rho^2\frac{\om{d-1}}{\om{d}C_k^{(d-1)/2}(1)}\frac{(d-1)^2}{d} \int_{-1}^1 \frac{C_{k-1}^{(d+1)/2}(t)^2(1-t^2)^{d/2}}{C_k^{(d-1)/2}(1)+\rho\, C_k^{(d-1)/2}(t)} \,\rd t.
  \end{align*}
  When $d=2$ and $k=1$,
  \begin{align*}
    \sigma^2_{\mathrm{ML}}(\rho)&=\frac{\rho^3}{\tanh^{-1}(\rho)-\rho},\quad \sigma^2_{\mathrm{ML}}(\bmu)= \frac{2\rho}{\rho - (1 - \rho^2)\tanh^{-1}(\rho)}.
  \end{align*}
\end{enumerate}
\end{theorem}

For $d=1$ and $k=1$, Theorem \ref{thm:mle}\ref{thm:mle:2} gives the statements in \citet[p. 5]{Pewsey2026} after adjusting for the different parametrization of $\rho$.
The boundary case $\rho=1$ is excluded from Theorem~\ref{thm:mle}, with \cite{Self1987} indicating a non-Gaussian limit distribution to be expected in that case.

In practice, the ML estimator is obtained by numerically maximizing \eqref{eq:llxi} using standard routines (e.g., \texttt{nlm} or \texttt{optim} in R). Since $\hat{\rho}_{\mathrm{ML}}=|\hat{\bxi}_{\mathrm{ML}}|\leq 1$ by construction, ML estimation does not require truncation. When $k$ is odd, $\rho$ can take either sign; the sign can be determined by comparing the log-likelihood evaluated at $\hat{\bxi}_{\mathrm{ML}}$ and $-\hat{\bxi}_{\mathrm{ML}}$. When $k$ is even, the density \eqref{eq:sphcard} is symmetric, so no sign selection of $\rho$ is needed.

An advantage of the MM over ML estimators is the direct computation of the former, without requiring numerical optimization. Appendix~\ref{sec:ares} of the SM contains the asymptotic relative efficiencies of the MM estimators relative to the ML estimators.

\section{Goodness-of-fit}
\label{sec:gof}

%
We now turn our attention to goodness-of-fit testing for the $\mathrm{C}_k(\bmu,\rho)$ distribution. Given an iid sample $\bX_1,\ldots,\bX_n\sim \mathrm{P}$, we aim at testing the spherical cardioidness of $\mathrm{P}$ through the null hypothesis
\begin{align*}
  H_0\colon\mathrm{P}\in\lrb{\mathrm{C}_k(\bmu,\rho):(\bmu,\rho)\in\Sd\times[-1,1]},
\end{align*}
for a fixed $k\geq 1$. We build on the integrated projected-ecdf approach employed in \cite{Garcia-Portugues2020b} by using a test statistic that integrates, across projecting directions, the quadratic discrepancies between the null cumulative distribution function (cdf) of the projected data and the empirical cdf (ecdf) of these projections.

Let $F_{\bga}$ be the cdf of the spherical cardioid distribution projected along direction $\bga\in\Sd$. That is, if $\bX\sim \mathrm{C}_k(\bmu,\rho)$, then $T=\bga^\top \bX\sim F_{\bga}$. Let $F_{n,\bga}(x)=(1/n)\sum_{i=1}^n 1_{\{\bga^\top \bX_i \leq x\}}$ denote the ecdf of the projected data $\{\bga^\top \bX_i\}_{i=1}^n$ and $\hat{F}_{\bga}$ the projected cdf $F_{\bga}$ featuring the estimates $(\hat\bmu,\hat\rho)$ obtained under $H_0$. We consider the test statistic
\begin{align}
    P_{n}^{W,\lambda}\defin n\int_{\Sd}\int_{-1}^1 (F_{n,\bga}(x)-\hat{F}_{\bga}(x))^2\,\rd W(\hat{F}_{\bga}(x))\,\lambda(\rd \bga),\label{eq:projstat}
\end{align}
where $\lambda$ is a distribution on $\Sd$ and $W$ a measure on $[0,1]$. Large values of $P_{n}^{W,\lambda}$ provide evidence against $H_0$.

\subsection{Test statistic}
\label{sec:teststat}

%
We discuss two main choices for the distribution $\lambda$ in \eqref{eq:projstat}. First, let us consider $\bga\sim\mathrm{Unif}(\Sd)$, that is, $\lambda=\sigma_d/\om{d}$. This is the choice in \cite{Garcia-Portugues2020b} for testing $H_0\colon \mathrm{P}=\mathrm{Unif}(\Sd)$ and naturally aggregates deviations from the null hypothesis for all possible directions. It is also a conservative choice, as it does not favor any particular direction to detect departures from the null hypothesis. The test statistic becomes
\begin{align}
    P_n^{W,\,\mathrm{Unif}}\defin\frac{n}{\om{d}}\int_{\Sd}\int_{-1}^1 (F_{n,\bga}(x)-\hat{F}_{\bga}(x))^2\,\rd W(\hat{F}_{\bga}(x))\,\sigma_d(\rd \bga).\label{eq:Unstat}
\end{align}

Alternatively, let us consider $\bga$ distributed as the empirical distribution $\mathrm{P}_n$ of the sample $\bX_1,\ldots,\bX_n$:
\begin{align}
    P_n^{W,\,\mathrm{P}_n}&\defin n\int_{\Sd}\int_{-1}^1 (F_{n,\bga}(x)-\hat{F}_{\bga}(x))^2\,\rd W(\hat{F}_{\bga}(x))\,\mathrm{P}_n(\rd\bga)\nonumber\\
    &=\frac{1}{n}\sum_{i=1}^n n\int_{-1}^1 (F_{n,\bX_i}(x)-\hat{F}_{\bX_i}(x))^2\,\rd W(\hat{F}_{\bX_i}(x)).\label{eq:Pnstatn}
\end{align}
This choice weights the directions to detect departures from the null hypothesis according to $\mathrm{P}$. Its main advantage is the simpler expression it provides for the test statistic. The version of \eqref{eq:Pnstatn} with $P_n^{W,\,\mathrm{P}_n}$ is equivalent to \eqref{eq:Unstat} for testing uniformity, but not necessarily for testing spherical cardioidness.

A third alternative is to consider $\lambda$ as $\mathrm{C}_k(\bmu,\rho)$, i.e., weighting projecting directions according to the null distribution. This statistic, $P_n^{W,\,\mathrm{C}_k}$, is aligned with the $\rd W(\hat{F}_{\bga}(x))$ weight but gives a test statistic at least as complex as \eqref{eq:Unstat}.

If $(\hat{\bmu},\hat{\rho})$ are rotation-equivariant estimators, as those in Sec.~\ref{sec:estimation} are, then both statistics \eqref{eq:Unstat} and \eqref{eq:Pnstatn} are rotationally invariant. To see this, consider an orthogonal matrix $\bO$ and the rotated sample $\tilde{\bX}_i=\bO\bX_i$, $i=1,\ldots,n$. We denote with tilde the objects computed from the rotated sample, and set $\tilde{\bga}\defin \bO^\top\bga$. Then
\begin{align*}
  \tilde{F}_{n,\bga}(x)
  =\frac{1}{n}\sum_{i=1}^n 1_{\{\bga^\top\tilde{\bX}_i\leq x\}}
  =\frac{1}{n}\sum_{i=1}^n 1_{\{(\bO^\top\bga)^\top\bX_i\leq x\}}
  =F_{n,\tilde{\bga}}(x)
\end{align*}
and, because of Theorem \ref{thm:projgamma},
\begin{align}
  \tilde{\hat{F}}_{\bga}(x)
  &=F_d(x)-\tilde{\hat{\rho}}\eta_k(\bga^\top\tilde{\hat{\bmu}})\, G_k(x)
  =F_d(x)-\hat{\rho}\eta_k((\bO^\top\bga)^\top\hat{\bmu})\, G_k(x)
  =\hat{F}_{\tilde{\bga}}(x).\label{eq:tildeF}
\end{align}
Therefore,
\begin{align*}
  \tilde{P}^{W,\lambda}_n=&\;n\int_{\Sd}\int_{-1}^1 (\tilde{F}_{n,\bga}(x)-\tilde{\hat{F}}_{\bga}(x))^2\,\rd W(\tilde{\hat{F}}_{\bga}(x))\,\lambda(\rd \bga)\\
  =&\;n\int_{\Sd}\int_{-1}^1 (F_{n,\bO^\top\bga}(x)-\hat{F}_{\bO^\top\bga}(x))^2\,\rd W(\hat{F}_{\bO^\top\bga}(x))\,\lambda(\rd \bga),
\end{align*}
implying that $\tilde{P}^{W,\,\mathrm{Unif}}_n=P^{W,\,\mathrm{Unif}}_n$. This argument actually shows that the uniform distribution is the only distribution $\lambda$ independent from the sample that makes $P^{W,\lambda}_n$ rotation-invariant. Similarly to \eqref{eq:tildeF}, $\tilde{F}_{n,\tilde{\bX}_i}(x)=F_{n,\bX_i}(x)$ and $\tilde{\hat{F}}_{\tilde{\bX}_i}(x)=\hat{F}_{\bX_i}(x)$, so
\begin{align*}
  \tilde{P}^{W,\mathrm{P}_n}_n=\frac{1}{n}\sum_{i=1}^n n\int_{-1}^1 (\tilde{F}_{n,\tilde{\bX}_i}(x)-\tilde{\hat{F}}_{\tilde{\bX}_i}(x))^2\,\rd W(\tilde{\hat{F}}_{\tilde{\bX}_i}(x))%
  =&\;P^{W,\mathrm{P}_n}_n.
\end{align*}
The test statistic with $\lambda$ as $\mathrm{C}_k(\hat\bmu,\hat\rho)$ is also rotationally invariant.

The evaluation of $P_n^{W,\lambda}$ can be carried out using the well-known closed-form expressions for the Cram\'er--von Mises (CvM) and Anderson--Darling (AD) statistics. Specifically, the weights $\rd W^{\mathrm{CvM}}(u)=\rd u$ and $\rd W^{\mathrm{AD}}(u)=[u(1-u)]^{-1}\,\rd u$, $u\in[0,1]$, yield the CvM and AD statistics, respectively:
\begin{align}
  W_{n,\bga}^2&=n\int_{-1}^1 (F_{n,\bga}(x)-\hat{F}_{\bga}(x))^2\,\rd \hat{F}_{\bga}(x)=\sum_{i=1}^n \lrb{U_{(i)}^{(\bga)}-\frac{2i-1}{2n}}^2+\frac{1}{12n},\nonumber\\
  A_{n,\bga}^2&=n\int_{-1}^1 \frac{(F_{n,\bga}(x)-\hat{F}_{\bga}(x))^2}{\hat{F}_{\bga}(x)(1-\hat{F}_{\bga}(x))}\,\rd \hat{F}_{\bga}(x)\nonumber\\
  &=-n-\frac{1}{n}\sum_{i=1}^n \lrb{(2i-1)\log\lrpbig{U_{(i)}^{(\bga)}}+(2(n-i)+1)\log\lrpbig{1-U_{(i)}^{(\bga)}}},\label{eq:AD}
\end{align}
where $\{U_{(i)}^{(\bga)}\}_{i=1}^n$ are the ($\bga$-dependent) ordered values of $U_{i}^{(\bga)}=\hat{F}_{\bga}(\bga^\top\bX_i)$, $i=1,\ldots,n$. These expressions can be combined with a Monte Carlo approximation of the integral over $\Sd$ in \eqref{eq:Unstat}: given a Monte Carlo sample $\bga_1,\ldots,\bga_K\sim\lambda$, then
\begin{align}
    \hat{P}_n^{\mathrm{CvM},\lambda}=\frac{1}{K}\sum_{j=1}^K W_{n,\bga_j}^2\quad\text{and}\quad \hat{P}_n^{\mathrm{AD},\lambda}=\frac{1}{K}\sum_{j=1}^K A_{n,\bga_j}^2\label{eq:MCstat}
\end{align}
efficiently approximate $P_n^{\mathrm{CvM},\lambda}$ and $P_n^{\mathrm{AD},\lambda}$. If using $\mathrm{P}_n$ as $\lambda$, then \eqref{eq:Pnstatn} is exactly computable as
\begin{align}
    P_n^{\mathrm{CvM},\mathrm{P}_n}=\frac{1}{n}\sum_{i=1}^n W_{n,\bX_i}^2\quad\text{and}\quad P_n^{\mathrm{AD},\mathrm{P}_n}=\frac{1}{n}\sum_{i=1}^n A_{n,\bX_i}^2.\label{eq:PnCvMAD}
\end{align}
For the $P_n^{\mathrm{AD},\mathrm{P}_n}$ statistic, care must be taken because $U_i^{(\bX_i)}=U_{(n)}^{(\bX_i)}=1$, which causes $A_{n,\bX_i}^2=\infty$, $i=1,\ldots,n$. A simple solution is to omit the addend $i=n$ in \eqref{eq:AD}.

The exact computational form for $P_n^{W,\lambda}$ for a general weight $W$ and distribution $\lambda$ is more involved. The next result provides a $V$-statistic expression for $P_n^{W,\lambda}$ for a finite measure $W$ and distribution $\lambda$. The follow-up corollary covers the Anderson--Darling weight, which is not integrable.

\begin{theorem}[$V$-statistic form for $P_n^{W,\lambda}$] \label{thm:stat}
Let $\lambda$ be a distribution on $\Sd$ and $W$ a finite measure on $[0,1]$. Denote $W(u)\equiv W([0,u])$, $u\in[0,1]$, $W_1(u)\defin\int_{0}^u v\,\rd W(v)$, and $W_2(u)\defin\int_{0}^u v^2\,\rd W(v)$. Then,
\begin{align}
    P_n^{W,\lambda}=&\;n(W(1)+W_2(1)-W_2(0)-2W_1(1))\nonumber\\
    &+2\sum_{i=1}^n\mathrm{E}_{\bga}\left[W_1(\hat{F}_{\bga}(\bga^\top\bX_i))\right] -\frac{1}{n}\sum_{i, j=1}^n\mathrm{E}_{\bga}\left[W(\hat{F}_{\bga}(\max(\bga^\top\bX_i,\bga^\top\bX_j)))\right].\label{eq:UnstatU}%
\end{align}
\end{theorem}

\begin{corollary}[$V$-statistic form for $P_n^{\mathrm{AD},\lambda}$] \label{cor:adstat}
Let $|\rho|<1$. Let $\lambda$ be an absolutely continuous distribution on $\Sd$ with bounded density. Then,
\begin{align*}
    P_n^{\mathrm{AD},\lambda}=&-n -2\sum_{i=1}^n\mathrm{E}_{\bga}\left[\log(1-\hat{F}_{\bga}(\bga^\top\bX_i))\right]\\
    & -\frac{1}{n}\sum_{i, j=1}^n\mathrm{E}_{\bga}\bigg[\log\lrpbigg{\frac{\hat{F}_{\bga}(\max(\bga^\top\bX_i,\bga^\top\bX_j))}{1-\hat{F}_{\bga}(\max(\bga^\top\bX_i,\bga^\top\bX_j))}}\bigg].
\end{align*}
\end{corollary}

The previous results are not computationally advantageous with respect to \eqref{eq:projstat}, as they require evaluating the expectations with respect to $\bga\sim \lambda$, which are complicated because of their dependence on the distribution of $(\bga^\top\bmu,\bga^\top\bX_i,\bga^\top\bX_j)$. We therefore focus on obtaining closed forms for specific cases of the statistic $P_n^{\mathrm{CvM},\,\mathrm{Unif}}$.

\begin{theorem}[Computational form for $P_n^{\mathrm{CvM},\,\mathrm{Unif}}$] \label{thm:cvmstat}
For the CvM weight and $\bga\sim\mathrm{Unif}(\Sd)$, the test statistic in \eqref{eq:UnstatU} is
\begin{align}
  P_n^{\mathrm{CvM},\,\mathrm{Unif}}=\frac{3-2n}{6}-\sum_{i=1}^n\varphi(\bX_i) +\frac{2}{n}\sum_{i<j} \psi(\bX_i,\bX_j). \label{eq:Ustatcvm}
\end{align}
Denote $\mu_i\defin\bX_i^\top\bmu$ and $t_{ij}\defin \bX_i^\top\bX_j$. The kernels $\varphi$ and $\psi$ are as follows:
\begin{enumerate}[label=(\roman{*})., ref=(\roman{*})]
  \item If $d=1$ and $k\geq 1$, \label{thm:cvmstat:i}
  \begin{align*}
    \varphi(\bX_i)=&\;\frac{\hat{\rho}}{2\pi^2 k^2}\lrb{T_k(\hat{\mu}_i)-\frac{\hat{\rho}}{4}(2-T_{2k}(\hat{\mu}_i))},\\
    \psi(\bX_i,\bX_j)=&\;\psi^\mathrm{CvM}_1(\cos^{-1}(t_{ij}))\\
  &-\hat{\rho}\lrb{\frac{\pi-\cos^{-1}(t_{ij})}{2\pi^2 k}T_k\lrp{\frac{\hat{\mu}_i+\hat{\mu}_j}{\sqrt{2(1+t_{ij})}}} \sin\lrp{\frac{k\cos^{-1}(t_{ij})}{2}}}.
  \end{align*}
  \item If $d=2$ and $k=1$, \label{thm:cvmstat:ii}
  \begin{align*}
    \varphi(\bX_i)=&\;\frac{\hat{\rho}}{30}\hat{\mu}_i-\frac{\hat{\rho}^2}{2}\lrb{\frac{1}{35}-\frac{2\hat{\mu}_i^2}{105}},\\
    \psi(\bX_i,\bX_j)=&\;\psi^\mathrm{CvM}_2(\cos^{-1}(t_{ij})) -\frac{\hat{\rho}}{32} \sqrt{\frac{1-t_{ij}}{2}}(\hat{\mu}_i+\hat{\mu}_j).
  \end{align*}
  \item If $d=2$ and $k=2$, \label{thm:cvmstat:iii}
  \begin{align*}
    \varphi(\bX_i)=&\;\frac{\hat{\rho}}{420}(3\hat{\mu}_i^2-1)-\frac{\hat{\rho}^2}{4}\lrb{\frac{1}{330}+\frac{3\hat{\mu}_i^2}{385}-\frac{\hat{\mu}_i^4}{110}},\\
    \psi(\bX_i,\bX_j)=&\;\psi^\mathrm{CvM}_2(\cos^{-1}(t_{ij}))\\
    &+\frac{\hat{\rho}}{128}\sqrt{\frac{1-t_{ij}}{2}}\Bigg\{\frac{1+t_{ij}}{2}+\frac{3(3t_{ij}-1)}{4(1-t_{ij})}\lrcbig{\hat{\mu}_{i}^2+\hat{\mu}_{j}^2} +\frac{3(t_{ij}-3)}{2(1-t_{ij})}\hat{\mu}_{i} \hat{\mu}_{j}\Bigg\}.
  \end{align*}
\end{enumerate}
In the above expressions,
\begin{align*}
  \psi^\mathrm{CvM}_d(\theta) = \begin{cases}
    \displaystyle\frac{1}{2}+\frac{\theta}{2\pi}\left(\frac{\theta}{2\pi}-1\right), & d=1,\\
    \displaystyle\frac{1}{2}-\frac{1}{4}\sin\lrp{\frac{\theta}{2}}, & d=2.
    \end{cases}
\end{align*}
\end{theorem}

Evaluating \eqref{eq:Ustatcvm} exactly has computational complexity order $O(n^2)$. The Monte Carlo approximations in \eqref{eq:MCstat} have order $O(K n\log n)$, and \eqref{eq:PnCvMAD} have order $O(n^2\log n)$.

\subsection{Bootstrapping}
\label{sec:boot}

%
The projection-based test statistics $P_n^{W,\lambda}$ can be readily used to assess the goodness-of-fit of $\mathrm{C}_k(\bmu,\rho)$ through parametric bootstrapping. The procedure is standard and builds on the contents of Secs. \ref{sec:simulation}, \ref{sec:estimation}, and \ref{sec:teststat}. It is detailed in Algorithm \ref{algo:boot}.

\begin{algorithm}
\caption{\small Parametric bootstrap for testing the goodness-of-fit of $\mathrm{C}_k(\bmu,\rho)$\label{algo:boot}}
\begin{enumerate}
  \item Compute $(\hat\bmu,\hat\rho)$ from $\bX_1,\ldots,\bX_n$ using any of the estimators in Sec.~\ref{sec:estimation}.\label{stepb1}
  \item Compute $P_n^{W,\lambda}$ from $\bX_1,\ldots,\bX_n$ and $(\hat\bmu,\hat\rho)$, either from the exact forms \eqref{eq:Ustatcvm} and \eqref{eq:PnCvMAD}, or the Monte Carlo approximation in \eqref{eq:MCstat}. \label{stepb2}
  \item For $b=1,\ldots,B$:
  \begin{enumerate}
      \item Simulate $\bX_1^{*b},\ldots,\bX_n^{*b}$ from $\mathrm{C}_k(\hat\bmu,\hat\rho)$ using Algorithm~\ref{algo:sim1} or~\ref{algo:sim2}.\label{stepb3a}
      \item Compute $(\hat\bmu^{*b},\hat\rho^{*b})$ from $\bX_1^{*b},\ldots,\bX_n^{*b}$. \label{stepb3b}
      \item Compute $P_n^{W,\lambda,*b}$ from $\bX_1^{*b},\ldots,\bX_n^{*b}$ and $(\hat\bmu^{*b},\hat\rho^{*b})$ as in step \ref{stepb2}. \label{stepb3c}
  \end{enumerate}
  \item Obtain the bootstrap $p$-value approximation
  \begin{align*}
  p\text{-value}\approx\frac{1+\sum_{b=1}^B1_{\{P_n^{W,\lambda,*b}> P_n^{W,\lambda}\}}}{B+1}.
  \end{align*}
\end{enumerate}
\end{algorithm}

The algorithm can be easily adapted to test the simple null hypothesis $H_0\colon \mathrm{P}=\mathrm{C}_k(\bmu_0,\rho_0)$ for fixed $(\bmu_0,\rho_0)$. In this case, simply replace $(\hat\bmu,\hat\rho)$ and $(\hat\bmu^{*b},\hat\rho^{*b})$ with $(\bmu_0,\rho_0)$, and skip the estimation steps. If $\rho_0=0$, then the simple null hypothesis corresponds to uniformity and $P_n^{\mathrm{CvM},\,\mathrm{Unif}}=P_n^{\mathrm{CvM},\,\mathrm{C}_k}$ and $P_n^{\mathrm{AD},\,\mathrm{Unif}}=P_n^{\mathrm{AD},\,\mathrm{C}_k}$ coincide with the projected CvM and AD statistics in \cite{Garcia-Portugues2020b}.

Percentile bootstrap $100(1-\alpha)\pct$-confidence intervals for $(\bmu,\rho)$ can be obtained as a direct by-product of running Algorithm \ref{algo:boot}. The percentile bootstrap confidence interval for $\rho$ is $(\hat{\rho}^{*(\lceil (B+1)\alpha/2 \rceil)},\hat{\rho}^{*(\lfloor (B+1)(1-\alpha/2)\rfloor)})$, while, for odd $k$, a percentile bootstrap confidence spherical cap for $\bmu$ is $\{\bmu\in\Sd:\bmu^\top \hat{\bmu}\geq t^{*(\lceil \alpha (B+1)\rceil)}\}$, with $t^{*b}\defin\hat{\bmu}^{*b \top} \hat{\bmu}$, $b=1,\ldots,B$. For even $k$, the confidence region is the union of two spherical caps: $\{\bmu\in\Sd:|\bmu^\top \hat{\bmu}|\geq |t|^{*(\lceil \alpha (B+1)\rceil)}\}$, with $|t|^{*b}\defin|\hat{\bmu}^{*b \top} \hat{\bmu}|$.

\section{Numerical experiments}
\label{sec:experiments}

\subsection{Asymptotic distributions of estimators}
\label{sec:experiments:asymp}

%
We validated empirically the asymptotic distributions of $(\hat{\bmu}_{\mathrm{MM},k},\hat{\rho}_{\mathrm{MM},k})$, $k=1,2$, $\hat{\rho}_{\mathrm{GM}}$, and $(\hat{\bmu}_{\mathrm{ML}},\hat{\rho}_{\mathrm{ML}})$ obtained in Sec.~\ref{sec:estimation}. We considered $d=2$ and the models $\mathrm{C}_1(\bmu,\rho)$ and $\mathrm{C}_2(\bmu,\rho)$ with $\bmu=(0, 0, 1)^\top$ and $\rho=0.5$. For each model, we simulated $M=10^4$ samples of size $n=1000$, computed the estimators $\{(\hat{\bmu}^{(j)},\hat{\rho}^{(j)})\}_{j=1}^M$, and constructed their standardized versions $\{\sqrt{n}(\hat{\mu}^{(j)}_1-\mu_1)\}_{j=1}^M$ (first entries of the estimators) and $\{\sqrt{n}(\hat{\rho}^{(j)}-\rho)\}_{j=1}^M$.

\begin{figure}[h!]
\centering
\begin{subfigure}{0.45\textwidth}
    \centering
    \includegraphics[width=0.5\textwidth]{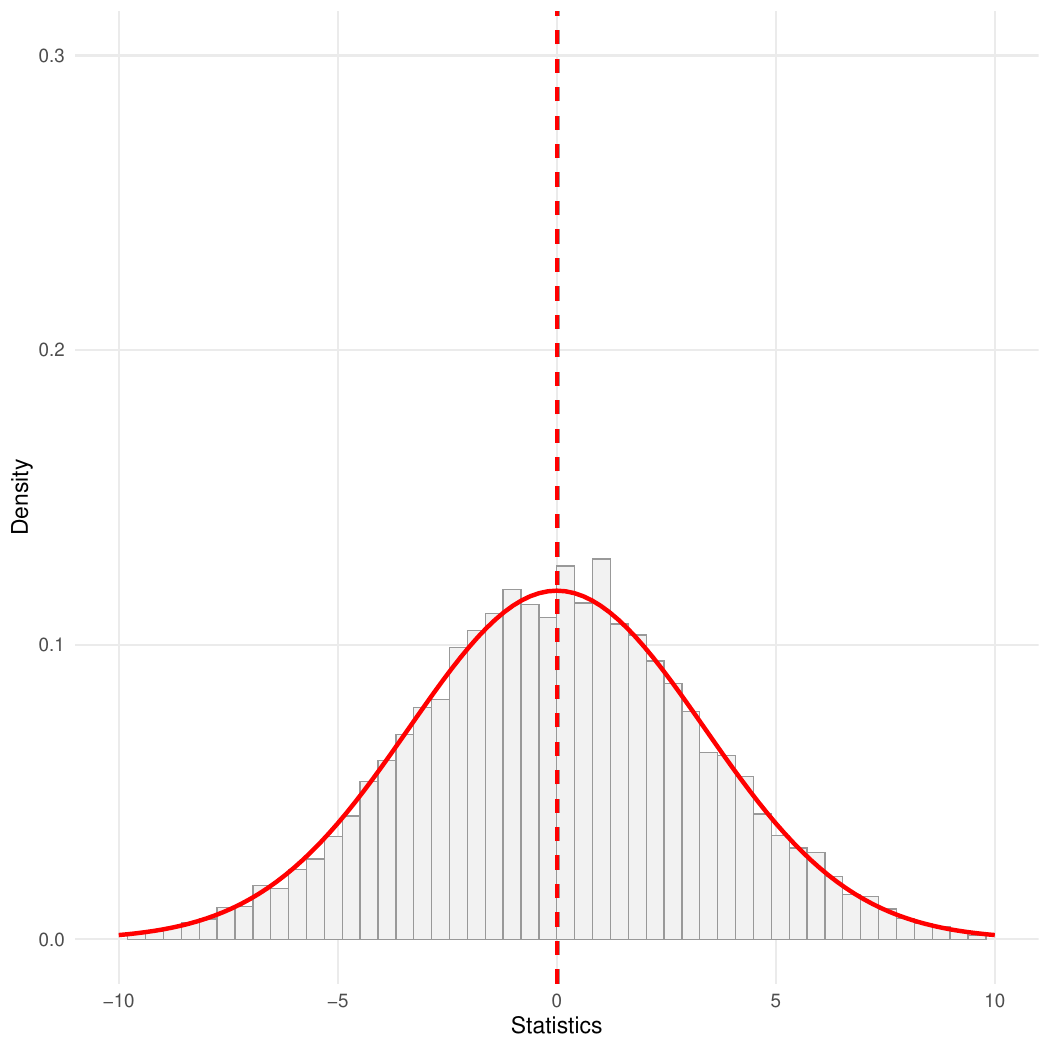}\includegraphics[width=0.5\textwidth]{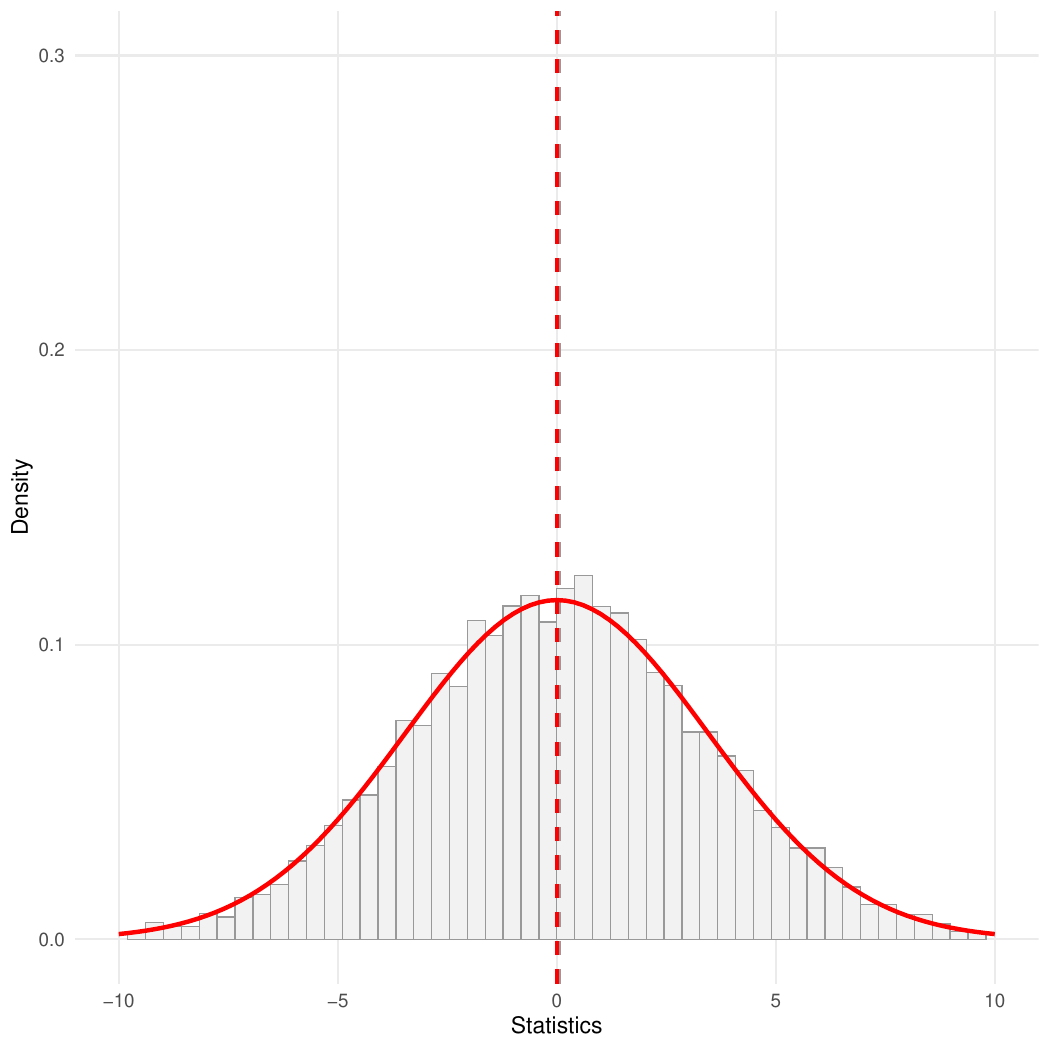}
    \caption{\small $\sqrt{n}(\hat{\mu}^{(j)}_1-\mu_1)$ for $k=1$}
    \label{fig:muk1}
\end{subfigure}
\hspace{0.5cm}
\begin{subfigure}{0.45\textwidth}
    \includegraphics[width=0.5\textwidth]{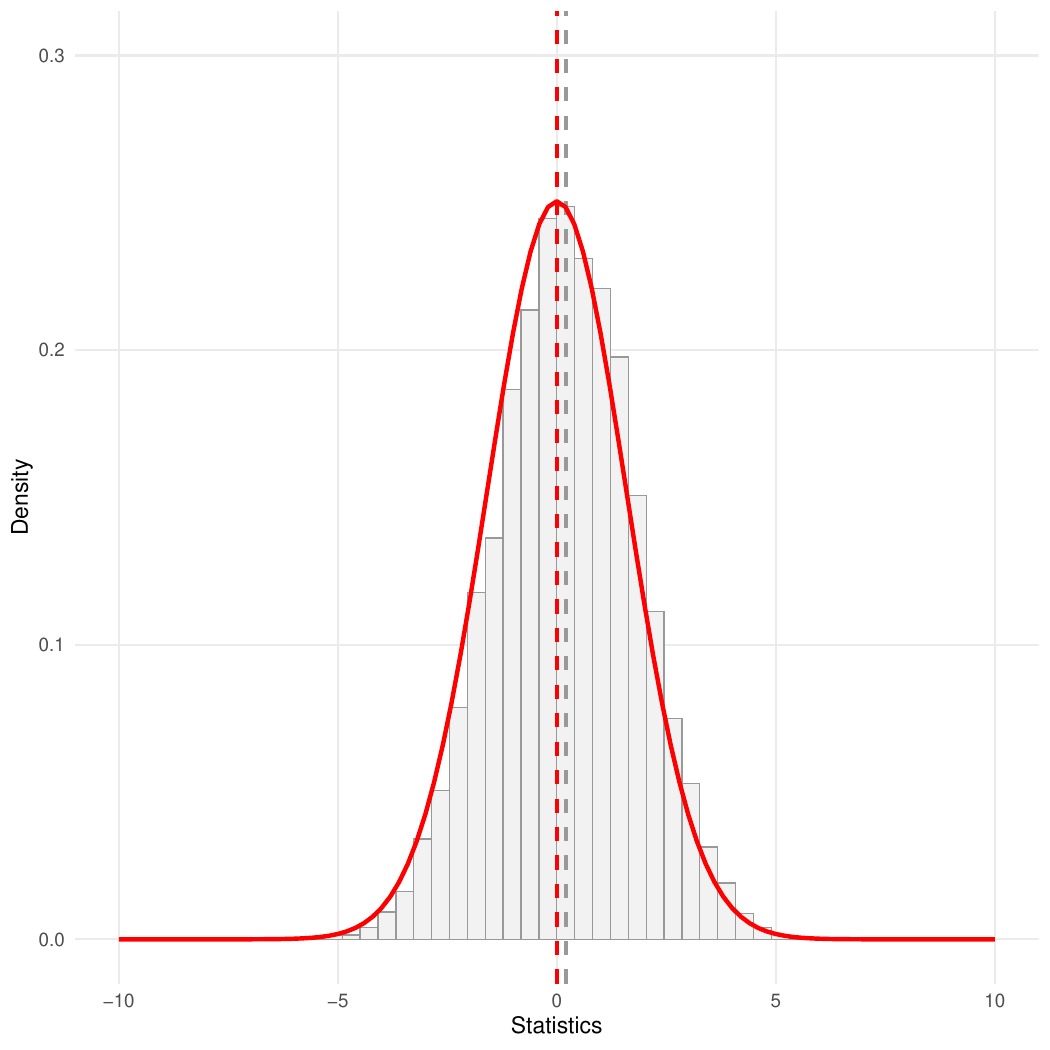}\includegraphics[width=0.5\textwidth]{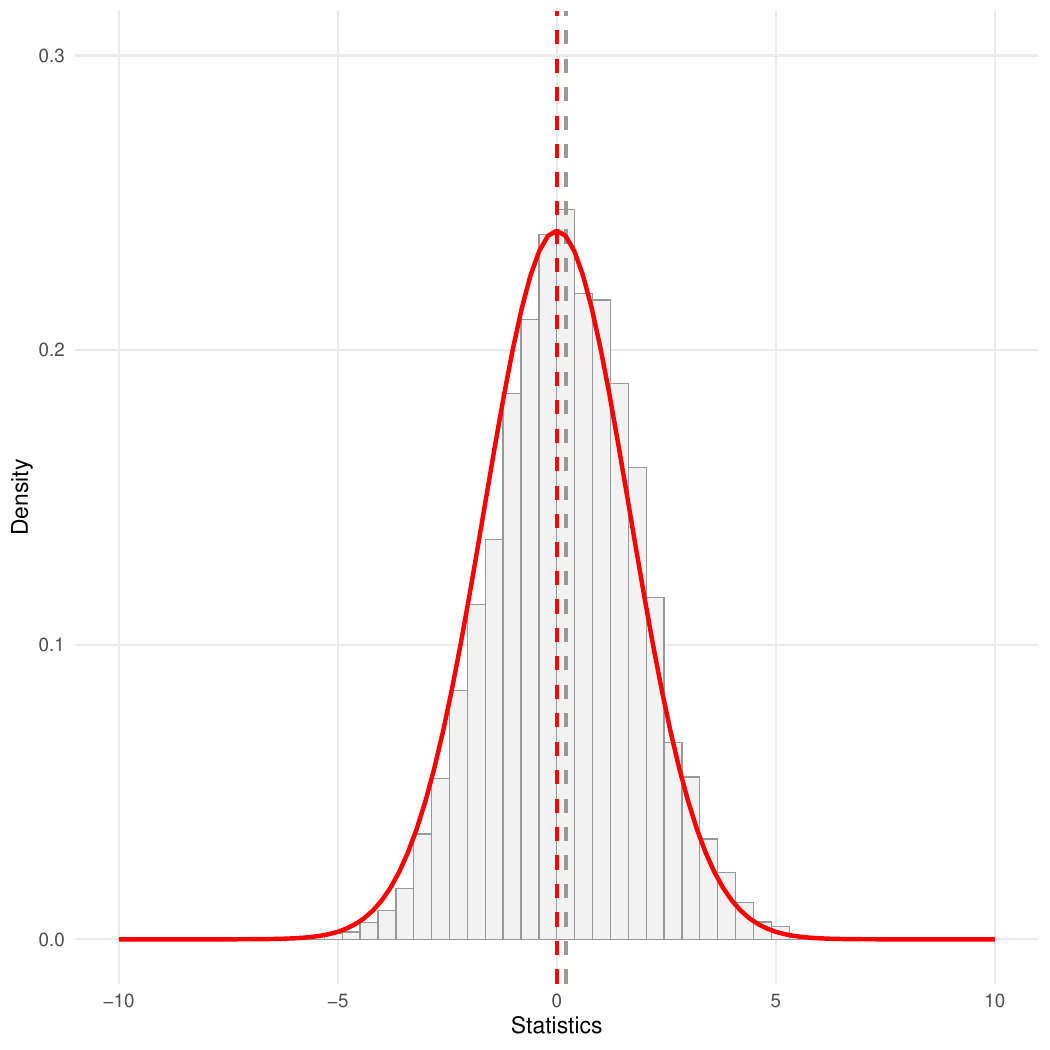}
    \caption{\small $\sqrt{n}(\hat{\rho}^{(j)}-\rho)$ for $k=1$}
    \label{fig:rhok1}
\end{subfigure}\\
\begin{subfigure}{0.45\textwidth}
    \centering
    \includegraphics[width=0.5\textwidth]{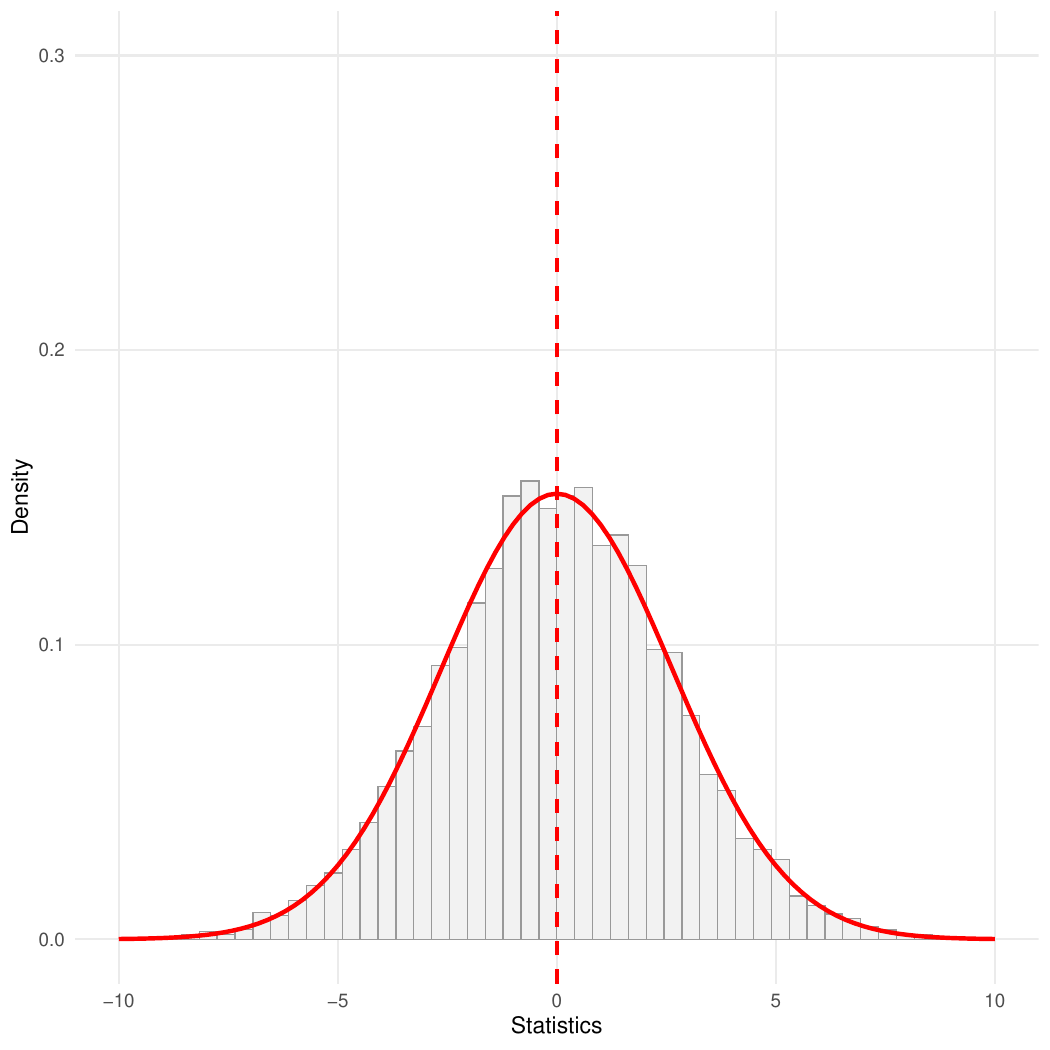}\includegraphics[width=0.5\textwidth]{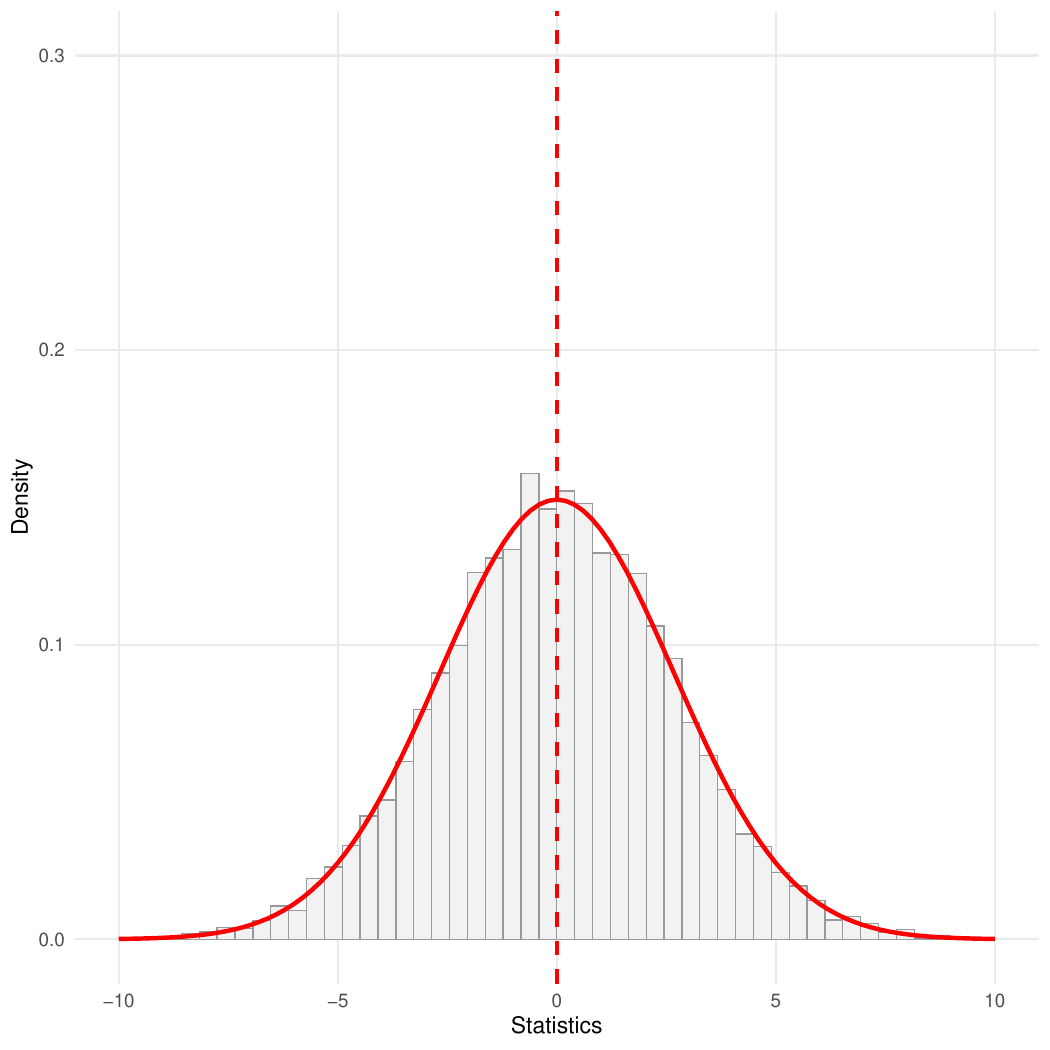}
    \caption{\small $\sqrt{n}(\hat{\mu}^{(j)}_1-\mu_1)$ for $k=2$}
    \label{fig:muk2}
\end{subfigure}
\hspace{0.5cm}
\begin{subfigure}{0.45\textwidth}
    \includegraphics[width=0.5\textwidth]{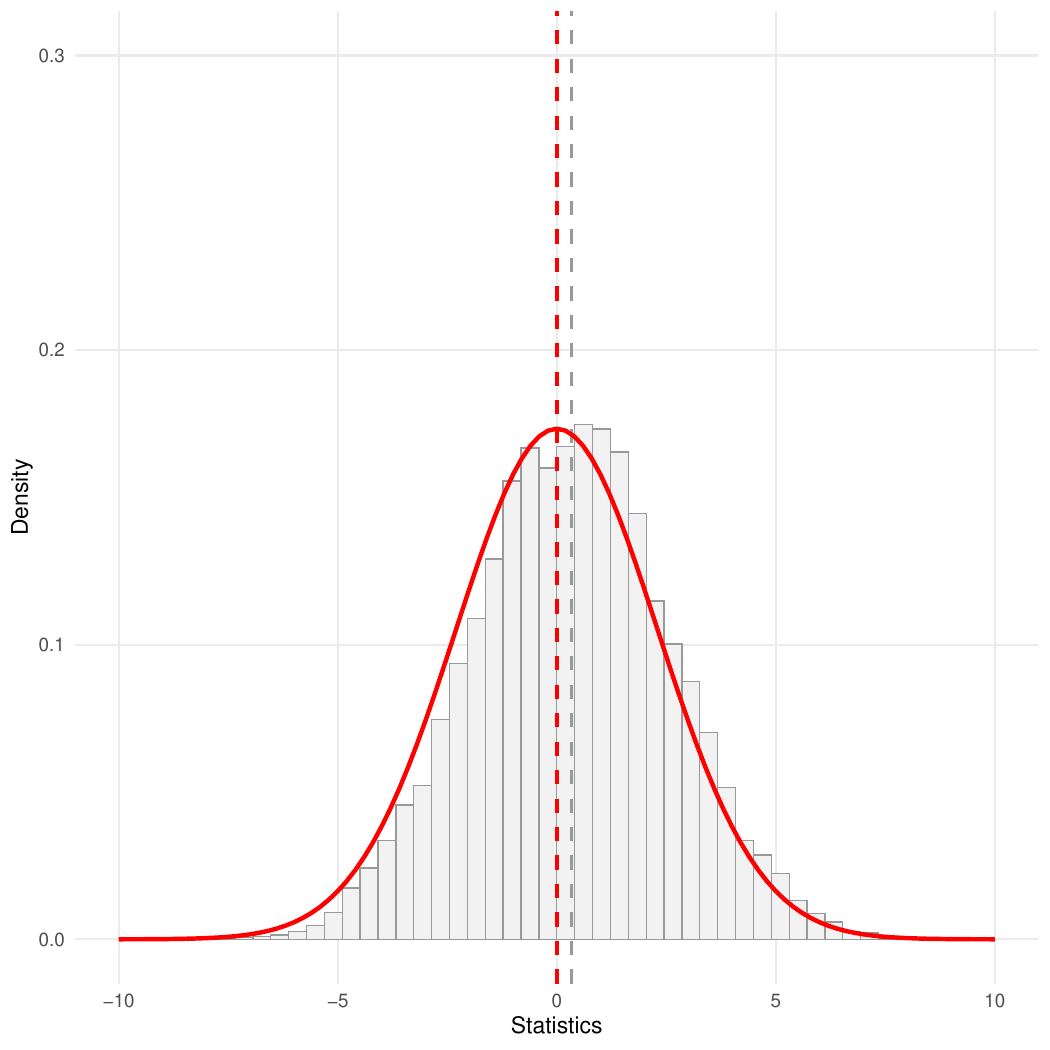}\includegraphics[width=0.5\textwidth]{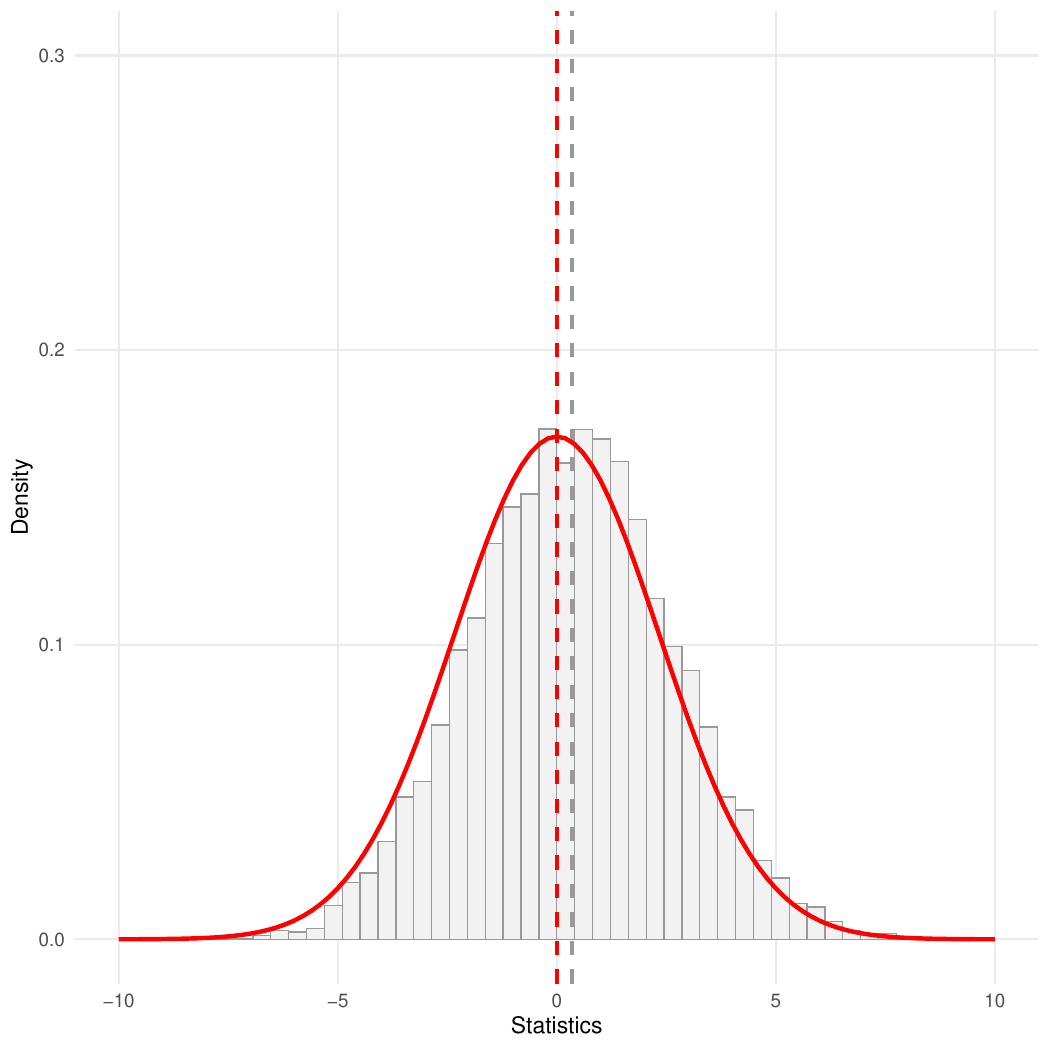}
    \caption{\small $\sqrt{n}(\hat{\rho}^{(j)}-\rho)$ for $k=2$}
    \label{fig:rhok2}
\end{subfigure}%
\caption{\small Histograms of $\{\sqrt{n}(\hat{\mu}^{(j)}_1-\mu_1)\}_{j=1}^M$ and $\{\sqrt{n}(\hat{\rho}^{(j)}-\rho)\}_{j=1}^M$ against their asymptotic normal density, for $k=1,2$ and $d=2$. Inside each panel, the left plot corresponds to the maximum likelihood estimator and the right plot corresponds to the method of moments estimator. The gray/red dashed lines indicate the empirical/asymptotic means.}
\label{fig:asymp}
\end{figure}

Figure \ref{fig:asymp} shows the resulting histograms overlaid with a centered normal density with standard deviation equal to $\sigma_{\mathrm{MM},k}(\bmu)$, $\sigma_{\mathrm{MM},k}(\rho)$, $\sigma_{\mathrm{ML}}(\bmu)$, or $\sigma_{\mathrm{ML}}(\rho)$, depending on the considered estimator. The match between the empirical and asymptotic standard deviations of the estimators is tight, with relative errors smaller than $2\pct$ for $\mu_1$ and $\rho$, for $k=1,2$. The estimated standardized bias $\sqrt{n}(\bar{\hat{\mu}}_1-\mu_1)$ has order $10^{-2}$, but $\sqrt{n}(\bar{\hat{\rho}}-\rho)$ is still significant: approximately $0.20$ for $k=1$ and $0.33$ for $k=2$ (see gray dashed lines), in line with the positive asymptotic bias of the non-truncated moment estimator noted just after \citet[Eq.~(10)]{Pewsey2026}.

A more comprehensive finite-sample comparison of the RMSE of the ML, MM, and GM estimators, across varying $k$, $d$, $\rho$, and $n$, is provided in Appendix~\ref{sec:rmse} of the SM.

\subsection{Finite-sample behavior of goodness-of-fit tests}
\label{sec:experiments:gof}

%
We carried out two numerical experiments to assess the finite-sample behavior of the goodness-of-fit test for $H_0\colon\mathrm{P}\in\lrb{\mathrm{C}_k(\bmu,\rho):(\bmu,\rho)\in\Sd\times[-1,1]}$ using Algorithm \ref{algo:boot}. We considered the tests based on the test statistics $P_n^{W,\lambda}$, for $W\in\{\mathrm{CvM},\mathrm{AD}\}$ and $\lambda\in\{\mathrm{Unif},\mathrm{P}_n,\mathrm{C}_k\}$, computed using closed-form expressions or the Monte Carlo approximation \eqref{eq:MCstat} with $K=1000$ random directions. The parameters $(\bmu,\rho)$ were estimated by $(\hat{\bmu}_{\mathrm{MM}},\hat{\rho}_{\mathrm{MM}})$, which are available in closed form for the orders $k=1,2$ considered here. We used $M=1000$ Monte Carlo replications per scenario and $B=1000$ bootstrap samples. The main text reports the representative sample sizes $n=200$ for size and $n=100$ for power; the complete results for $n\in\{50,100,200,400\}$ are gathered in Appendix~\ref{sec:gofextra} of the SM.

To evaluate the nominal size of the test under $H_0$, we generated samples from $\mathrm{C}_k(\bmu,\rho)$ with $k\in\{1,2\}$, $\rho\in\{0.25, 0.5, 0.75\}$, and $d\in\{1,2\}$. Table \ref{tab:h0} reports the rejection percentages at the nominal level $\alpha=5\pct$ for $n=200$. The test respects the nominal size well, with only a mild conservativeness for $(k=1,d=2,\rho=0.75)$ at small $n$ that fades as $n$ increases; this can be explained by the fact that the Monte Carlo samples are shared within each row. The six tests behave very similarly.

\begin{table}[h!]
\centering
\scriptsize
\begin{tabular}{lll|>{\raggedleft\arraybackslash}p{1.3cm}>{\raggedleft\arraybackslash}p{1.3cm}|>{\raggedleft\arraybackslash}p{1.3cm}>{\raggedleft\arraybackslash}p{1.3cm}|>{\raggedleft\arraybackslash}p{1.3cm}>{\raggedleft\arraybackslash}p{1.3cm}}
\toprule
$k$ & $d$ & $\rho$ & $P_n^{\mathrm{CvM},\,\mathrm{Unif}}$ & $P_n^{\mathrm{AD},\,\mathrm{Unif}}$ & $P_n^{\mathrm{CvM},\,\mathrm{P}_n}$ & $P_n^{\mathrm{AD},\,\mathrm{P}_n}$ & $P_n^{\mathrm{CvM},\,\mathrm{C}_k}$ & $P_n^{\mathrm{AD},\,\mathrm{C}_k}$\\ \midrule
1 & 1 & 0.25 & 4.9 & 5.0 & 4.7 & 4.9 & 5.0 & 5.5 \\
  &   & 0.50 & 4.5 & 4.3 & 4.4 & 4.1 & 4.2 & 4.8 \\
  &   & 0.75 & 4.6 & 4.4 & 4.6 & 4.3 & 4.3 & 4.1 \\ \midrule
  & 2 & 0.25 & 4.4 & 4.2 & 4.1 & 3.9 & 4.2 & 4.0 \\
  &   & 0.50 & 3.8 & 3.8 & 4.1 & \underline{3.4} & 4.1 & 3.9 \\
  &   & 0.75 & \underline{3.6} & \underline{3.6} & \underline{3.4} & \underline{3.1} & \underline{3.5} & 3.8 \\\midrule
2 & 1 & 0.25 & 4.4 & 4.6 & 4.5 & 4.5 & 4.5 & 4.7 \\
  &   & 0.50 & 4.0 & 4.1 & 4.1 & 4.5 & 4.2 & 4.4 \\
  &   & 0.75 & 3.8 & 4.0 & 4.1 & 4.3 & 3.9 & 4.3 \\\midrule
  & 2 & 0.25 & 4.2 & 4.0 & 4.1 & 4.0 & 4.0 & 4.8 \\
  &   & 0.50 & 4.2 & 4.4 & 3.9 & 4.2 & 3.8 & 4.6 \\
  &   & 0.75 & 4.4 & 4.4 & 4.2 & 4.1 & 4.2 & 4.4 \\
\bottomrule
\end{tabular}
\caption{\small Rejection percentages of the goodness-of-fit test for the null hypothesis of spherical cardioidness of order $k$ at the significance level $\alpha=5\pct$, for $n=200$. The null hypothesis holds: the data-generating process is $\mathrm{C}_k(\bmu,\rho)$. Underlined values indicate empirical sizes outside the equal-tail $95\pct$ prediction interval for $\mathrm{Bin}(M, \alpha)\times 100/M$.}
\label{tab:h0}
\end{table}

The second experiment assessed the power of the test against three families of alternatives, all with $n=100$. First, a misspecified spherical cardioid order: the data follow $\mathrm{C}_{k}(\bmu,\rho)$, while the null hypothesis fixes the order $k_0\neq k$, with $(k_0,k)\in\{(1,2),(2,1)\}$ and $\rho\in\{0.10,0.25,0.5\}$. Second, a von Mises--Fisher (vMF) alternative with $\kappa\in\{0.50,1.50,2\}$; and third, a Watson (W) alternative with $\kappa\in\{0.50,2.50,5\}$. Because of \eqref{eq:approxM} and \eqref{eq:approxW}, we only investigate the most challenging confrontations: $\mathrm{C}_1$ vs. $\mathrm{vMF}(\bmu,\kappa)$, and $\mathrm{C}_2$ vs. $\mathrm{W}(\bmu,\kappa)$. The four alternatives are reported in Table~\ref{tab:h1}. For the cardioid alternative, all tests quickly detect the misspecified order, with power increasing in $\rho$. The rejection rates for the vMF and W alternatives are close to the nominal level at low concentration, as expected, and increase to one as $\kappa$ grows. The six statistics have similar powers under the cardioid and vMF alternatives. Against the strongly concentrated Watson, however, the $\mathrm{AD}$-weight statistics are markedly more powerful than the $\mathrm{CvM}$ ones, reflecting the additional sensitivity of $\mathrm{AD}$ to deviations in the tails of the projected distributions. Also for the W alternative, the weight $\mathrm{P}_n$ shows improved performance over the choices $\mathrm{Unif}$ and $\mathrm{C}_{k_0}$.

As shown in Appendix~\ref{sec:gofextra} of the SM, the tests conducted with the MM estimators yield better finite-sample size control than ML estimators for small-to-moderate sample sizes. The choice of estimator has little effect on power over most of the design.

\begin{table}[h!]
\centering
\scriptsize
\begin{tabular}{lll|>{\raggedleft\arraybackslash}p{1.3cm}>{\raggedleft\arraybackslash}p{1.3cm}|>{\raggedleft\arraybackslash}p{1.3cm}>{\raggedleft\arraybackslash}p{1.3cm}|>{\raggedleft\arraybackslash}p{1.3cm}>{\raggedleft\arraybackslash}p{1.3cm}}
\toprule
$H_1$ & $d$ & $\rho/\kappa$ & $P_n^{\mathrm{CvM},\,\mathrm{Unif}}$ & $P_n^{\mathrm{AD},\,\mathrm{Unif}}$ & $P_n^{\mathrm{CvM},\,\mathrm{P}_n}$ & $P_n^{\mathrm{AD},\,\mathrm{P}_n}$ & $P_n^{\mathrm{CvM},\,\mathrm{C}_{k_0}}$ & $P_n^{\mathrm{AD},\,\mathrm{C}_{k_0}}$\\ \midrule
$\mathrm{C}_2$ & 1 & 0.10 & 7.5 & 7.7 & 7.3 & 7.0 & 7.6 & 8.1 \\
 &  & 0.25 & 30.8 & 31.0 & 29.8 & 29.7 & 30.8 & 30.2 \\
 &  & 0.50 & 89.4 & 88.7 & 87.9 & 86.3 & 89.6 & 88.2 \\
 & 2 & 0.10 & 5.6 & 6.0 & 5.4 & 4.3 & 5.7 & 5.2 \\
 &  & 0.25 & 10.0 & 10.2 & 10.4 & 8.7 & 10.1 & 9.4 \\
 &  & 0.50 & 33.4 & 33.6 & 35.3 & 30.5 & 32.5 & 32.9 \\ \midrule
$\mathrm{C}_1$ & 1 & 0.10 & 8.5 & 8.5 & 8.1 & 7.9 & 8.8 & 8.0 \\
 &  & 0.25 & 33.2 & 32.9 & 33.1 & 31.9 & 32.6 & 32.9 \\
 &  & 0.50 & 91.4 & 91.4 & 89.9 & 89.7 & 90.0 & 90.5 \\
 & 2 & 0.10 & 5.6 & 5.8 & 5.8 & 4.8 & 5.6 & 5.8 \\
 &  & 0.25 & 17.6 & 17.8 & 16.8 & 14.8 & 15.9 & 16.6 \\
 &  & 0.50 & 68.0 & 68.0 & 65.1 & 62.0 & 66.1 & 65.5 \\ \midrule
$\mathrm{vMF}$ & 1 & 0.50 & 5.2 & 5.9 & 5.4 & 5.2 & 5.2 & 6.4 \\
 &  & 1.50 & 77.0 & 76.5 & 77.1 & 74.1 & 76.9 & 76.5 \\
 &  & 2.00 & 99.9 & 99.8 & 99.9 & 99.7 & 99.8 & 99.8 \\
 & 2 & 0.50 & 4.7 & 4.7 & 4.9 & 4.2 & 4.1 & 4.8 \\
 &  & 1.50 & 77.0 & 75.5 & 79.7 & 76.2 & 76.9 & 76.1 \\
 &  & 2.00 & 100.0 & 99.9 & 99.9 & 99.8 & 99.9 & 100.0 \\ \midrule
$\mathrm{W}$ & 1 & 0.50 & 5.8 & 5.8 & 5.4 & 5.2 & 5.5 & 6.1 \\
 &  & 2.50 & 7.3 & 8.0 & 8.0 & 8.2 & 7.8 & 8.7 \\
 &  & 5.00 & 85.0 & 96.9 & 92.3 & 96.7 & 75.5 & 91.1 \\
 & 2 & 0.50 & 6.7 & 6.3 & 7.0 & 6.5 & 6.5 & 6.9 \\
 &  & 2.50 & 18.9 & 26.3 & 31.7 & 41.1 & 21.7 & 32.0 \\
 &  & 5.00 & 100.0 & 100.0 & 100.0 & 100.0 & 100.0 & 100.0 \\
\bottomrule
\end{tabular}
\caption{\small Rejection percentages of the goodness-of-fit test for the null hypothesis of spherical cardioidness of order $k_0$ at the significance level $\alpha=5\pct$, for $n=100$. The first column gives the data-generating alternative $H_1$: spherical cardioids $\mathrm{C}_k(\bmu,\rho)$ of orders $k=2,1$ for $k_0=1,2$, $\mathrm{vMF}(\bmu,\kappa)$ for $k_0=1$, and $\mathrm{W}(\bmu,\kappa)$ for $k_0=2$.}
\label{tab:h1}
\end{table}

As a takeaway from this limited simulation study, the test based on $P_n^{\mathrm{AD},\,\mathrm{P}_n}$ is recommended for its increased power, as long as the sample size allows its computation. If not, the test based on $P_n^{\mathrm{AD},\,\mathrm{Unif}}$ evaluated through Monte Carlo offers a cost-effective alternative. The tests based on $\lambda=\mathrm{C}_k$ are not recommended, as they are more computationally expensive and do not offer any apparent power advantage.

\section{Spherical cardioidness of comet orbits}
\label{sec:application}

%
Long-period comets are hypothesized to originate in the Oort cloud, a conjectured roughly spherical reservoir of icy planetesimals surrounding the Solar System, whereas short-period comets are associated with the flattened Kuiper belt. These different origins are reflected in orbital orientations: long-period comets exhibit an approximately isotropic distribution, in sharp contrast with the ecliptic-concentrated orientations of short-period comets (see, e.g., \citet[Sections 5 and 7.2]{Dones2015} and references therein). Orbital orientations can be represented by the directed unit normal vectors to the orbital planes on $\mathbb{S}^2$. An orbit with \emph{inclination} $i\in[0,\pi]$ and \emph{longitude of the ascending node} $\Omega\in[0,2\pi)$ \citep[see][]{Jupp2003} has directed normal vector $(\sin(i)\sin(\Omega),-\sin(i)\cos(\Omega),\cos(i))^\top$ to the orbit's plane (see the illustrative graphs in Figure~\ref{fig:comets}). The sign encodes prograde (northern hemisphere) versus retrograde (southern hemisphere) motion.

The \texttt{comets} object in the \texttt{sphunif} R package \citep{Garcia-Portugues:sphunif} contains a dataset of comet orbits sourced from the JPL Small-Body Database Search. We analyze two subsets of comets with elliptical orbits: a set of $n=610$ long-period comets (orbital period exceeding $200$ years) and a set of $n=784$ short-period comets (orbital period below $200$ years). Records corresponding to comet fragments are excluded to avoid spurious clustering. Further details on data retrieval and preprocessing are given in \citet[Sec.~7]{Garcia-Portugues2025}.

\begin{figure}[h!]
\centering
\begin{subfigure}{0.45\textwidth}
    \centering
    \includegraphics[width=0.5\textwidth,trim={1.1cm 0cm 0.75cm 0cm},clip]{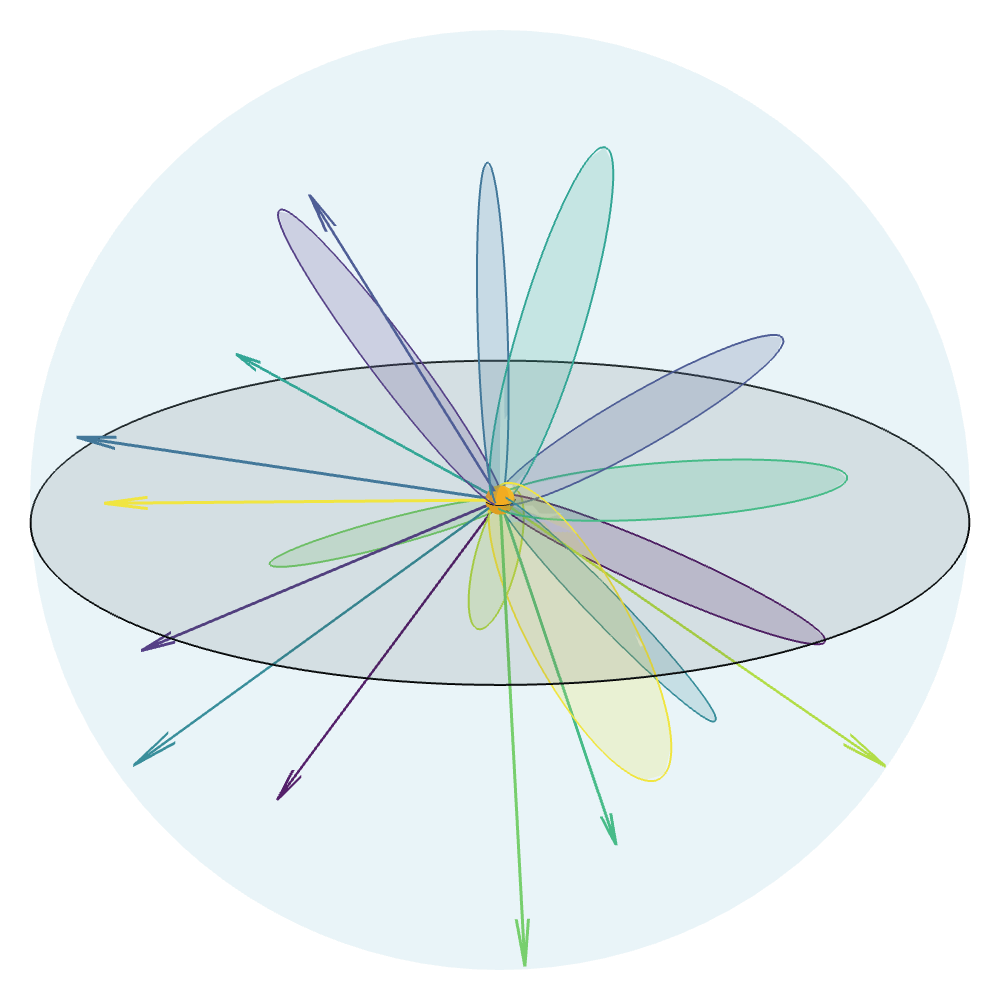}\includegraphics[width=0.5\textwidth,trim={0.75cm 0cm 1.1cm 0cm},clip]{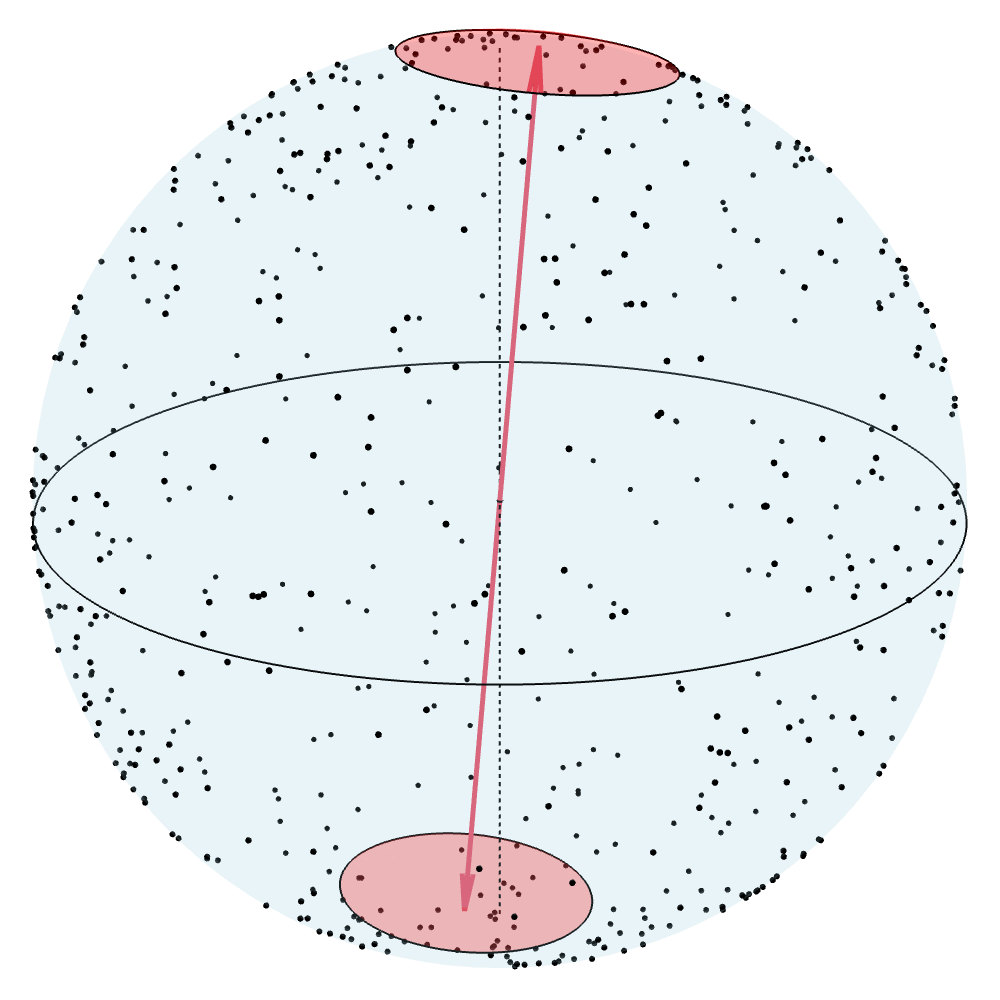}
    \caption{\small Long-period comets}
    \label{fig:oort}
\end{subfigure}
\hspace{0.5cm}
\begin{subfigure}{0.45\textwidth}
    \includegraphics[width=0.5\textwidth,trim={1.1cm 0cm 0.75cm 0cm},clip]{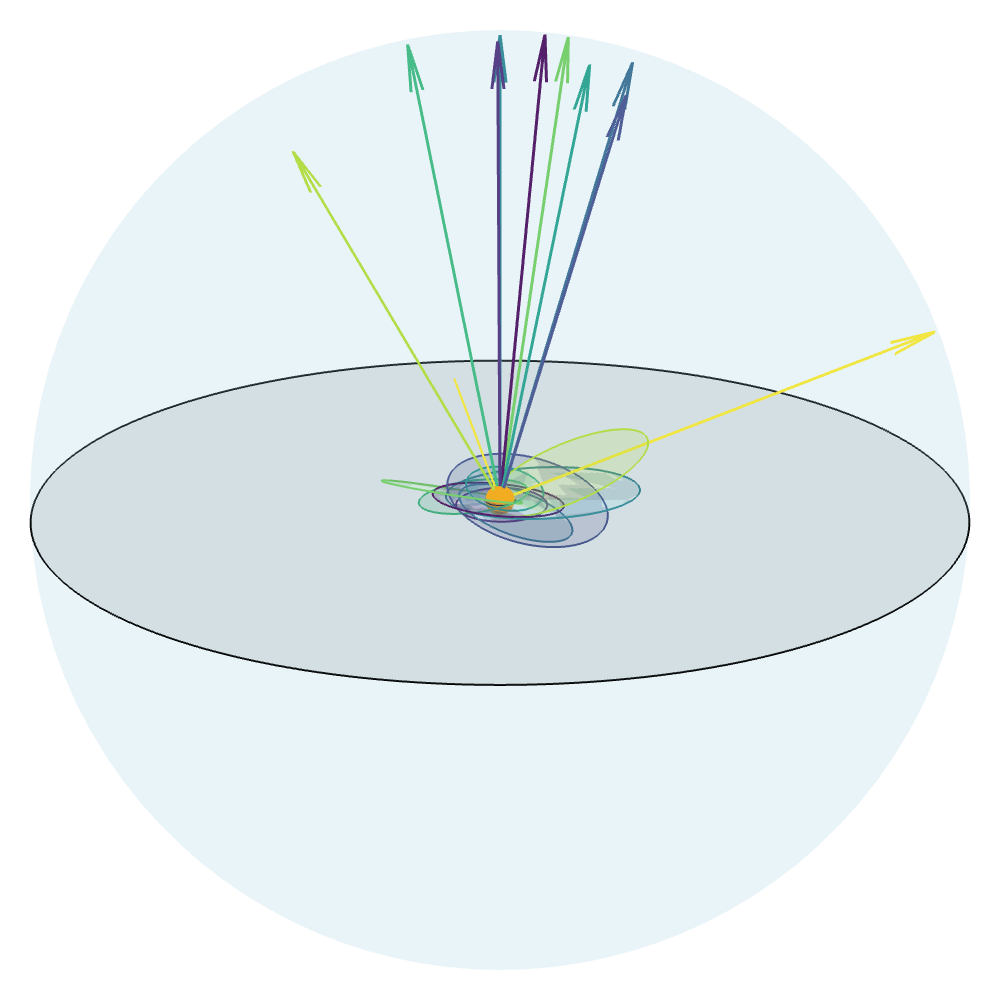}\includegraphics[width=0.5\textwidth,trim={0.75cm 0cm 1.1cm 0cm},clip]{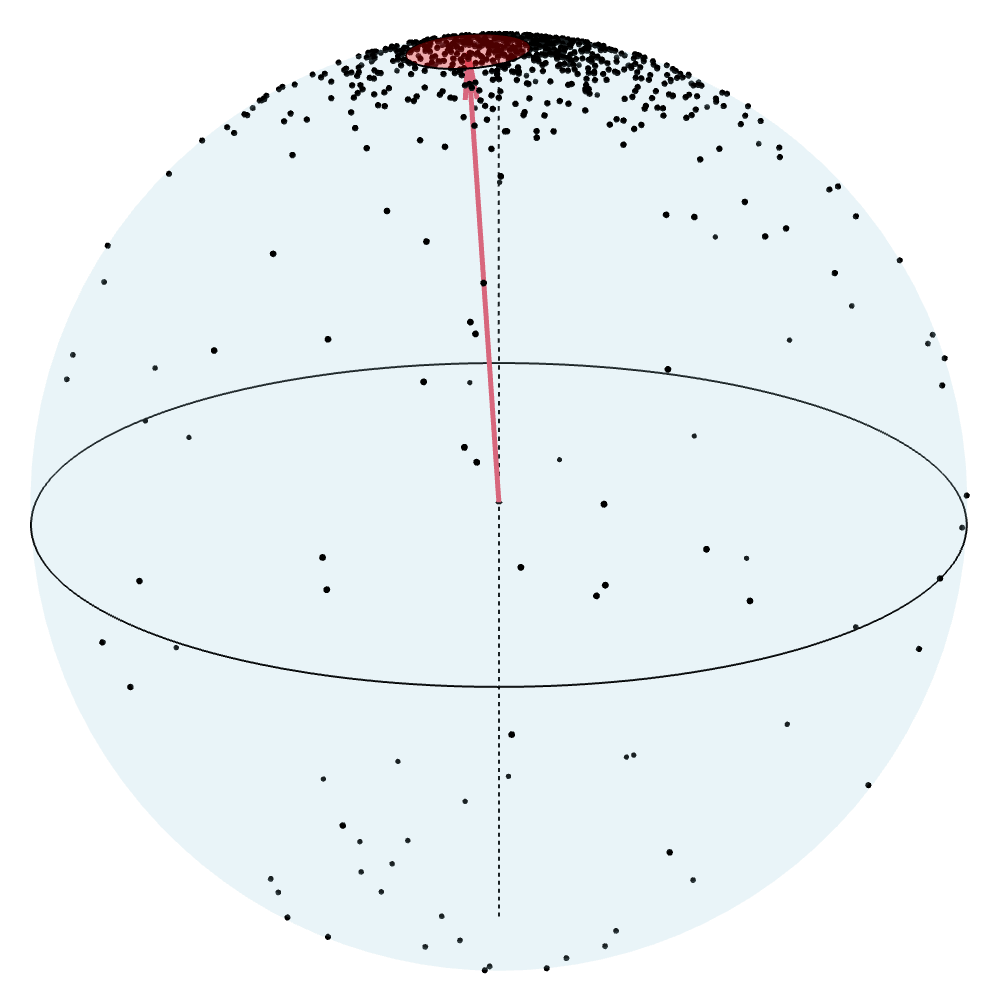}
    \caption{\small Short-period comets}
    \label{fig:kuiper}
\end{subfigure}%
\caption{\small Orbits of long- and short-period comets and their normal vectors. Within each figure, the left plot displays ten illustrative elliptical orbits and their associated normal vectors, with the ecliptic plane shown in gray and the Sun represented as an orange sphere (one focus of the elliptical orbits). The right plot in each figure shows the full dataset of normal orbit vectors on~$\mathbb{S}^2$. Red indicates the $\hat{\bmu}_{\mathrm{ML}}$ direction/axis and the bootstrap $95\pct$-confidence spherical cap for $\bmu$. This figure is an enhancement of \citet[Fig.~4]{Garcia-Portugues2025}.}
\label{fig:comets}
\end{figure}

Earlier symmetry diagnostics for these orbital normals suggested that long-period comets depart from uniformity mainly through a low-order axial effect, whereas rotational symmetry about an (unspecified) axis $\bmu$ is not rejected at the $10\pct$ significance level \citep[Sec.~7]{Garcia-Portugues2025}. A plausible explanation for this combination of ``non-uniform yet symmetric'' behavior is observational selection: comet surveys concentrate a substantial fraction of their sky coverage and cadence at small ecliptic latitudes \citep{Jupp2003}. This preferential sampling increases the incidence of near-ecliptic orbital planes, which translates into approximately antipodal (north--south) accumulations of the corresponding normal vectors on~$\mathbb{S}^2$. These considerations motivate assessing the family $\mathrm{C}_k(\bmu,\rho)$ for small order $k$ as a parsimonious model that retains rotational symmetry about~$\bmu$ while allowing moderate, low-order deviations from uniformity, particularly for long-period comets.

Table~\ref{tab:comets} reports the bootstrap $p$-values obtained by Algorithm \ref{algo:boot} for testing the goodness-of-fit of $\mathrm{C}_k(\bmu,\rho)$, $k=1,2,3,4$, for long-period comets. We considered the statistics $P_n^{W,\lambda}$, for $W\in\{\mathrm{CvM},\mathrm{AD}\}$ and $\lambda\in\{\mathrm{Unif},\mathrm{P}_n,\mathrm{C}_k\}$, computed using closed-form expressions or the Monte Carlo approximation \eqref{eq:MCstat} with $K=10^4$ random directions. We used $B=10^4$ bootstrap samples and the ML estimates $(\hat\bmu_{\mathrm{ML}},\hat\rho_{\mathrm{ML}})$. Maximum likelihood is adopted here because the comparison spans the orders $k=1,\ldots,4$. The test for the simple null hypothesis of uniformity was also conducted for reference.

\begin{table}[h!]
\centering
\scriptsize
\begin{tabular}{c|p{1.3cm}p{1.3cm}|p{1.3cm}p{1.3cm}|p{1.3cm}p{1.3cm}}
\toprule
$H_0$ & $P_n^{\mathrm{CvM},\,\mathrm{Unif}}$ & $P_n^{\mathrm{AD},\,\mathrm{Unif}}$ & $P_n^{\mathrm{CvM},\,\mathrm{P}_n}$ & $P_n^{\mathrm{AD},\,\mathrm{P}_n}$ & $P_n^{\mathrm{CvM},\,\mathrm{C}_k}$ & $P_n^{\mathrm{AD},\,\mathrm{C}_k}$\\ \midrule
$\mathrm{Unif}$ & 0.026 & 0.009 & 0.026 & 0.008 & 0.029 & 0.012\\
$\mathrm{C}_1$  & 0.000 & 0.000 & 0.000 & 0.000 & 0.000 & 0.000\\
$\mathrm{C}_2$  & 0.283 & 0.116 & 0.184 & 0.139 & 0.146 & 0.130\\
$\mathrm{C}_3$  & 0.026 & 0.012 & 0.024 & 0.008 & 0.029 & 0.011\\
$\mathrm{C}_4$  & 0.028 & 0.013 & 0.027 & 0.011 & 0.023 & 0.009\\
\bottomrule
\end{tabular}
\caption{\small Bootstrap $p$-values of the goodness-of-fit test of spherical cardioidness applied to the normal vectors of the orbits of long-period comets. The tests are conducted for the orders $k=1,2,3,4$ and the simple null hypothesis of uniformity ($\rho_0=0$).}
\label{tab:comets}
\end{table}

The spherical cardioid model with $k=2$ is not rejected at the $10\pct$ significance level as the candidate model for the distribution of orbital normals of long-period comets, for all tests. Higher- and lower-order models, including uniformity, are found incompatible for long-period comets at the $5\pct$ significance level. This supports the suitability of the spherical cardioid family to capture the moderate non-uniformity of long-period comet orbits while accommodating their marked rotational symmetry. For short-period comets, all models are emphatically rejected, with minimal bootstrap $p$-values $1/(B+1)\approx 10^{-4}$; among them, the model with $k=1$ has the smallest test statistics.

Figure~\ref{fig:projs} gives additional insights into the outcomes of the goodness-of-fit tests. It displays the ecdf $F_{n,\hat{\bmu}_{\mathrm{ML}}}$ of the projected sample $\{\hat{\bmu}_{\mathrm{ML}}^\top\bX_i\}_{i=1}^n$ versus the projected cdf $\hat{F}_{\hat{\bmu}_{\mathrm{ML}}}$ of the fitted $\mathrm{C}_k(\hat\bmu_{\mathrm{ML}},\hat\rho_{\mathrm{ML}})$, for long- ($k=2$) and short-period ($k=1$) comets. For long-period comets, the agreement between the ecdf and cdf is remarkable. For short-period comets, the fitted cdf departs strongly from the ecdf, as the projected cdf cannot capture the heavy concentration on the right tail of the ecdf, caused by the cluster near the north pole.

The fitted parameters for long-period orbits with $k=2$ are $\hat{\bmu}_{\mathrm{ML},\,\text{long}}=(0.0804, -0.0067, 0.9967)^\top$ and $\hat{\rho}_{\mathrm{ML},\,\text{long}}=0.4727$. The axis estimate $\hat{\bmu}_{\mathrm{ML},\,\text{long}}$ is slightly tilted with respect to the normal axis of the ecliptic plane (north--south axis). The percentile bootstrap $95\pct$-confidence interval for $\rho_{\mathrm{long}}$ is $(0.3121, 0.6747)$, coherent with the significant non-uniformity. The percentile bootstrap $95\pct$-confidence region for $\bmu_{\mathrm{long}}$ is $\{\bmu\in\mathbb{S}^2:|\bmu^\top \hat{\bmu}_{\mathrm{ML},\,\text{long}}|\geq 0.9571\}$ (see Figure~\ref{fig:oort}). For short-period comets with $k=1$, $\hat{\bmu}_{\mathrm{ML},\,\text{short}}=(-0.0626, -0.0662, 0.9958)^\top$ and $\{\bmu\in\mathbb{S}^2: \bmu^\top \hat{\bmu}_{\mathrm{ML},\,\text{short}} \geq 0.9922\}$ (Figure~\ref{fig:kuiper}). The concentration estimate $\hat{\rho}_{\mathrm{ML},\,\text{short}}=1$ and a percentile interval that contains only the boundary value $1$ indicate the lack of fit of the spherical cardioid model for the distribution of short-period orbits.

\begin{figure}[h!]
\centering
\begin{subfigure}{0.45\textwidth}
    \centering
    \includegraphics[width=0.8\textwidth]{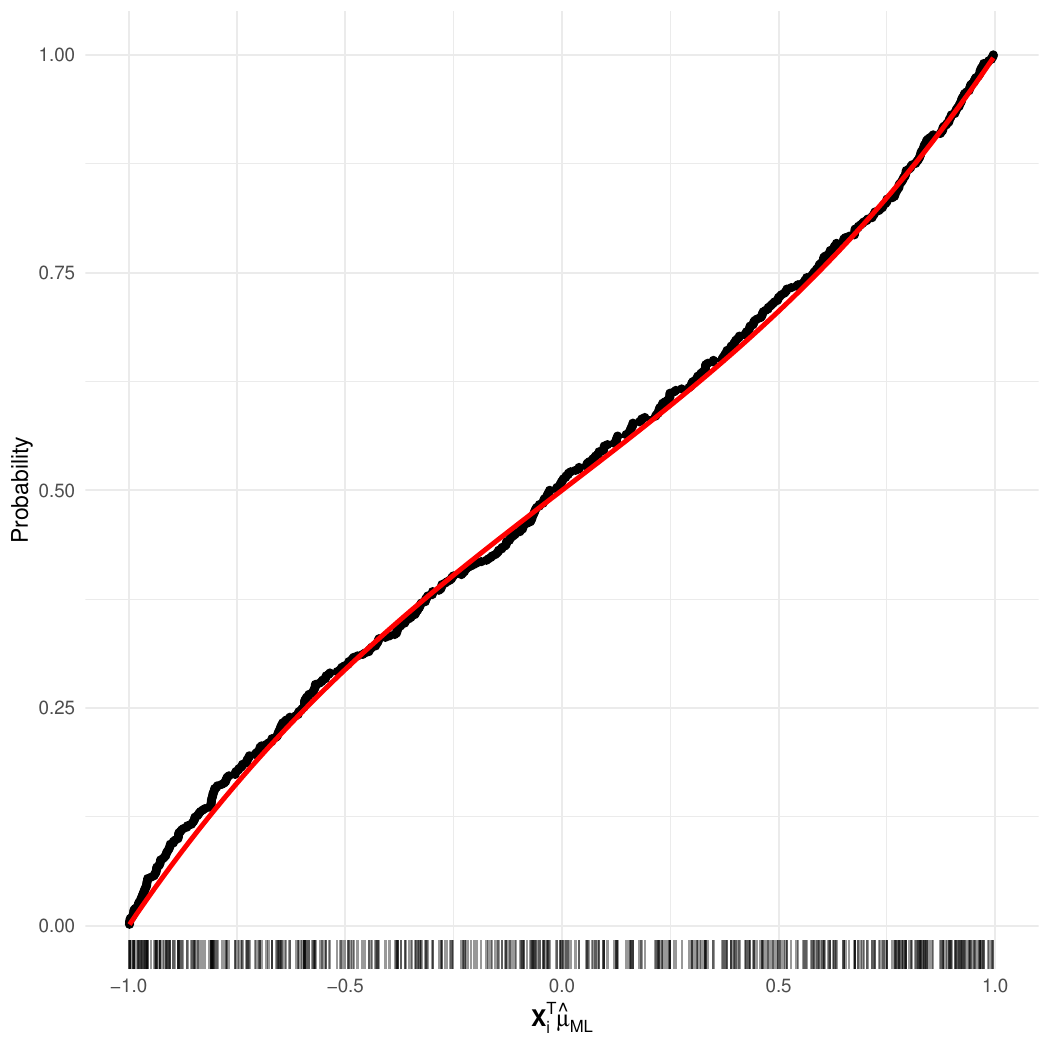}
    \caption{\small Long-period comets ($k=2$)}
    \label{fig:ecdflong}
\end{subfigure}
\hspace{0.5cm}
\begin{subfigure}{0.45\textwidth}
    \includegraphics[width=0.8\textwidth]{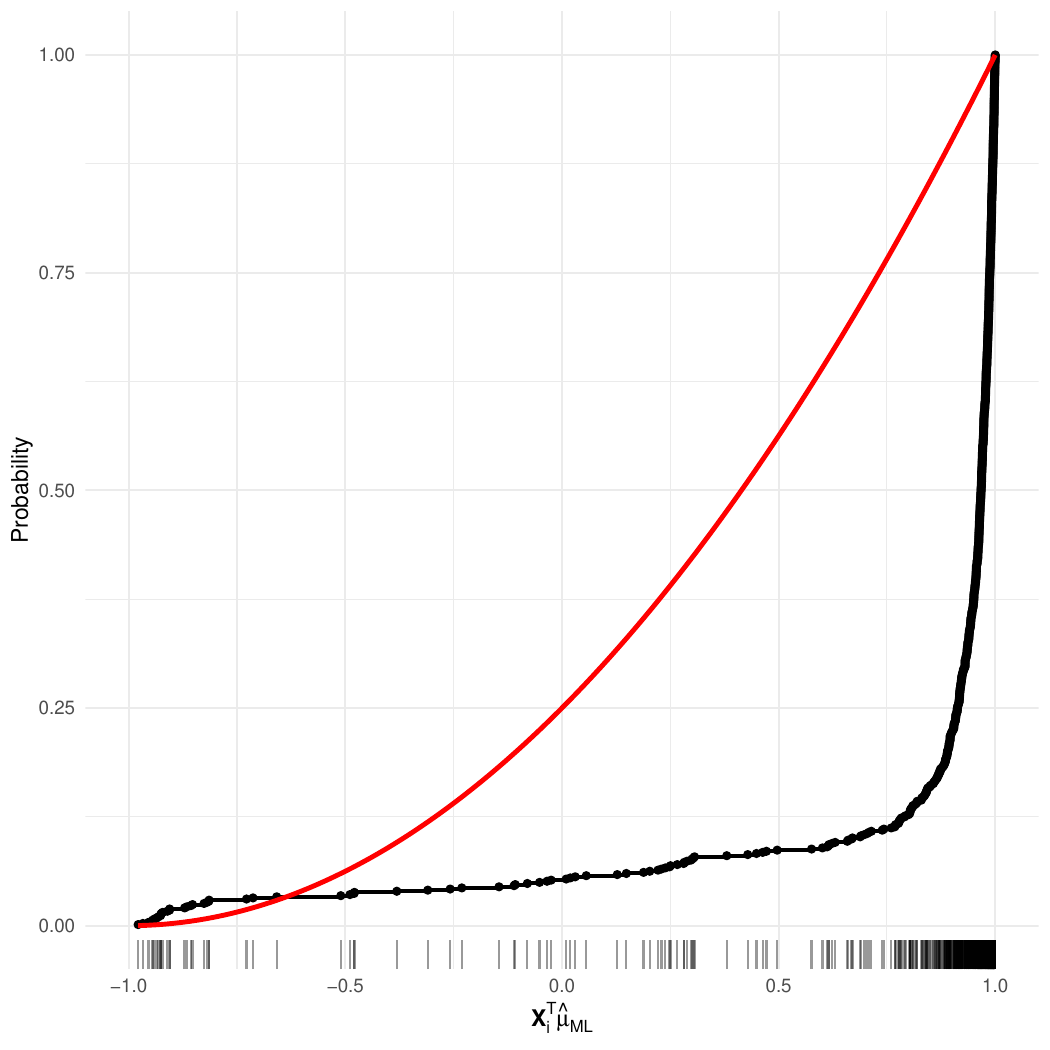}
    \caption{\small Short-period comets ($k=1$)}
    \label{fig:ecdfshort}
\end{subfigure}%
\caption{\small Comparison of the ecdf $F_{n,\hat{\bmu}_{\mathrm{ML}}}$ of the projected sample $\{\hat{\bmu}_{\mathrm{ML}}^\top\bX_i\}_{i=1}^n$ (black curve) versus the projected cdf $\hat{F}_{\hat{\bmu}_{\mathrm{ML}}}$ of the fitted $\mathrm{C}_k(\hat\bmu_{\mathrm{ML}},\hat\rho_{\mathrm{ML}})$ (red curve).}
\label{fig:projs}
\end{figure}

\section{Discussion}
\label{sec:discussion}

%
We have studied a generalization of the well-known circular cardioid distribution that allows for spherical data and multimodal and girdle-like patterns. Among its most attractive features, the spherical cardioid family enjoys closed-form estimators, asymptotic inference, efficient simulation, explicit projected distributions, and tractable goodness-of-fit testing. The distribution is close to the uniform distribution on the sphere, as reflected in its moment structure, making it suitable for modeling mildly non-uniform spherical data. A good example of such data is provided by the orbital normals of long-period comets, for which the spherical cardioid distribution of order two provides a satisfactory fit.

The goodness-of-fit methodology developed here, based on integrating quadratic discrepancies of the projected ecdf, provides a natural and computationally tractable testing approach for spherical cardioidness. This approach complements the empirical characteristic function-based tests of \cite{Ebner2024} for related hypotheses on the sphere. While neither testing approach dominates the other in all settings, our projection-based tests have the advantage of producing graphical, direction-specific diagnostics, as evidenced in the real data application.

Generalizations of the spherical cardioid distribution to product spaces like the torus $(\mathbb{S}^1)^d$ or the polysphere $\mathbb{S}^{d_1}\times \cdots\times \mathbb{S}^{d_r}$ seem possible based on \eqref{eq:sphcard} and could be relevant for constructing tractable dependence structures on these spaces. These extensions will be explored in future work. Further research could investigate whether alternatives exist for which the choice of the distribution $\lambda$ has a significant impact on the power of the goodness-of-fit tests in Sec.~\ref{sec:experiments}.

\section*{Acknowledgments}

The author acknowledges support by grant PID2021-124051NB-I00, funded by MCIN/\-AEI/\-10.13039/\-501100011033 and ERDF/EU. The author also acknowledges the valuable feedback from two anonymous reviewers.



\fi

\ifsupplement

\newpage

\title{Supplementary materials for ``On the spherical cardioid distribution and its goodness-of-fit''}
\setlength{\droptitle}{-1cm}
\predate{}%
\postdate{}%
\date{}

\author{Eduardo Garc\'ia-Portugu\'es$^{1}$}
\footnotetext[1]{Department of Statistics, Universidad Carlos III de Madrid (Spain).}
\maketitle

\begin{abstract}
	These supplementary materials contain four appendices. Appendix~\ref{sec:proofs} provides the proofs of the results of the paper. Appendix~\ref{sec:ares} analyzes the asymptotic relative efficiencies of the moment estimators with respect to the maximum likelihood estimators. Appendix~\ref{sec:rmse} reports additional finite-sample simulations comparing the estimators. Appendix~\ref{sec:gofextra} gathers experiments for the goodness-of-fit tests for additional sample sizes and estimators.
\end{abstract}

\appendix

\section{Proofs}
\label{sec:proofs}

\subsection{Proofs of results in Section~\ref{sec:cardioid}}
\label{sec:proofs:cardioid}

\subsubsection{Proof of result in Section~\ref{sec:genesis}}
\label{sec:proofs:genesis}

\begin{proof}[Proof of Proposition \ref{prp:conv}]
A direct computation using \eqref{eq:ckd2} and \eqref{eq:addgegen} gives
\begin{align*}
  \int_{\Sd} &f_{\mathrm{C}_{k_1}}(\bx;\bmu_1,\rho_1)f_{\mathrm{C}_{k_2}}(\bx;\bmu_2,\rho_2)\,\sigma_d(\rd\bx)\\
  &= \frac{1}{\om{d}^2}\int_{\Sd} \lrb{1+\rho_1\rho_2\, \tilde{C}_{k_1}^{(d-1)/2}(\bx^\top\bmu_1) \tilde{C}_{k_2}^{(d-1)/2}(\bx^\top\bmu_2)}\,\sigma_d(\rd\bx)\\
  &=\frac{1}{\om{d}}+\frac{\delta_{k_1,k_2}\rho_1\rho_2}{\om{d}C_{k_1}^{(d-1)/2}(1)^2} \tau_{k_2,d}^{-1} C_{k_1}^{(d-1)/2}(\bmu_1^\top\bmu_2)\\
  &=\frac{1}{\om{d}}+\frac{\delta_{k_1,k_2}\rho_1\rho_2/d_{k_1,d}}{\om{d}} \tilde{C}_{k_1}^{(d-1)/2}(\bmu_1^\top\bmu_2)\\
  &= f_{\mathrm{C}_{k_1}}(\bmu_1;\bmu_2,\delta_{k_1,k_2}\rho_1\rho_2/d_{k_1,d}).
\end{align*}

An application of this result with $\bmu_1=\bx$ yields
\begin{align*}
    \int_{\Sd} f_{\mathrm{C}_{k_1}}(\bx;\by,\rho_1)f_{\mathrm{C}_{k_2}}(\by;\bmu,\rho_2) \,\sigma_d(\rd \by)=f_{\mathrm{C}_{k_1}}\lrp{\bx;\bmu,\delta_{k_1,k_2}\frac{\rho_1\rho_2}{d_{k_1,d}}},
\end{align*}
thus giving the density of $\bX$ by the law of total probability.
\end{proof}

\subsubsection{Proofs of results in Section~\ref{sec:moments}}
\label{sec:proofs:moments}

\begin{proof}[Proof of Theorem \ref{thm:mom}]
By construction,
\begin{align*}
  \Ebig{\bX^{\otimes m}}%
  &=\Ebig{\bU^{\otimes m}}+\frac{\rho}{\om{d}\, C_{k}^{(d-1)/2}(1)}\int_{\Sd} \bx^{\otimes m} C_k^{(d-1)/2}(\bx^\top\bmu)\,\sigma_{d}(\rd\bx)\\
  &\indef \bmu_{d+1,m}+\frac{\rho}{\om{d}\, C_{k}^{(d-1)/2}(1)}I_{m,k}(\bmu).
\end{align*}
The uniform moments \eqref{eq:unifmom} follow from Lemma 5 in \cite{Chacon2026|SM}.

We next focus on computing $I_{m,k}(\bmu)$.

\emph{Proof of \ref{thm:mom:1}}. The vectorized monomial $\bx^{\otimes m}$ is a homogeneous polynomial of order $m$. Its entries are not necessarily harmonic, i.e., such that the entrywise Laplacian $\Delta \bx^{\otimes m}$ is null. The projection of $\bx^{\otimes m}$ onto $\mathcal{H}_k^d$, the space of spherical harmonics of degree $k$ on $\Sd$ (i.e., homogeneous and harmonic polynomials) is carried out by the projection operator $\mathrm{proj}_k: L^2(\Sd)\to\mathcal{H}_k^d$. For $f\in L^2(\Sd)$, this linear operator is
\begin{align*}
  \mathrm{proj}_k\, f(\bx)=\frac{1}{\om{d}}\int_{\Sd} f(\by) Z_k(\bx, \by)\,\sigma_d(\rd\by),\quad Z_k(\bx, \by)=\tau_{k,d}\,C_k^{(d-1)/2}(\bx^\top\by),%
\end{align*}
see \citet[Lemma 1.2.4, Theorem 1.2.6]{Dai2013|SM} for $d\geq2$.

When $m<k$, $\mathrm{proj}_k\, \bx^{\otimes m}=\zero$ since $\bx^{\otimes m}$ decomposes into spherical harmonics of degree at most $m$, and hence the projection onto $\mathcal{H}_k^d$ is null. Thus,
\begin{align*}
  I_{m,k}(\bmu)&=\tau_{k,d}^{-1}\int_{\Sd} \bx^{\otimes m} Z_k(\bx, \bmu)\,\sigma_d(\rd\bx) =\tau_{k,d}^{-1}\,\om{d}\, \mathrm{proj}_k\, (\bx^{\otimes m})\big\vert_{\bx=\bmu}=\zero
\end{align*}
and the result follows.

\emph{Proof of \ref{thm:mom:3}}. On the one hand, $(-\bx)^{\otimes m}=(-1)^m \bx^{\otimes m}$ because $\bx^{\otimes m}$ is a homogeneous polynomial of degree $m$. On the other hand, $\mathrm{proj}_k\, f(-\bx)=(-1)^k \mathrm{proj}_k\, f(\bx)$ for any $f\in L^2(\Sd)$ since $Z_k(-\bx, \by)=(-1)^k Z_k(\bx, \by)$. Therefore, $\mathrm{proj}_k\, (-\bx)^{\otimes m}=(-1)^m \mathrm{proj}_k\, \bx^{\otimes m}$ and $\mathrm{proj}_k\, (-\bx)^{\otimes m}=(-1)^k \mathrm{proj}_k\, \bx^{\otimes m}$. When $m>k$ and $m-k$ is odd, then it must be that $\mathrm{proj}_k\, \bx^{\otimes m}=\zero$.

\emph{Proof of \ref{thm:mom:2}}. The projection operator is relatively explicit for a homogeneous polynomial of degree $k$ \citep[Lemma 1.2.1]{Dai2013|SM}, from which it follows that
\begin{align}
  \mathrm{proj}_k\, \bx^{\otimes k}%
  &=\sum_{j=0}^{\lfloor k/2\rfloor} \frac{(-1)^j}{4^j j! (k+(d-1)/2-j)_j} \Delta^j \bx^{\otimes k},\label{eq:projk}
\end{align}
where $(a)_j$ is the Pochhammer symbol and $\Delta^j f=\big(\sum_{i=1}^{d+1}\partial_i^2\big)^j f$ is the iterated Laplacian operator with $\Delta^0 f=f$, and the operator is applied entrywise on $\bx^{\otimes k}$. (Note that this projection operator is given in Lemma 1.2.1 of \cite{Dai2013|SM} for $d\geq2$ but it extends to $d=1$.) From the coordinate computation of the Laplacian, we have that
\begin{align*}
  \Delta \bx^{\otimes k}= k(k-1) \Sym_{d+1,k}(\operatorname{vec} \bI_{d+1}\otimes \bx^{\otimes k-2}),\quad k\geq 2,
\end{align*}
where $\Sym_{d+1,k}$ is the symmetrizer matrix. More generally,
\begin{align}
  \Delta^j\bx^{\otimes k}=\frac{k!}{(k-2j)!} \Sym_{d+1,k}\big((\operatorname{vec} \bI_{d+1})^{\otimes j}\otimes \bx^{\otimes k-2j}\big), \quad 0\leq j\leq \lfloor k/2\rfloor, \label{eq:lapm}
\end{align}
and $\Delta^j\bx^{\otimes k}=\zero$ for $j>\lfloor k/2\rfloor$. Plugging \eqref{eq:lapm} into \eqref{eq:projk} leads to
\begin{align*}
  \mathrm{proj}_k\, \bx^{\otimes k}=&\;\sum_{j=0}^{\lfloor k/2\rfloor} \frac{(-1)^j}{4^j j! (k+(d-1)/2-j)_j} \lrb{\frac{k!}{(k-2j)!} \Sym_{d+1,k}\big((\operatorname{vec} \bI_{d+1})^{\otimes j}\otimes \bx^{\otimes k-2j}\big)}\\
  =&\;\Sym_{d+1,k}\sum_{j=0}^{\lfloor k/2\rfloor} a_{k,j} (\operatorname{vec} \bI_{d+1})^{\otimes j}\otimes \bx^{\otimes k-2j},
\end{align*}
where
\begin{align*}
  a_{k,j}=(-1)^j\frac{k!}{(k-2j)!}\frac{\Gamma(k+(d-1)/2-j)}{4^j j! \Gamma(k+(d-1)/2)}=(-1)^j\frac{k!}{2^j(k-2j)!j!}\frac{1}{\prod_{r=1}^{j} (2(k-r)+d-1)}
\end{align*}
with the convention that $a_{k,0}=1$. Therefore,
\begin{align*}
  \frac{1}{\om{d}\, C_{k}^{(d-1)/2}(1)}I_{k,k}(\bmu)&=\frac{1}{\om{d}\, C_{k}^{(d-1)/2}(1)}\lrp{1+\frac{2k}{d-1}}^{-1}\om{d}\, \mathrm{proj}_k\, (\bx^{\otimes k})\big\vert_{\bx=\bmu}\\
  &= \frac{1}{d_{k,d}}\, \Sym_{d+1,k} \sum_{j=0}^{\lfloor k/2\rfloor} a_{k,j} (\operatorname{vec} \bI_{d+1})^{\otimes j}\otimes \bmu^{\otimes k-2j},
\end{align*}
proving the result.

\emph{Proof of \ref{thm:mom:4}}. For $m>k$, we follow a different approach, given the lack of an explicit expression for the projection of $\bx^{\otimes m}$ onto $\mathcal{H}_k^d$.

Consider the tangent-normal change of variables $\bx=t\bmu+(1-t^2)^{1/2}\bB_{\bmu}\bxi$, with $t=\bx^\top \bmu\in[-1,1]$, $\bxi\in\Sdm$, and $\bB_{\bmu}$ a semi-orthogonal $(d+1)\times d$ matrix such that $\bB_{\bmu}\bB_{\bmu}^\top=\bI_{d+1}-\bmu\bmu^\top$ and $\bB^\top_{\bmu}\bB_{\bmu}=\bI_{d}$. Then, $\sigma_{d}(\rd\bx)=(1-t^2)^{d/2-1}\,\rd t\,\sigma_{d-1}(\rd \bxi)$. Using this change and the binomial theorem for the Kronecker product, we have
\begin{align}
    I_{m,k}(\bmu)=&\;\int_{\Sdm}\int_{-1}^1 \big(t\bmu +(1-t^2)^{1/2}\bB_{\bmu}\bxi\big)^{\otimes m} C_k^{(d-1)/2}(t) (1-t^2)^{d/2-1}\,\rd t\,\sigma_{d-1}(\rd \bxi)\nonumber\\
    =&\; \Sym_{d+1,m}\int_{\Sdm} \int_{-1}^1 \sum_{\ell=0}^m \binom{m}{\ell} \bmu^{\otimes \ell}\otimes (\bB_{\bmu}\bxi)^{\otimes m-\ell} \nonumber\\
    &\times t^\ell (1-t^2)^{(m-\ell)/2} C_k^{(d-1)/2}(t)(1-t^2)^{d/2-1}\,\rd t \,\sigma_{d-1}(\rd \bxi)\nonumber\\
    =&\; \Sym_{d+1,m}\sum_{\ell=0}^m \binom{m}{\ell} \bmu^{\otimes \ell}\otimes \lrb{\bB_{\bmu}^{\otimes m-\ell} \int_{\Sdm} \bxi^{\otimes m-\ell}\,\sigma_{d-1}(\rd \bxi)} \nonumber\\
    &\times \int_{-1}^1 t^\ell (1-t^2)^{(d+m-\ell-2)/2} C_k^{(d-1)/2}(t)\,\rd t \nonumber\\
    =&\; \om{d-1}\Sym_{d+1,m}\sum_{\ell=0}^m c_{\ell,m,k} \binom{m}{\ell} \bmu^{\otimes \ell}\otimes \lrb{\bB_{\bmu}^{\otimes m-\ell}\, \bmu_{d,m-\ell}},\label{eq:Imk}
\end{align}
where we have used \eqref{eq:unifmom} and
\begin{align*}
  c_{\ell,m,k}=\int_{-1}^1 t^\ell (1-t^2)^{(d+m-\ell-2)/2} C_k^{(d-1)/2}(t)\,\rd t.
\end{align*}

We plug \eqref{eq:unifmom} into \eqref{eq:Imk} and apply Lemma \ref{lem:Bmuotimes} in Appendix~\ref{sec:aux}, using that the symmetrizer matrix commutes with the Kronecker powers, so
\begin{align*}
  \bB_{\bmu}^{\otimes p}\,\Sym_{d,p} = \Sym_{d+1,p}\,\bB_{\bmu}^{\otimes p}.
\end{align*}
This gives
\begin{align*}
  I_{m,k}(\bmu)=&\; \om{d-1}\Sym_{d+1,m}\sum_{\ell=0}^m c_{\ell,m,k} \binom{m}{\ell} \frac{(m-\ell-1)!!}{\prod_{r=0}^{(m-\ell)/2-1} (d+2r)} \\
  &\times \bmu^{\otimes \ell}\otimes \lrb{\bB_{\bmu}^{\otimes m-\ell} \lrc{\Sym_{d,m-\ell}(\operatorname{vec}\bI_{d})^{\otimes (m-\ell)/2}1_{\{m-\ell\;\text{even}\}}}}\\
  =&\; \om{d-1}\Sym_{d+1,m}\sum_{\ell=0}^m c_{\ell,m,k} \binom{m}{\ell} \frac{(m-\ell-1)!!}{\prod_{r=0}^{(m-\ell)/2-1} (d+2r)} \\
  &\times \bmu^{\otimes \ell}\otimes \lrb{\Sym_{d+1,m-\ell} \,\lrpbig{\operatorname{vec}\lrcbig{\bI_{d+1}-\bmu\bmu^\top}}^{\otimes (m-\ell)/2}}1_{\{m-\ell\;\text{even}\}}.
\end{align*}
The inner symmetrizer can be absorbed into the outer one, yielding
\begin{align*}
  I_{m,k}(\bmu)=&\; \om{d-1}\Sym_{d+1,m}\sum_{\substack{\ell=0\\m-\ell\;\text{even}}}^m c_{\ell,m,k} \binom{m}{\ell} \frac{(m-\ell-1)!!}{\prod_{r=0}^{(m-\ell)/2-1} (d+2r)} \\
  &\times \bmu^{\otimes \ell}\otimes \lrb{\lrpbig{\operatorname{vec}\bI_{d+1}-\bmu^{\otimes 2}}^{\otimes (m-\ell)/2}}
\end{align*}
and then the parity of the indexes can be enforced with the summation index $j=(m-\ell)/2$, $j=0,\ldots,\lfloor m/2\rfloor$, leading to
\begin{align}
  I_{m,k}(\bmu)=&\;\om{d-1}\,\Sym_{d+1,m} \sum_{j=0}^{\lfloor m/2\rfloor} c_{m-2j,m,k}\,\binom{m}{2j} \frac{(2j-1)!!}{\prod_{r=0}^{j-1}(d+2r)}\nonumber\\
    &\times \bmu^{\otimes (m-2j)}\otimes \lrb{\lrpbig{\operatorname{vec}\bI_{d+1}-\bmu^{\otimes 2}}^{\otimes j}}.\label{eq:Imk2}
\end{align}

It remains only to compute the coefficients
\begin{align*}
  c_{m-2j,m,k} %
  &=\int_{-1}^1 t^{m-2j} (1-t^2)^{b} C_k^{(d-1)/2}(t)\,\rd t
\end{align*}
where $b \defin d/2-1+j$. We use the series expansion of the Gegenbauer and Chebyshev polynomials \citep[see][Eqs.~18.5.10 and 18.5.11\_1]{NIST|SM}, unified for $d\geq 1$ as
\begin{align}
  C_k^{(d-1)/2}(x) =&\; g_{k,d} \sum_{s=0}^{\lfloor k/2\rfloor} (-1)^s\frac{\Gamma\lrp{(d-1)/2+k-s}}{s!(k-2s)!}(2x)^{k-2s}, \label{eq:gegenexp}\\
   g_{k,d}\defin&\;\begin{cases}
    k/2,& d=1,\\
    [\Gamma\lrp{(d-1)/2}]^{-1},& d\geq 2.
  \end{cases}\nonumber
\end{align}

Substituting \eqref{eq:gegenexp} into the definition of $c_{m-2j,m,k}$ gives
\begin{align}
  c_{m-2j,m,k} &=  g_{k,d} \int_{-1}^1 t^{m-2j} (1-t^2)^{b} \sum_{s=0}^{\lfloor k/2\rfloor} (-1)^s\frac{\Gamma\lrp{(d-1)/2+k-s}}{s!(k-2s)!}(2t)^{k-2s}\rd t \nonumber\\
  &= g_{k,d}\sum_{s=0}^{\lfloor k/2\rfloor} (-1)^s\frac{2^{k-2s}\Gamma\lrp{(d-1)/2+k-s}}{s!(k-2s)!} \int_{-1}^1 t^{a_s}(1-t^2)^{b}\,\rd t\label{eq:ints}
\end{align}
with $a_s \defin m-2j+k-2s=m+k-2(j+s)$.

The parity of the integrands in \eqref{eq:ints} depends on $a_s$, which in turn depends on $m+k$. If $m+k$ is odd, then all $a_s$ are odd and $c_{m-2j,m,k}=0$. If $m+k$ is even, then all $a_s$ are even and the integrals are beta-type integrals that can be computed explicitly. Assume henceforth that $m+k$ is even and write $a_s=2r_s$ with $r_s \defin (m+k)/2 - (j+s)$. Then
\begin{align}
  \int_{-1}^1 t^{2r}(1-t^2)^{b}\,\rd t %
  =\mathrm{B}(r+1/2,b+1)%
  =\frac{\Gamma(r+1/2)\Gamma(b+1)}{\Gamma(r+b+3/2)}.\label{eq:beta}
\end{align}
Substituting this expression back into \eqref{eq:ints}, and recalling the definitions of $b$ and $r_s$, we obtain
\begin{align*}
  c_{m-2j,m,k} &= g_{k,d}\sum_{s=0}^{\lfloor k/2\rfloor} (-1)^s\frac{2^{k-2s}\Gamma((d-1)/2+k-s)}{\Gamma((d-1)/2)s!(k-2s)!} \frac{\Gamma(r_s+1/2)\Gamma(b+1)}{\Gamma(r_s+b+3/2)}\\
  &=\frac{\Gamma(d/2+j)}{g_{k,d}} \sum_{s=0}^{\lfloor k/2\rfloor} (-1)^s\frac{2^{k-2s}\Gamma((d-1)/2+k-s)}{s!(k-2s)!} \frac{\Gamma((m+k+1)/2-j-s)}{\Gamma((d+k+m+1)/2-s)}.
\end{align*}

Coming back to \eqref{eq:Imk2}, we have proved that
\begin{align*}
    \frac{1}{\om{d}C_k^{(d-1)/2}(1)}&I_{m,k}(\bmu)\\
    =\frac{\om{d-1}}{\om{d}}&\frac{1}{C_k^{(d-1)/2}(1)}\,\Sym_{d+1,m} \sum_{j=0}^{\lfloor m/2\rfloor} e_{j,k,m}\, \bmu^{\otimes (m-2j)}\otimes \lrb{\lrpbig{\operatorname{vec}\bI_{d+1}-\bmu^{\otimes 2}}^{\otimes j}},
\end{align*}
where
\begin{align*}
  e_{j,k,m}\defin\binom{m}{2j} \frac{(2j-1)!!}{\prod_{r=0}^{j-1}(d+2r)} f_{j,m,k} 1_{\{m+k\;\text{even}\}}
\end{align*}
and $f_{j,m,k}\defin c_{m-2j,m,k}$.
\end{proof}

\begin{proof}[Proof of Corollary \ref{cor:momspecial}]
We derive the moments using Theorem \ref{thm:mom}\ref{thm:mom:2}. First, we need the coefficients $a_{k,j}$, $j=0,\ldots,\lfloor k/2\rfloor$, for $k=1,\ldots,4$:
\begin{align*}
  a_{1,0}&=1;\quad
  a_{2,0}=1,\quad a_{2,1}=-\frac{1}{d+1};\quad
  a_{3,0}=1,\quad a_{3,1}=-\frac{3}{d+3};\\
  a_{4,0}&=1,\quad a_{4,1}=-\frac{6}{d+5},\quad a_{4,2}=\frac{3}{(d+3)(d+5)}.
\end{align*}

Second, we compute $\bmu_{d+1,k}$ for $k=1,\ldots,4$ using \eqref{eq:unifmom}:
\begin{align*}
  \bmu_{d+1,1}&=\mathbf{0},\quad
  \bmu_{d+1,2}=\frac{1}{d+1}\,\operatorname{vec}\bI_{d+1},\quad
  \bmu_{d+1,3}=\mathbf{0},\\
  \bmu_{d+1,4}&=\frac{3}{(d+1)(d+3)}\,\Sym_{d+1,4}(\operatorname{vec}\bI_{d+1})^{\otimes 2}.
\end{align*}

Third, we obtain $d_{k,d}$ using \eqref{eq:dkd}:
\begin{align*}
  d_{1,d}&=d+1, \quad d_{2,d}=\frac{d(d+3)}{2}, \quad d_{3,d}=\frac{d(d+1)(d+5)}{6},\\
  d_{4,d}&=\frac{d(d+1)(d+2)(d+7)}{24}.
\end{align*}

We can now plug in these values into Theorem \ref{thm:mom}\ref{thm:mom:2} to obtain the desired results:
\begin{align*}
  \Ebig{\bX_1}=&\;\frac{\rho}{d_{1,d}}\, \Sym_{d+1,1} \bmu=\frac{\rho}{d+1}\bmu,\\
  \Ebig{\bX^{\otimes 2}}=&\;\bmu_{d+1,2}+\frac{\rho}{d_{2,d}}\, \Sym_{d+1,2} \lrb{\bmu^{\otimes 2}+a_{2,1} \operatorname{vec} \bI_{d+1}}\\
  =&\;\frac{1}{d+1}\operatorname{vec}\bI_{d+1}+\frac{2\rho}{d(d+3)}\lrb{\bmu^{\otimes 2}-\frac{1}{d+1}\operatorname{vec}\bI_{d+1}}\\
  \Ebig{\bX^{\otimes 3}}=&\;\frac{\rho}{d_{3,d}}\, \Sym_{d+1,3} \lrb{\bmu^{\otimes 3}+a_{3,1} (\operatorname{vec} \bI_{d+1})\otimes \bmu}\\
  =&\;\frac{6\rho}{d(d+1)(d+5)}\, \Sym_{d+1,3} \lrb{\bmu^{\otimes 3}-\frac{3}{d+3} (\operatorname{vec} \bI_{d+1})\otimes \bmu}\\
  \Ebig{\bX^{\otimes 4}}=&\;\bmu_{d+1,4}+\frac{\rho}{d_{4,d}}\, \Sym_{d+1,4} \lrb{\bmu^{\otimes 4}+a_{4,1} (\operatorname{vec} \bI_{d+1})\otimes \bmu^{\otimes 2}+a_{4,2} (\operatorname{vec} \bI_{d+1})^{\otimes 2}}\\
  =&\;\frac{3}{(d+1)(d+3)} \Sym_{d+1,4}(\operatorname{vec}\bI_{d+1})^{\otimes 2}\\
  & + \frac{24\rho}{d(d+1)(d+2)(d+7)}\Sym_{d+1,4}\bigg\{\bmu^{\otimes 4}-\frac{6}{d+5}\,(\operatorname{vec} \bI_{d+1})\otimes \bmu^{\otimes 2}\\
  &+\frac{3}{(d+3)(d+5)}(\operatorname{vec}\bI_{d+1})^{\otimes 2}\bigg\}.
\end{align*}
\end{proof}

\begin{proof}[Proof of Corollary \ref{cor:scatter}]
Using that $\operatorname{vec}(\bba\bb^\top)=\bba\otimes\bb$, we have
\begin{align*}
  \Vbig{\bX^{\otimes m}}%
  &=\operatorname{vec}_{(d+1)^{m},(d+1)^{m}}^{-1}\lrp{\Ebig{\operatorname{vec}\lrpbig{\bX^{\otimes m}\bX^{\otimes m\top}}}-\operatorname{vec}\lrp{\Ebig{\bX^{\otimes m}}\Ebig{\bX^{\otimes m}}^\top}}\\
  &=\operatorname{vec}_{(d+1)^{m},(d+1)^{m}}^{-1}\lrp{\Ebig{\bX^{\otimes 2m}}-\Ebig{\bX^{\otimes m}}^{\otimes 2}}.
\end{align*}
Hence, covariance matrices follow from Theorem \ref{thm:mom}. In particular, Theorem \ref{thm:mom}\ref{thm:mom:3} leads to $\Ebig{\bX^{\otimes 2m}}=\bmu_{d+1,2m}$ when $2m+k$ is odd, i.e., when $k$ is odd. For $k=1$, the result follows from $\bmu_{d+1,2}=(d+1)^{-1}\operatorname{vec}\bI_{d+1}$ and Corollary \ref{cor:momspecial}.
\end{proof}

\subsubsection{Proofs of results in Section~\ref{sec:characteristic}}
\label{sec:proofs:cf}

\begin{proof}[Proof of Proposition \ref{prp:mgf}]
To derive the moment generating function, we consider the second equation in \citet[p. 227]{Magnus1966|SM} and the arguments from the proof of Proposition 3 in \cite{Fernandez-de-Marcos2023|SM} to obtain
\begin{align}
	e^{\kappa x}=\lrp{\frac{2}{\kappa}}^{(d-1)/2} \sum_{\ell=0}^{\infty} e_{\ell,d}\, \mathcal{I}_{(2\ell+d-1)/2}(\kappa) C_\ell^{(d-1)/2}(x),\quad x\in(-1,1),\quad \kappa> 0, \label{eq:ekx}
\end{align}
where
\begin{align*}
e_{\ell,d}=\begin{cases}
  \Gamma((d-1)/2) (\ell+(d-1)/2), & d\geq2,\\
  2-\delta_{\ell,0}, & d=1.
\end{cases}
\end{align*}
Plugging \eqref{eq:ekx} into the definition of the moment generating function and using \eqref{eq:addgegen}, we have
\begin{align}
  M_{\bX}(\bt)%
  =&\;\frac{1}{\om{d}}\int_{\Sd} e^{\|\bt\| \bx^\top (\bt/\|\bt\|)} \lrb{1+\frac{\rho}{C_k^{(d-1)/2}(1)}C_k^{(d-1)/2}(\bx^\top\bmu)}\,\sigma_d(\rd\bx)\nonumber\\
  =&\;\frac{1}{\om{d}}\lrp{\frac{2}{\|\bt\|}}^{(d-1)/2}\sum_{\ell=0}^{\infty} e_{\ell,d} \,\mathcal{I}_{(2\ell+d-1)/2}(\|\bt\|) \nonumber\\
  &\times \int_{\Sd} C_\ell^{(d-1)/2}\lrpbigg{\frac{\bx^\top \bt}{\|\bt\|}} \lrb{1+\frac{\rho}{C_k^{(d-1)/2}(1)}C_k^{(d-1)/2}(\bx^\top\bmu)}\,\sigma_d(\rd\bx)\nonumber\\
  =&\;\lrp{\frac{2}{\|\bt\|}}^{(d-1)/2} \bigg\{e_{0,d} \,\mathcal{I}_{(d-1)/2}(\|\bt\|)\nonumber\\
  &+ \frac{\rho}{d_{k,d} } e_{k,d} \,\mathcal{I}_{(2k+d-1)/2}(\|\bt\|)\, C_k^{(d-1)/2}\lrpbigg{\frac{\bmu^\top \bt}{\|\bt\|}} \bigg\}.\label{eq:mgf1}
\end{align}

To derive the characteristic function, we use an analogous argument as in \eqref{eq:ekx} but now using the third equation in \citet[p. 227]{Magnus1966|SM} to obtain
\begin{align*}
	e^{\mathrm{i}\kappa x}=\lrp{\frac{2}{\kappa}}^{(d-1)/2} \sum_{\ell=0}^{\infty} e_{\ell,d}\, \mathrm{i}^\ell \mathcal{J}_{(2\ell+d-1)/2}(\kappa) C_\ell^{(d-1)/2}(x),\quad x\in(-1,1),\quad \kappa> 0. %
\end{align*}
Then,
\begin{align}
  \varphi_{\bX}(\bt)%
  =&\;\frac{1}{\om{d}}\int_{\Sd} e^{\mathrm{i}\|\bt\| \bx^\top (\bt/\|\bt\|)} \lrb{1+\frac{\rho}{C_k^{(d-1)/2}(1)}C_k^{(d-1)/2}(\bx^\top\bmu)}\,\sigma_d(\rd\bx)\nonumber\\
  =&\;\lrp{\frac{2}{\|\bt\|}}^{(d-1)/2} \bigg\{e_{0,d} \,\mathcal{J}_{(d-1)/2}(\|\bt\|)\nonumber\\
  &+ \frac{\rho}{d_{k,d} } e_{k,d} \,\mathrm{i}^k\mathcal{J}_{(2k+d-1)/2}(\|\bt\|)\, C_k^{(d-1)/2}\lrpbigg{\frac{\bmu^\top \bt}{\|\bt\|}} \bigg\}.\label{eq:cf1}
\end{align}

For $\bt=\zero$, both \eqref{eq:mgf1} and \eqref{eq:cf1} reduce to $1$ by continuity using that, for any $\bt\in\mathbb{R}^{d+1}$,
\begin{align*}
  \left|C_k^{(d-1)/2}\lrpbigg{\frac{\bmu^\top \bt}{\|\bt\|}}\right|\leq C_k^{(d-1)/2}(1)
\end{align*}
and that, as $x\to0$, $\mathcal{I}_{(d-1)/2}(x)\sim \mathcal{J}_{(d-1)/2}(x)\sim (x/2)^{(d-1)/2}/\Gamma((d+1)/2)$ \citep[Eqs. 10.30.1 and 10.7.3]{NIST|SM}. %
\end{proof}

\subsubsection{Proof of result in Section~\ref{sec:projections}}
\label{sec:proofs:projections}

\begin{proof}[Proof of Theorem \ref{thm:projgamma}]
Let us consider a rotationally symmetric density on $\Sd$ about $\bmu$, $\bx\mapsto c_d(g)g(\bx^\top\bmu)$, where $g:[-1,1]\to\mathbb{R}_{\geq0}$ is an angular function and $c_d(g)$ the normalizing constant. Note that \eqref{eq:sphcard} belongs to this setup.

Consider the tangent-normal decomposition $\bx=t\bga+(1-t^2)^{1/2}\bB_{\bga}\bxi$, where $t=\bx^\top \bga$, $\bxi\in \Sdm$, and $\bB_{\bga}^\top\bB_{\bga}=\bI_{d}$ and $\bB_{\bga}\bB_{\bga}^\top=\bI_{d+1}-\bga\bga^\top$. Then
\begin{align*}
    \int_{\Sd} &c_d(g) g(\bx^\top\bmu)\,\sigma_{d}(\rd\bx)\\
    &=\int_{\Sdm}\int_{-1}^1 c_d(g)g([t\bga+(1-t^2)^{1/2}\bB_{\bga}\bxi]^\top\bmu)(1-t^2)^{d/2-1}\,\rd t\,\sigma_{d-1}(\rd\bxi)\\
    &=\int_{-1}^1 c_d(g) (1-t^2)^{d/2-1}\lrb{\int_{\Sdm}g([t\bga+(1-t^2)^{1/2}\bB_{\bga}\bxi]^\top\bmu)\,\sigma_{d-1}(\rd\bxi)}\,\rd t\\
    &\indef\int_{-1}^1 c_d(g) (1-t^2)^{d/2-1}g_{\bga,\bmu}(t)\,\rd t
\end{align*}
and $T=\bga^\top\bX$ has density $c_d(g) (1-t^2)^{d/2-1}g_{\bga,\bmu}(t)$ on $[-1,1]$. Let us compute $g_{\bga,\bmu}$ in a simpler form under \eqref{eq:sphcard} (in this case $c_d(g)=1$) to obtain the density \eqref{eq:projpdf}.

We tackle first the case $d=1$, where $\xi\in\mathbb{S}^0=\{-1,1\}$ and $\bB_{\bga}=(\gamma_2,\,-\gamma_1)^\top$. Thus,
\begin{align*}
    g_{\bga,\bmu}(t)=&\;\frac{1}{2}\lrb{g([t\bga+(1-t^2)^{1/2}\bB_{\bga}]^\top\bmu)+g([t\bga-(1-t^2)^{1/2}\bB_{\bga}]^\top\bmu)}\\
    =&\;\frac{1}{4\pi}\lrb{2+\rho\, T_k([t\bga+(1-t^2)^{1/2}\bB_{\bga}]^\top\bmu)+\rho\, T_k([t\bga-(1-t^2)^{1/2}\bB_{\bga}]^\top\bmu)}.
\end{align*}
Denote $t=\cos\alpha$ and $\bga^\top\bmu=\cos\beta$, so that
\begin{align*}
  [t\bga\pm(1-t^2)^{1/2}\bB_{\bga}]^\top\bmu=&\;\cos\alpha\cos\beta \pm \sin\alpha\sin\beta=\cos(\alpha\mp\beta).
\end{align*}
Then
\begin{align*}
    g_{\bga,\bmu}(t)=&\;\frac{1}{4\pi}\lrb{2+\rho\, T_k(\cos(\alpha-\beta))+\rho\, T_k(\cos(\alpha+\beta))}\\
    =&\;\frac{1}{4\pi}\lrb{2+2\rho \cos(k\beta)\cos(k\alpha)}\\
    =&\;\frac{1}{2\pi}\lrb{1+\rho\, T_k(\bga^\top\bmu)T_k(t)}.%
\end{align*}
The density of $T=\bga^\top\bmu$ for $d=1$ is then
\begin{align}
    f_{\bga}(t)&=\frac{1}{2\pi}\lrb{1+\rho\, T_k(\bga^\top\bmu)T_k(t)} (1-t^2)^{-1/2}\nonumber\\
    &=f_1(t)\lrb{1+\rho\, \tilde{C}_{k}^{0}(\bga^\top\bmu)\tilde{C}_{k}^{0}(t)}.\label{eq:fT1}
\end{align}

We now tackle the case $d\geq 2$:
\begin{align*}
    g_{\bga,\bmu}(t)=&\;\int_{\Sdm}g([t\bga+(1-t^2)^{1/2}\bB_{\bga}\bxi]^\top\bmu)\,\sigma_{d-1}(\rd\bxi)\\
    =&\;\int_{\Sdm}\frac{1}{\om{d}}\lrb{1+\rho\, \tilde{C}_k^{(d-1)/2}([t\bga+(1-t^2)^{1/2}\bB_{\bga}\bxi]^\top\bmu)}\,\sigma_{d-1}(\rd\bxi)\\
    =&\;\frac{\om{d-1}}{\om{d}} +\frac{\rho}{\om{d}C_k^{(d-1)/2}(1)}\\
    &\times\int_{\Sdm} C_k^{(d-1)/2}\lrp{t[\bga^\top\bmu]+(1-t^2)^{1/2}(1-[\bga^\top\bmu]^2)^{1/2}\bxi^\top\lrc{\frac{\bB_{\bga}^\top\bmu}{\|\bB_{\bga}^\top\bmu\|}}}\,\sigma_{d-1}(\rd\bxi)\\
    =&\;\frac{\om{d-1}}{\om{d}} +\frac{\rho\,\om{d-2}}{\om{d}C_k^{(d-1)/2}(1)}\\
    &\times\int_{-1}^1 C_k^{(d-1)/2}\lrp{t[\bga^\top\bmu]+(1-t^2)^{1/2}(1-[\bga^\top\bmu]^2)^{1/2}s}(1-s^2)^{(d-3)/2}\,\rd s.
\end{align*}
Denote $t=\cos\alpha$ and $\bga^\top\bmu=\cos\beta$ and use Eq. ET II 283(20) in \cite{Gradshteyn2014|SM}:
\begin{align*}
   \int_{-1}^{1}(1-x^{2})^{\nu-1} &C_{k}^{\nu}(\cos \alpha \cos \beta+x \sin \alpha \sin \beta) \,\rd x\nonumber\\
   &=\frac{2^{2 \nu-1} k![\Gamma(\nu)]^{2}}{\Gamma(2 \nu+k)} C_{k}^{\nu}(\cos \alpha) C_{k}^{\nu}(\cos \beta).%
\end{align*}
Then,
\begin{align*}
    g_{\bga,\bmu}(t)=&\;\frac{\om{d-1}}{\om{d}} +\frac{\rho\,\om{d-2}}{\om{d}C_k^{(d-1)/2}(1)}\int_{-1}^1 C_k^{(d-1)/2}\lrp{\cos\alpha\cos\beta+\sin\alpha\sin\beta s}(1-s^2)^{(d-3)/2}\,\rd s\\
    =&\;\frac{\om{d-1}}{\om{d}} +\frac{\rho\,\om{d-2}}{\om{d}C_k^{(d-1)/2}(1)} \frac{2^{d-2} k![\Gamma((d-1)/2)]^{2}}{\Gamma(d-1+k)} C_{k}^{(d-1)/2}(t) C_{k}^{(d-1)/2}(\bga^\top\bmu)\\
    =&\;\frac{\om{d-1}}{\om{d}}\lrb{1 +\rho\frac{ \om{d-2} 2^{d-2} [\Gamma((d-1)/2)]^{2}}{\om{d-1}\Gamma(d-1) } \frac{C_{k}^{(d-1)/2}(\bga^\top\bmu) }{\lrcbig{C_k^{(d-1)/2}(1)}^2}C_{k}^{(d-1)/2}(t)},
\end{align*}
since $C_k^{(d-1)/2}(1)=\Gamma(d-1+k)/(\Gamma(d-1)k!)$. Using the expression for $\om{d}$ and the Legendre duplication formula %
$\Gamma\left((d-1)/2\right)\Gamma\left(d/2\right) = 2^{2-d}\sqrt{\pi}\Gamma(d-1)$, we have
\begin{align*}
    \frac{\om{d-2}2^{d-2} [\Gamma((d-1)/2)]^{2}}{\om{d-1}\Gamma(d-1)}=\frac{\pi^{-1/2} 2^{d-2} \Gamma((d-1)/2) }{\Gamma(d-1)/\Gamma(d/2)}
    =%
    1.
\end{align*}
The density of $T=\bga^\top\bmu$ for $d\geq 2$ is then
\begin{align}
    f_{\bga}(t)&=\frac{\om{d-1}}{\om{d}}\left\{1 +\rho\frac{C_{k}^{(d-1)/2}(\bga^\top\bmu)}{\lrcbig{C_k^{(d-1)/2}(1)}^2}C_{k}^{(d-1)/2}(t) \right\}(1-t^2)^{d/2-1}\nonumber\\
    &=f_d(t)\lrb{1+\rho\, \tilde{C}_{k}^{(d-1)/2}(\bga^\top\bmu)\tilde{C}_{k}^{(d-1)/2}(t)}.\label{eq:fT}
\end{align}
Expressions \eqref{eq:fT1} and \eqref{eq:fT} prove \eqref{eq:projpdf} for $d\geq 1$.

To obtain the cdf \eqref{eq:projcdf} we use
\begin{align*}
    G_k(x)\defin&-\int_{-1}^x C_{k}^{(d-1)/2}(t) (1-t^2)^{d/2-1}\,\rd t\nonumber\\
    =&\;\begin{cases}
    \displaystyle\frac{1}{k}\sin(k\cos^{-1}(x)),&d=1,\\
    \displaystyle\frac{d-1}{k(k+d-1)}C_{k-1}^{(d+1)/2}(x)(1-x^2)^{d/2},&d\geq 2.
    \end{cases}%
\end{align*}
(Lemma B.5 in \cite{Garcia-Portugues2020b|SM}), and \eqref{eq:fT1} and \eqref{eq:fT}:
\begin{align*}
    F_{\bga}(x)=&\;\int_{-1}^x f_d(t)+\lrb{\rho \frac{\om{d-1}}{\om{d}}\frac{C_k^{(d-1)/2}(\bga^\top\bmu)}{\lrcbig{C_k^{(d-1)/2}(1)}^2}}C_{k}^{(d-1)/2}(t)(1-t^2)^{d/2-1}\,\rd t\\
    =&\;F_d(x)+\rho\,\eta_k(\bga^\top\bmu)\int_{-1}^x C_{k}^{(d-1)/2}(t) (1-t^2)^{d/2-1}\,\rd t\\
    =&\;F_d(x)-\rho\,\eta_k(\bga^\top\bmu) G_k(x).
\end{align*}
\end{proof}

\subsubsection{Proof of result in Section~\ref{sec:simulation}}
\label{sec:proofs:simulation}

\begin{proof}[Proof of Proposition \ref{prp:algo2}]
We show that the density of $T=SR$ is $f_{\bmu}(t)=f_d(t)\lrbbig{1+\rho\tilde{C}_k^{(d-1)/2}(t)}$, as given in \eqref{eq:projpdf}.

By construction, the density of $T$ is
\begin{align*}
  f_{T}(t)&= 1_{\{t \in [-1,0)\}} f_{R}(|t|) \Prob{S=-1\mid R=|t|}+1_{\{t \in [0,1]\}} f_{R}(|t|) \Prob{S=1\mid R=|t|}\\
  &= f_{R}(|t|)\lrb{1_{\{t \in [-1,0)\}} \frac{1+\rho\tilde{C}_k^{(d-1)/2}(t)}{2} +1_{\{t \in [0,1]\}}  \frac{1+\rho\tilde{C}_k^{(d-1)/2}(t)}{2}}\\
  &= f_{R}(|t|)\frac{1+\rho\tilde{C}_k^{(d-1)/2}(t)}{2},
\end{align*}
where we have used the oddness of $\tilde{C}_k^{(d-1)/2}$.

Since $X=\bU^\top\be_1\sim F_{d}$ is symmetric, $f_d$ is even and the density of $R=|X|$ is $f_{R}(r)=f_{d}(r)+f_{d}(-r)=2f_d(r)$, $r\in[0,1]$. Therefore,
\begin{align*}
  f_{T}(t)=f_d(|t|)\lrb{1+\rho\tilde{C}_k^{(d-1)/2}(t)}=f_{\bmu}(t),
\end{align*}
proving the result.
\end{proof}

\subsection{Proofs of results in Section~\ref{sec:estimation}}
\label{sec:proofs:estimation}

\begin{proof}[Proof of Theorem \ref{thm:mm1}]
Let us denote $\bxi=\rho\bmu$. Then,
\begin{align*}
  \E{\bX}=\frac{1}{d+1}\bxi,\quad \V{\bX}=\frac{1}{d+1}\bI_{d+1}-\frac{1}{(d+1)^2}\bxi\bxi^\top.
\end{align*}

\textit{Proof of \ref{thm:mm1:1}}. By the Strong Law of Large Numbers (SLLN), $\bar{\bX}\stackrel{\text{a.s.}}{\longrightarrow} (d+1)^{-1}\bxi$. The continuous mapping theorem applied to $\bx\mapsto ((d+1)\|\bx\|,\bx/\|\bx\|)$, $\bx\neq\zero$, gives the almost sure convergence of the estimators.

\textit{Proof of \ref{thm:mm1:2}}. By the multivariate CLT,
\begin{align*}
  \sqrt{n}\lrp{\bar{\bX} - \E{\bX}}\inlaw \mathcal{N}_{d+1}\lrp{\mathbf{0}, \frac{1}{(d+1)^2}\big((d+1)\bI_{d+1}-\bxi\bxi^\top\big)}.
\end{align*}
We consider the transformation $g:\R^{d+1}\to \R \times \Sd$ defined by $g(\bx)\defin (\|\bx\|,\bx/\|\bx\|)$ to apply the delta method:
\begin{align*}
  \sqrt{n}(g(\bar{\bX}) - g(\E{\bX}))\inlaw \mathcal{N}_{d+2}\lrp{\mathbf{0},\mathsf{D} g(\E{\bX})\V{\bX}\mathsf{D} g(\E{\bX})^\top}.
\end{align*}

The function $g$ is differentiable at all $\bx\neq\zero$, with Jacobian matrix
\begin{align*}
  \mathsf{D} g(\bx)=\begin{pmatrix}
    \|\bx\|^{-1}\bx^\top \\
    \|\bx\|^{-1}(\bI_{d+1} - \|\bx\|^{-2}\bx\bx^\top)
\end{pmatrix}_{(d+2)\times (d+1)}
\end{align*}
that, when applied at $\E{\bX}=[\rho/(d+1)]\bmu$, becomes
\begin{align*}
  \mathsf{D} g(\E{\bX})=\begin{pmatrix}
    \bmu^\top \\
    \frac{d+1}{\rho}(\bI_{d+1} - \bmu\bmu^\top)
\end{pmatrix}_{(d+2)\times (d+1)}.
\end{align*}

The asymptotic covariance matrix becomes
\begin{align*}
  \mathsf{D} g(\E{\bX})&\V{\bX}\mathsf{D} g(\E{\bX})^\top\\
  &=\frac{1}{(d+1)^2}\mathsf{D} g(\E{\bX})\lrc{(d+1)\bI_{d+1}-\rho^2\bmu\bmu^\top}\mathsf{D} g(\E{\bX})^\top\\
  &=\frac{1}{(d+1)^2}\begin{pmatrix}
    \bmu^\top \lrcbig{(d+1)\bI_{d+1}-\rho^2\bmu\bmu^\top}\\
    \frac{d+1}{\rho}(\bI_{d+1} - \bmu\bmu^\top) \lrcbig{(d+1)\bI_{d+1}-\rho^2\bmu\bmu^\top}
  \end{pmatrix} \mathsf{D} g(\E{\bX})^\top\\
  &=\frac{1}{(d+1)^2}\begin{pmatrix}
    d+1-\rho^2 & \zero^\top\\
    \zero & (d+1)^3\rho^{-2}(\bI_{d+1} - \bmu\bmu^\top)\\
  \end{pmatrix}.
\end{align*}

Since $g(\E{\bX})=(\|\E{\bX}\|,\E{\bX}/\|\E{\bX}\|)=(\rho/(d+1),\bmu)$, then
\begin{align*}
\sqrt{n}
\begin{pmatrix}
    \|\bar{\bX}\|-\frac{\rho}{d+1} \\
    \frac{\bar{\bX}}{\|\bar{\bX}\|}-\bmu
\end{pmatrix}
\inlaw \mathcal{N}_{d+2}\lrp{\mathbf{0},
\frac{1}{(d+1)^2}\begin{pmatrix}
    d+1-\rho^2 & \zero^\top\\
    \zero & (d+1)^3\rho^{-2}(\bI_{d+1} - \bmu\bmu^\top)\\
  \end{pmatrix}}
\end{align*}
and therefore
\begin{align*}
\sqrt{n}
\begin{pmatrix}
    \hat{\rho}_{\mathrm{MM},1}-\rho \\
    \hat{\bmu}_{\mathrm{MM},1}-\bmu
\end{pmatrix}
\inlaw \mathcal{N}_{d+2}\lrp{\mathbf{0},
\begin{pmatrix}
    d+1-\rho^2 & \zero^\top\\
    \zero & (d+1)\rho^{-2}(\bI_{d+1} - \bmu\bmu^\top)\\
  \end{pmatrix}}.
\end{align*}
Reordering the components of this statement concludes the proof.
\end{proof}

\begin{proof}[Proof of Theorem \ref{thm:mm2}]
 We assume without loss of generality that $\rho>0$ and therefore $\bu(\cdot)=\bu_+(\cdot)$ and $\lambda(\cdot)=\lambda_-(\cdot)$ represent the first eigenpair.

The multivariate CLT
\begin{align*}
  \sqrt{n}\lrp{\operatorname{vec}(\bS) - \Ebig{\bX^{\otimes 2}}}\inlaw \mathcal{N}_{(d+1)^2}\lrp{\mathbf{0}, \Vbig{\bX^{\otimes 2}}}
\end{align*}
readily follows. We seek to apply the multivariate delta method to the map
\begin{align*}
  g:\operatorname{vec}(\bS)\in\R^{(d+1)^2}\mapsto (\lambda(\bS),\bu(\bS))\in \R_+\times\Sd.
\end{align*}
Note that $g$ is differentiable at any $\operatorname{vec}(\bS)$ such that $\lambda(\bS)$ is a simple eigenvalue of $\bS$, which is guaranteed as long as $\rho\neq 0$.

Theorem~7 in \citet[p. 158]{Magnus1999|SM} states that the differentials of a simple eigenpair $(\lambda,\bu)$ of a real symmetric matrix $\bS$ with respect to its entries are
\begin{align*}
  \rd\lambda = \bu^\top (\rd\bS)\,\bu, \quad \rd\bu = (\lambda\bI_{d+1}-\bS)^+ (\rd\bS)\,\bu.
\end{align*}
where $(\cdot)^+$ denotes the Moore--Penrose pseudoinverse. These differentials can be expressed as gradients/Jacobians with respect to $\operatorname{vec}(\bS)$. For that, we use $\mathrm{tr}(\bA^\top \bB)=\operatorname{vec}(\bA)^\top\operatorname{vec}(\bB)$ in $\rd\lambda$ and \eqref{eq:Kvec} in $\rd\bu$, to obtain
\begin{align*}
  \rd\lambda &= \mathrm{tr}\big(\bu\bu^\top\,\rd\bS\big) = \operatorname{vec}(\bu\bu^\top)^\top \operatorname{vec}(\rd\bS)=\bu^{\otimes 2\top} \operatorname{vec}(\rd\bS),\\
  \rd\bu &= (\lambda\bI_{d+1}-\bS)^+ [(\rd\bS)\,\bu] = (\lambda\bI_{d+1}-\bS)^+ [(\bu^\top\otimes \bI_{d+1})\,\operatorname{vec}(\rd\bS)].
\end{align*}
These differentials yield the Jacobian matrix
\begin{align*}
  \mathsf{D} g(\operatorname{vec}(\bS))=\begin{pmatrix}
    \bu^{\otimes 2\top}\\
    (\lambda\bI_{d+1}-\bS)^+ (\bu^\top\otimes \bI_{d+1})
  \end{pmatrix}_{(1+(d+1))\times (d+1)^2}.
\end{align*}

We evaluate the Jacobian at $\bSigma=a(\rho)\bmu\bmu^\top + b(\rho)(\bI_{d+1}-\bmu\bmu^\top)$. First, note that
\begin{align*}
a(\rho) \bI_{d+1} - \bSigma
&= a(\rho)\bI_{d+1} - \big[b(\rho)\bI_{d+1} + (a(\rho)-b(\rho))\bmu\bmu^\top\big]\\
&= (a(\rho)-b(\rho))\big(\bI_{d+1}-\bmu\bmu^\top\big)\\
&=\frac{2\rho}{d(d+3)}\big(\bI_{d+1}-\bmu\bmu^\top\big)
\end{align*}
so the Moore--Penrose pseudoinverse is
\begin{align*}
  (a(\rho)\bI_{d+1}-\bSigma)^+= \frac{d(d+3)}{2\rho}(\bI_{d+1}-\bmu\bmu^\top).
\end{align*}
Therefore,
\begin{align*}
  \mathsf{D} g(\operatorname{vec}(\bSigma))&=
  \begin{pmatrix}
    \boldsymbol{\partial}_{\lambda} g^\top\\
    \boldsymbol{\partial}_{\bu}g
  \end{pmatrix}=\begin{pmatrix}
    \bmu^{\otimes 2\top}\\
    \frac{d(d+3)}{2\rho}(\bI_{d+1}-\bmu\bmu^\top) (\bmu^\top\otimes \bI_{d+1})
  \end{pmatrix}_{(1+(d+1))\times (d+1)^2}
\end{align*}
and
\begin{align*}
\sqrt{n}
\begin{pmatrix}
    \lambda(\bS)-a(\rho) \\
    \bu(\bS)-\bmu
\end{pmatrix}
\inlaw \mathcal{N}_{d+2}\lrp{\mathbf{0},
\mathsf{D} g(\operatorname{vec}(\bSigma))\Vbig{\bX^{\otimes 2}}\mathsf{D} g(\operatorname{vec}(\bSigma))^\top}.
\end{align*}

The asymptotic covariance matrix has the following block structure
\begin{align*}
  \mathsf{D} g(\operatorname{vec}(\bSigma))\Vbig{\bX^{\otimes 2}}\mathsf{D} g(\operatorname{vec}(\bSigma))^\top=\begin{pmatrix}
    \boldsymbol{\partial}_{\lambda} g^\top \, \Vbig{\bX^{\otimes 2}} \boldsymbol{\partial}_{\lambda} g & \boldsymbol{\partial}_{\lambda} g^\top \, \Vbig{\bX^{\otimes 2}} \boldsymbol{\partial}_{\bu} g^\top \\[0.2em]
    \boldsymbol{\partial}_{\bu} g \, \Vbig{\bX^{\otimes 2}} \boldsymbol{\partial}_{\lambda} g & \boldsymbol{\partial}_{\bu} g \, \Vbig{\bX^{\otimes 2}} \boldsymbol{\partial}_{\bu} g^\top
  \end{pmatrix}.
\end{align*}
We proceed to evaluate each block separately.

\emph{Block $\boldsymbol{\partial}_{\lambda} g^\top \, \Vbig{\bX^{\otimes 2}} \boldsymbol{\partial}_{\lambda} g$.}

We will use that, for a $(d+1)\times(d+1)$ matrix $\bA$ and the random vector $\bX$, due to
\begin{align*}
  \bX^\top \bA\bX=\mathrm{tr}(\bA \bX\bX^\top)=\operatorname{vec}(\bA)^\top \operatorname{vec}(\bX\bX^\top)=\operatorname{vec}(\bA)^\top \bX^{\otimes 2},
\end{align*}
we have that
\begin{align*}
  \operatorname{vec}(\bA)^\top \Vbig{\bX^{\otimes 2}} \operatorname{vec}(\bB)=&\;\Covbig{\bX^\top \bA\bX}{\bX^\top \bB \bX}.
\end{align*}
We also use unvectorization operator $\operatorname{vec}^{-1}(\cdot)$ mapping to $(d+1)\times (d+1)$ matrices:
\begin{align*}
  \operatorname{vec}^{-1}(\boldsymbol{\partial}_{\lambda} g)= \operatorname{vec}^{-1}\lrpbig{\bmu^{\otimes 2\top}}=\bmu\bmu^\top.
\end{align*}
Using these facts, we compute
\begin{align*}
  \boldsymbol{\partial}_{\lambda} g^\top \, \Vbig{\bX^{\otimes 2}} \boldsymbol{\partial}_{\lambda} g&= \Covbig{\bX^\top\operatorname{vec}^{-1}(\boldsymbol{\partial}_{\lambda} g) \bX}{\bX^\top \operatorname{vec}^{-1}(\boldsymbol{\partial}_{\lambda} g)\bX}\\
  &= \Covbig{\bX^\top \bmu\bmu^\top \bX}{\bX^\top \bmu\bmu^\top \bX}\\
  &= \Vbig{(\bmu^\top \bX)^2}=\Ebig{(\bmu^\top \bX)^4} - \Ebig{(\bmu^\top \bX)^2}^2\\
  &\indef v_1(\rho).
\end{align*}

We compute these projected moments using $\tilde{C}_2^{(d-1)/2}(t)=d^{-1}\big((d+1)t^2 -1\big)$ and \eqref{eq:beta}:
\begin{align*}
  \Ebig{(\bmu^\top \bX)^{2m}}%
  &=\frac{\om{d-1}}{\om{d}}\int_{-1}^1 t^{2m} \lrb{1 + \rho \, \tilde{C}_2^{(d-1)/2}(t)} (1-t^2)^{d/2-1}\, \rd t\\
  &=\frac{\om{d-1}}{\om{d}}\lrb{\frac{d-\rho}{d}\int_{-1}^1 t^{2m} (1-t^2)^{d/2-1}\, \rd t+\frac{\rho(d+1)}{d}\int_{-1}^1 t^{2m+2} (1-t^2)^{d/2-1}\, \rd t}\\
  &=\frac{\om{d-1}}{\om{d}\,d}\lrb{(d-\rho)\mathrm{B}(m+1/2,d/2)+\rho(d+1)\mathrm{B}(m+3/2,d/2)}.
\end{align*}
For $m=1$ this yields
\begin{align*}
  \Ebig{(\bmu^\top \bX)^2}%
  &=\frac{d+3+2\rho}{(d+1)(d+3)}
\end{align*}
and
\begin{align*}
  \Ebig{(\bmu^\top \bX)^4}
  &=\frac{\om{d-1}}{\om{d}\,d}\lrb{(d-\rho)\frac{\Gamma(5/2)\Gamma(d/2)}{\Gamma(d/2+5/2)} +\rho(d+1)\frac{\Gamma(7/2)\Gamma(d/2)}{\Gamma(d/2+7/2)}}\\
  &=\frac{3(d+5+4\rho)}{(d+1)(d+3)(d+5)}.
\end{align*}
Then
\begin{align*}
  v_1(\rho)=&\; \Ebig{(\bmu^\top \bX)^4} - \Ebig{(\bmu^\top \bX)^2}^2\\
  =&\;\frac{3(d+5+4\rho)}{(d+1)(d+3)(d+5)} - \lrb{\frac{d+3+2\rho}{(d+1)(d+3)}}^2\\
  =&\;2\frac{d(d+3)(d+5) + 4(d-1)(d+3)\rho -2(d+5)\rho^2}{(d+1)^2(d+3)^2(d+5)}.
\end{align*}

\emph{Blocks $\boldsymbol{\partial}_{\lambda} g^\top \, \Vbig{\bX^{\otimes 2}} \boldsymbol{\partial}_{\bu} g^\top$ and $\boldsymbol{\partial}_{\bu} g \, \Vbig{\bX^{\otimes 2}} \boldsymbol{\partial}_{\lambda} g$.}

We show that these two blocks are zero matrices. By symmetry, it suffices to consider $\boldsymbol{\partial}_{\lambda} g^\top \, \Vbig{\bX^{\otimes 2}} \boldsymbol{\partial}_{\bu} g^\top$. We check that
\begin{align*}
  \lrp{\boldsymbol{\partial}_{\lambda} g^\top \, \Vbig{\bX^{\otimes 2}} \boldsymbol{\partial}_{\bu} g^\top}\bw=0,\quad \text{for all } \bw\in\R^{d+1}.
\end{align*}
First, we unvectorize to $(d +1)\times (d +1)$ matrices:
\begin{align*}
  \operatorname{vec}^{-1}(\boldsymbol{\partial}_{\bu} g^\top\bw)&=\frac{d(d+3)}{2\rho}\operatorname{vec}^{-1}\lrp{\lrcbig{(\bI_{d+1}-\bmu\bmu^\top) (\bmu^\top\otimes \bI_{d+1})}^\top \bw}\\
  &=\frac{d(d+3)}{2\rho}\operatorname{vec}^{-1}\lrp{(\bmu\otimes \bI_{d+1})  \lrcbig{(\bI_{d+1}-\bmu\bmu^\top) \bw}}\\
  &=\frac{d(d+3)}{2\rho} [(\bI_{d+1}-\bmu\bmu^\top) \bw] \bmu^\top,
\end{align*}
where we have used $\operatorname{vec}(\bba\bb^\top) = (\bb\otimes\bI_{d+1})\,\bba$.

Combining these results, we have
\begin{align*}
  (\boldsymbol{\partial}_{\lambda} g)^\top &\Vbig{\bX^{\otimes 2}} (\boldsymbol{\partial}_{\bu} g^\top\bw)\\
  =&\;\Covbig{\bX^\top\operatorname{vec}^{-1}(\boldsymbol{\partial}_{\lambda} g)\, \bX}{\bX^\top\operatorname{vec}^{-1}(\boldsymbol{\partial}_{\bu} g^\top\bw)\bX}\\
  =&\;\Covbig{(\bmu^\top\bX)^2}{\bX^\top\operatorname{vec}^{-1}(\boldsymbol{\partial}_{\bu} g^\top\bw)\bX}\\
  =&\;\frac{d(d+3)}{2\rho} \Covbig{(\bmu^\top\bX)^2}{[\bX^\top(\bI_{d+1}-\bmu\bmu^\top) \bw] (\bmu^\top\bX)}.%
\end{align*}
Using the tangent-normal decomposition $\bX=\bmu T + \bB_{\bmu}\sqrt{1-T^2}\,\bXi$ with $T=\bmu^\top\bX$ and $\bXi\sim \mathrm{Unif}(\mathbb{S}^{d-1})$ independent, it follows that
\begin{align}
  \Ebig{(\bmu^\top\bX)^3 [\bX^\top(\bI_{d+1}-\bmu\bmu^\top) \bw]}%
  &=\Ebig{T^3\sqrt{1-T^2}}\Ebig{\bXi^\top(\bB_{\bmu}^\top\bw)}=0\label{eq:tang0}
\end{align}
and hence the covariance is null. Consequently, $\boldsymbol{\partial}_{\lambda} g^\top \, \Vbig{\bX^{\otimes 2}} \boldsymbol{\partial}_{\bu} g^\top = \zero^\top$.

\emph{Block $\boldsymbol{\partial}_{\bu} g \, \Vbig{\bX^{\otimes 2}} \boldsymbol{\partial}_{\bu} g^\top$.}

We compute the $ij$th element of this block, for $i,j=1,\ldots,d+1$. First, note that
\begin{align*}
  \operatorname{vec}^{-1}(\boldsymbol{\partial}_{\bu} g^\top\be_i)&=\frac{d(d+3)}{2\rho} [(\bI_{d+1}-\bmu\bmu^\top) \be_i] \bmu^\top.
\end{align*}
We have
\begin{align}
  \be_i^\top \boldsymbol{\partial}_{\bu} g &\, \Vbig{\bX^{\otimes 2}} \boldsymbol{\partial}_{\bu} g^\top \be_j\nonumber\\
  =&\;\lrpbig{\boldsymbol{\partial}_{\bu} g^\top \be_i}^\top \, \Vbig{\bX^{\otimes 2}} (\boldsymbol{\partial}_{\bu} g^\top \be_j)\nonumber\\
  =&\;\Covbig{\bX^\top \operatorname{vec}^{-1}(\boldsymbol{\partial}_{\bu} g^\top\be_i)\bX}{\bX^\top\operatorname{vec}^{-1}(\boldsymbol{\partial}_{\bu} g^\top\be_j)\bX}\nonumber\\
  =&\;\frac{d^2(d+3)^2}{4\rho^2}\Covbig{(\bmu^\top\bX)[\be_i^\top (\bI_{d+1}-\bmu\bmu^\top) \bX]}{(\bmu^\top\bX) [\bX^\top (\bI_{d+1}-\bmu\bmu^\top) \be_j]}\nonumber\\
  =&\;\frac{d^2(d+3)^2}{4\rho^2}\be_i^\top \bigg\{\Ebig{(\bmu^\top\bX)^2[(\bI_{d+1}-\bmu\bmu^\top) \bX] [\bX^\top (\bI_{d+1}-\bmu\bmu^\top)]}\nonumber\\
  &-(\bI_{d+1}-\bmu\bmu^\top)\Ebig{(\bmu^\top\bX) \bX}\Ebig{(\bmu^\top\bX) \bX^\top}(\bI_{d+1}-\bmu\bmu^\top)\bigg\}\be_j\nonumber\\
  =&\;\frac{d^2(d+3)^2}{4\rho^2}\be_i^\top (\bI_{d+1}-\bmu\bmu^\top)\Ebig{(\bmu^\top\bX)^2\bX\bX^\top}(\bI_{d+1}-\bmu\bmu^\top) \be_j,\label{eq:eij}
\end{align}
where we have used $(\bI_{d+1}-\bmu\bmu^\top)\Ebig{(\bmu^\top\bX)\bX}=\zero$ due to an analogous argument as in \eqref{eq:tang0}.

Thus, it remains to compute the following matrix:
\begin{align*}
  (\bI_{d+1}&-\bmu\bmu^\top)\Ebig{(\bmu^\top\bX)^2\bX \bX^\top}\\
  =&\;(\bI_{d+1}-\bmu\bmu^\top)\frac{\om{d-1}}{\om{d}}\int_{-1}^1 t^2 \lrc{t^2\bmu\bmu^\top+\frac{1-t^2}{d}(\bI_{d+1}-\bmu\bmu^\top)} \lrb{1+\frac{\rho}{d}\,\big((d+1)t^2-1\big)}\\
  &\times (1-t^2)^{d/2-1}\,\rd t\\
  =&\;(\bI_{d+1}-\bmu\bmu^\top) \frac{\om{d-1}}{\om{d}\, d^2} \lrb{\rho(d+1)\int_{-1}^1 t^4 (1-t^2)^{d/2}\,\rd t+ (d-\rho)\int_{-1}^1 t^2 (1-t^2)^{d/2}\,\rd t}\\
  =&\;(\bI_{d+1}-\bmu\bmu^\top) \frac{\om{d-1}}{\om{d}\, d^2} \lrb{\rho(d+1)\mathrm{B}(5/2,d/2+1)+ (d-\rho)\mathrm{B}(3/2,d/2+1)},
\end{align*}
where the last line used \eqref{eq:beta}. Further simplification gives
\begin{align}
  (\bI_{d+1}&-\bmu\bmu^\top)\Ebig{(\bmu^\top\bX)^2\bX \bX^\top}\nonumber\\
  =&\;(\bI_{d+1}-\bmu\bmu^\top) \frac{\om{d-1}}{\om{d}\, d^2} \lrb{\rho(d+1)\frac{\Gamma(5/2)\Gamma(d/2+1)}{\Gamma(d/2+7/2)}+ (d-\rho)\frac{\Gamma(3/2)\Gamma(d/2+1)}{\Gamma(d/2+5/2)}}\nonumber\\
  =&\;(\bI_{d+1}-\bmu\bmu^\top) \frac{d(d+5)+2(d-1)\rho}{d(d+1)(d+3)(d+5)}.\label{eq:Imu2}
\end{align}
Replacing \eqref{eq:Imu2} into \eqref{eq:eij} gives
\begin{align*}
    \be_i^\top \boldsymbol{\partial}_{\bu} g &\, \Vbig{\bX^{\otimes 2}} \boldsymbol{\partial}_{\bu} g^\top \be_j\\
    &=\frac{d^2(d+3)^2}{4\rho^2}\be_i^\top (\bI_{d+1}-\bmu\bmu^\top) \frac{d(d+5)+2(d-1)\rho}{d(d+1)(d+3)(d+5)} \be_j\\
    &=\frac{d(d+3)[d(d+5)+2(d-1)\rho]}{4\rho^2(d+1)(d+5)} \be_i^\top (\bI_{d+1}-\bmu\bmu^\top) \be_j\\
    &\indef v_2(\rho) \be_i^\top (\bI_{d+1}-\bmu\bmu^\top) \be_j,
\end{align*}
in turn providing $\boldsymbol{\partial}_{\bu} g \, \Vbig{\bX^{\otimes 2}} \boldsymbol{\partial}_{\bu} g^\top = v_2(\rho)\,(\bI_{d+1}-\bmu\bmu^\top)$.

Putting together the three asymptotic covariance blocks gives
\begin{align*}
\sqrt{n}
\begin{pmatrix}
    \lambda(\bS)-a(\rho) \\
    \bu(\bS)-\bmu
\end{pmatrix}
\inlaw \mathcal{N}_{d+2}\lrp{\mathbf{0},
\begin{pmatrix}
    v_1(\rho) & \zero^\top\\
    \zero &  v_2(\rho) (\bI_{d+1} - \bmu\bmu^\top)\\
  \end{pmatrix}}.
\end{align*}

Finally, we apply the linear map $h(x)=((d+3)/2)((d+1)x-1)$ to the first entry, for which $h(\lambda(\bS))=\hat{\rho}_{\mathrm{MM},2}$ and $h(a(\rho))=\rho$. Another application of the delta method yields the stated result:
\begin{align}
\sqrt{n}
\begin{pmatrix}
    \hat{\rho}_{\mathrm{MM},2}-\rho \\
    \hat{\bmu}_{\mathrm{MM},2}-\bmu
\end{pmatrix}
\inlaw \mathcal{N}_{d+2}\lrp{\mathbf{0},
\begin{pmatrix}
    \sigma^2_{\mathrm{MM},2}(\rho) & \zero^\top\\
    \zero &  \sigma^2_{\mathrm{MM},2}(\bmu) (\bI_{d+1} - \bmu\bmu^\top)\\
  \end{pmatrix}},\label{eq:clt2}
\end{align}
where $\sigma^2_{\mathrm{MM},2}(\bmu)=v_2(\rho)$ and
\begin{align*}
  \sigma^2_{\mathrm{MM},2}(\rho)&=\lrp{\frac{(d+3)(d+1)}{2}}^2v_1(\rho)%
  = \frac{d(d+3)}{2}+ \frac{2(d-1)(d+3)}{d+5}\rho -\rho^2.
\end{align*}
For $d=1$, $\sigma_{\mathrm{MM},2}^2(\bmu)=2\rho^{-2}=\sigma_{\mathrm{MM},1}^2(\bmu)$ and $\sigma_{\mathrm{MM},2}^2(\rho)=2-\rho^2=\sigma_{\mathrm{MM},1}^2(\rho)$.

Reordering the components of \eqref{eq:clt2} concludes the proof.
\end{proof}

\begin{proof}[Proof of Theorem \ref{thm:gegen-moments}]
The estimator is motivated by the expectation of the Gegenbauer moment:
\begin{align*}
  \Ebig{C_k^{(d-1)/2}(\bX^\top\bmu)}%
  &=\frac{1}{\om{d}}\int_{\Sd} C_k^{(d-1)/2}(\bx^\top\bmu)\lrbbig{1+\rho\,\tilde{C}_k^{(d-1)/2}(\bx^\top\bmu)}\,\sigma_d(\rd\bx)\\
  &=\rho\frac{\om{d-1}}{\om{d}\,C_k^{(d-1)/2}(1)}\int_{-1}^1 \lrcbig{C_k^{(d-1)/2}(t)}^2 (1-t^2)^{d/2-1}\,\rd t\\
  &=\rho\frac{\om{d-1}}{\om{d}\,C_k^{(d-1)/2}(1)}c_{k,d}\\
  &=\frac{\rho}{\tau_{k,d}},
\end{align*}
where we have used the orthogonality of the Gegenbauer polynomials and \eqref{eq:ckd2}. Consequently, the unbiasedness \ref{thm:gegen-moments:i} follows immediately. Strong consistency \ref{thm:gegen-moments:ii} follows from the SLLN.

The asymptotic normality \ref{thm:gegen-moments:iii} follows from the CLT. The variance of $\hat{\rho}_{\mathrm{GM}}$ follows from
\begin{align*}
  \Ebig{&\lrbbig{C_k^{(d-1)/2}(\bX^\top \bmu)}^2}\\
  &=\frac{\om{d-1}}{\om{d}}\int_{-1}^1 \lrcbig{C_k^{(d-1)/2}(t)}^2\lrbbig{1+\rho\,\tilde{C}_k^{(d-1)/2}(t)}(1-t^2)^{d/2-1}\,\rd t\\
  &=\tau_{k,d}^{-2}\lrb{d_{k,d}+\rho\,\tau_{k,d}\frac{I_3(k)}{c_{k,d}}},
\end{align*}
where
\begin{align*}
  I_3(k)\defin\int_{-1}^1 \lrcbig{C_k^{(d-1)/2}(t)}^3(1-t^2)^{d/2-1}\,\rd t.
\end{align*}
We then have
\begin{align*}
  \Vbig{C_k^{(d-1)/2}(\bX^\top \bmu)}&=\Ebig{\lrbbig{C_k^{(d-1)/2}(\bX^\top \bmu)}^2} - \Ebig{C_k^{(d-1)/2}(\bX^\top \bmu)}^2\\
  &= \tau_{k,d}^{-2}\lrb{d_{k,d} + \rho\,\tau_{k,d}\frac{I_3(k)}{c_{k,d}} -\rho^2}.
\end{align*}
For $k$ odd, $I_3(k)=0$ by symmetry. Using Lemma 2 in \cite{Fernandez-de-Marcos2023|SM}, we can compute $I_3(k)$ for $k$ even. For $d\geq2$, $I_3(k)$ is given as the term $a_{k/2,k,k,(d-1)/2}$ defined therein:
\begin{align}
\lrp{1+\frac{2k}{d-1}}\frac{I_3(k)}{c_{k,d}}&=\lrp{1+\frac{2k}{d-1}}a_{k/2,k,k,(d-1)/2}\nonumber\\
  &=\lrp{1+\frac{2k}{d-1}}\frac{2k+d-1}{3k+d-1} \frac{1}{(k/2)!} \binom{k}{k/2}\nonumber\\
  &\quad\times \frac{\Gamma((d+k-1)/2) \mathrm{B}((d+k-1)/2,(d+k-1)/2)}{\Gamma((d-1)/2) \mathrm{B}((d+3k-1)/2,(d-1)/2)}\nonumber\\
  &=\frac{(2k+d-1)^2}{(3k+d-1)(d-1)} \frac{k!}{((k/2)!)^3}\nonumber\\
  &\quad \times \frac{\Gamma((d+k-1)/2)^3 \Gamma(d+3k/2-1)}{\Gamma(d+k-1) \Gamma((d-1)/2)^2 \Gamma((d+3k-1)/2)}\nonumber\\
  &\indef\eta_{k,d}.\label{eq:tauk}
\end{align}
For $d=1$, $I_3(k)=0\indef \eta_{k,1}$ for $k$ even.

The variance of $\hat{\rho}_{\mathrm{GM}}$ is then
\begin{align}
  \sigma^2_{\mathrm{GM},k}(\rho)=\tau_{k,d}^{2}\Vbig{C_k^{(d-1)/2}(\bX^\top \bmu)}=d_{k,d} + \rho\, \eta_{k,d}\, 1_{\{k \text{ even}\}} -\rho^2.\label{eq:sigma2gm}
\end{align}

That $\sigma^2_{\mathrm{GM},1}(\rho)=\sigma^2_{\mathrm{MM},1}(\rho)$ follows immediately from \eqref{eq:sigma2gm} and \eqref{eq:tauk}. For $k=2$ and $d\geq2$,
\begin{align*}
  \eta_{2,d}%
  &=\frac{8(d+3)}{(d+5)(d+1)(d-1)} \frac{\Gamma((d+1)/2)^2 (d+1)}{\Gamma((d-1)/2)^2}%
  =\frac{2 (d-1)(d+3)}{(d+5)},
\end{align*}
so that $\sigma^2_{\mathrm{GM},2}(\rho)=\sigma^2_{\mathrm{MM},2}(\rho)$ for $d\geq2$.
\end{proof}

\begin{proof}[Proof of Theorem \ref{thm:mle}]
The proof is split into computing the Fisher information matrix $\bIcal(\bxi)$, obtaining the asymptotic covariance matrix of $(\hat{\rho}_{\mathrm{MLE}}, \hat{\bmu}_{\mathrm{MLE}})$, and finally verifying conditions for the ML estimation asymptotics that ensure statements \ref{thm:mle}\ref{thm:mle:1}--\ref{thm:mle}\ref{thm:mle:3}.

\emph{Computation of $\bIcal(\bxi)$ for $d\geq 2$.}

We begin by deriving the score function $\bx\mapsto \dot{\ell}(\bxi)\defin\frac{\partial}{\partial \bxi} \ell(\bxi)$, where
\begin{align*}
  \ell(\bxi)\defin\log f_{\mathrm{C}_k}(\bx;\bxi)=-\log(\om{d}) + \log\lrbbig{1+\|\bxi\| \tilde{C}_k^{(d-1)/2}(\bx^\top\bxi/\|\bxi\|)}
\end{align*}
We proceed to differentiate with respect to $\bxi$, use that $[C_k^{(d-1)/2}(x)]'=(d-1) C_{k-1}^{(d+1)/2}(x)$, and rearrange terms:
\begin{align}
  \dot{\ell}(\bxi)%
  =&\;\frac{1}{C_k^{(d-1)/2}(1)+\|\bxi\| C_k^{(d-1)/2}(\bx^\top\bxi/\|\bxi\|)}\nonumber\\
  &\times\lrb{\|\bxi\|^{-1}\bxi C_k^{(d-1)/2}(\bx^\top\bxi/\|\bxi\|)
  +(d-1)C_{k-1}^{(d+1)/2}(\bx^\top \bxi/\|\bxi\|) (\bI_{d+1} - \|\bxi\|^{-2}\bxi\bxi^\top)\bx}\nonumber\\
  =&\;\frac{1}{C_k^{(d-1)/2}(1)+\rho\, C_k^{(d-1)/2}(\bx^\top\bmu)}\nonumber\\
  &\times\lrbBig{\!\lrc{C_k^{(d-1)/2}(\bx^\top\bmu) - (d-1)C_{k-1}^{(d+1)/2}(\bx^\top \bmu)(\bx^\top\bmu)}\bmu +\lrc{(d-1)C_{k-1}^{(d+1)/2}(\bx^\top \bmu)} \bx}.\nonumber%
\end{align}

Let us consider the tangent-normal change of variables $t=\bx^\top\bmu$, so that $\bx=t\bmu+(1-t^2)^{1/2}\bB_{\bmu}\bga$ for some $\bga\in\Sdm$ (see the proof of Theorem \ref{thm:mom}\ref{thm:mom:4}). Then,
\begin{align*}
  \dot{\ell}(\bxi)%
  =&\;\frac{1}{C_k^{(d-1)/2}(1)+\rho\, C_k^{(d-1)/2}(t)}\\
  &\times\lrb{C_k^{(d-1)/2}(t)\bmu+\lrc{(d-1)C_{k-1}^{(d+1)/2}(t)(1-t^2)^{1/2}} \bB_{\bmu}\bga}.
\end{align*}

From this expression we proceed to obtain the Fisher information matrix $\bIcal(\bxi)$. First,
\begin{align}
  \dot{\ell}(\bxi)\dot{\ell}&(\bxi)^\top\nonumber\\
  =&\;\frac{1}{\lrpbig{C_k^{(d-1)/2}(1)+\rho\, C_k^{(d-1)/2}(t)}^2}\nonumber\\
  &\times\bigg\{C_k^{(d-1)/2}(t)^2\bmu\bmu^\top+(d-1)^2C_{k-1}^{(d+1)/2}(t)^2(1-t^2) \bB_{\bmu}\bga\bga^\top\bB_{\bmu}^\top\nonumber\\
  &+C_k^{(d-1)/2}(t)\lrc{(d-1)C_{k-1}^{(d+1)/2}(t)(1-t^2)^{1/2}} \lrc{\bmu(\bB_{\bmu}\bga)^\top+(\bB_{\bmu}\bga)\bmu^\top}\bigg\}.\label{eq:ell2}
\end{align}
Second, using $\int_{\Sdm} \bga\,\sigma_{d-1}(\rd \bga)=\zero$, $\int_{\Sdm} \bga\bga^\top\,\sigma_{d-1}(\rd \bga)=(\om{d-1}/d)\bI_{d}$, $\bmu^\top\bB_{\bmu}=\zero^\top$, and also $\bB_{\bmu}\bB_{\bmu}^\top=\bI_{d+1}-\bmu\bmu^\top$, we have
\begin{align*}
  \bIcal(\bxi)=&\;\Ebig{\dot{\ell}(\bxi)\dot{\ell}(\bxi)^\top}\\
  =&\;\int_{\Sdm}\int_{-1}^1 \frac{1}{\lrpbig{C_k^{(d-1)/2}(1)+\rho\, C_k^{(d-1)/2}(t)}^2}\\
  &\times \lrb{C_k^{(d-1)/2}(t)^2\bmu\bmu^\top+(d-1)^2C_{k-1}^{(d+1)/2}(t)^2(1-t^2) \bB_{\bmu}\bga\bga^\top\bB_{\bmu}^\top} \\
  &\times \frac{1}{\om{d}}\lrb{1+\rho\,\tilde{C}_k^{(d-1)/2}(t)}(1-t^2)^{d/2-1}\,\rd t\,\sigma_{d-1}(\rd \bga)\\
  =&\;\frac{\om{d-1}}{\om{d}C_k^{(d-1)/2}(1)} \int_{-1}^1 \frac{\lrbbig{C_k^{(d-1)/2}(1)+\rho\,C_k^{(d-1)/2}(t)}}{\lrpbig{C_k^{(d-1)/2}(1)+\rho\, C_k^{(d-1)/2}(t)}^2}\\
  &\times \lrc{C_k^{(d-1)/2}(t)^2\bmu\bmu^\top+(d-1)^2C_{k-1}^{(d+1)/2}(t)^2\frac{1-t^2}{d}(\bI_{d+1}-\bmu\bmu^\top)} \\
  &\times (1-t^2)^{d/2-1}\,\rd t \\
  =&\;A_k(\rho) \bmu\bmu^\top + B_k(\rho) (\bI_{d+1}-\bmu\bmu^\top),
\end{align*}
where
\begin{align*}
  A_k(\rho)&= \frac{\om{d-1}}{\om{d}C_k^{(d-1)/2}(1)}\int_{-1}^1 \frac{C_k^{(d-1)/2}(t)^2(1-t^2)^{d/2-1}}{C_k^{(d-1)/2}(1)+\rho\, C_k^{(d-1)/2}(t)}\,\rd t,\\
  B_k(\rho)&= \frac{\om{d-1}}{\om{d}C_k^{(d-1)/2}(1)}\frac{(d-1)^2}{d} \int_{-1}^1 \frac{C_{k-1}^{(d+1)/2}(t)^2(1-t^2)^{d/2}}{C_k^{(d-1)/2}(1)+\rho\, C_k^{(d-1)/2}(t)} \,\rd t.
\end{align*}
For $d=2$ and $k=1$,
\begin{align*}
  A_1(\rho) &=\frac{1}{2} \int_{-1}^1 \frac{t^2}{1 + \rho t} \, \rd t= \frac{\tanh^{-1}(\rho)-\rho}{\rho^3},\\
  B_1(\rho) &=\frac{1}{4} \int_{-1}^1 \frac{1 - t^2}{1 + \rho t} \, \rd t=\frac{\rho - (1 - \rho^2)\tanh^{-1}(\rho)}{2\rho^3}.
\end{align*}

The matrix $\bIcal(\bxi)$ is positive definite since $A_k(\rho)>0$ and $B_k(\rho)>0$ for the specified ranges of $\rho$ (the integrands are positive). Its inverse follows straightforwardly by the spectral theorem:
\begin{align*}
  \bIcal(\bxi)^{-1}%
  =&\;\frac{1}{A_k(\rho)}\bmu\bmu^\top +\frac{1}{B_k(\rho)} (\bI_{d+1}-\bmu\bmu^\top).
\end{align*}

\emph{Computation of $\bIcal(\bxi)$ for $d=1$.}

When $d=1$ the derivation of $\dot{\ell}(\bxi)$ is analogous but using that $[T_k(x)]'=kU_{k-1}(x)$, with $U_k(x)=\sin((k+1)\cos^{-1}(x))/\sqrt{1-x^2}$ the Chebyshev polynomial of the second kind. We can directly replace in \eqref{eq:ell2} $C_k^{(d-1)/2}(t)$ with $T_k(t)$ and $(d-1)C_{k-1}^{(d+1)/2}(t)$ with $k U_{k-1}(t)$, giving
\begin{align*}
  \dot{\ell}(\bxi)\dot{\ell}(\bxi)^\top=&\;\frac{1}{\lrp{1+\rho\, T_k(t)}^2} \bigg\{T_k(t)^2\bmu\bmu^\top+k^2U_{k-1}(t)^2(1-t^2) \bB_{\bmu}\gamma\gamma^\top\bB_{\bmu}^\top\\
  &+T_k(t)\lrc{kU_{k-1}(t)(1-t^2)^{1/2}} \lrc{\bmu(\bB_{\bmu}\gamma)^\top+(\bB_{\bmu}\gamma)\bmu^\top}\bigg\},
\end{align*}
with the tangent-normal change of variables $t=\bx^\top\bmu$ now with $\gamma\in\mathbb{S}^0=\{-1,1\}$ and $\bB_{\bmu}=(\mu_2,\,-\mu_1)^\top$. Clearly, $\int_{\mathbb{S}^0} \gamma\,\sigma_{0}(\rd \gamma)=0$, $\int_{\mathbb{S}^0} \gamma^2\,\sigma_{0}(\rd \gamma)=2$, $\bmu^\top\bB_{\bmu}=\zero^\top$, and also $\bB_{\bmu}\bB_{\bmu}^\top=\bI_{2}-\bmu\bmu^\top$. Using these, we have
\begin{align*}
  \bIcal(\bxi)=&\;\Ebig{\dot{\ell}(\bxi)\dot{\ell}(\bxi)^\top}\\
  =&\;\frac{1}{\pi} \int_{-1}^1 \frac{1+\rho\,T_k(t)}{\lrb{1+\rho\, T_k(t)}^2} \lrc{T_k(t)^2\bmu\bmu^\top+k^2U_{k-1}(t)^2(1-t^2)(\bI_{2}-\bmu\bmu^\top)} (1-t^2)^{-1/2}\,\rd t \\
  =&\;\frac{1}{\pi} \int_{-1}^1 \frac{1}{1+\rho\, T_k(t)} \lrc{T_k(t)^2\bmu\bmu^\top+k^2U_{k-1}(t)^2(1-t^2)(\bI_{2}-\bmu\bmu^\top)} (1-t^2)^{-1/2}\,\rd t \\
  =&\;A_k(\rho) \bmu\bmu^\top + B_k(\rho) (\bI_{2}-\bmu\bmu^\top),
\end{align*}
where
\begin{align*}
  A_k(\rho)&= \frac{1}{\pi} \int_{-1}^1 \frac{T_k(t)^2(1-t^2)^{-1/2}}{1+\rho\, T_k(t)}\,\rd t%
  = \frac{1}{\pi} \int_{0}^\pi \frac{\cos(k\theta)^2}{1+\rho \cos(k\theta)}\,\rd \theta
  = \frac{1-\sqrt{1-\rho^2}}{\rho^2\sqrt{1-\rho^2}},\\
  B_k(\rho)&= \frac{k^2}{\pi} \int_{-1}^1 \frac{U_{k-1}(t)^2(1-t^2)^{1/2}}{1+\rho\, T_k(t)} \,\rd t%
  = \frac{k^2}{\pi} \int_{0}^\pi \frac{\sin(k\theta)^2}{1+\rho \cos(k\theta)} \,\rd \theta%
  = k^2\frac{1-\sqrt{1-\rho^2}}{\rho^2}.
\end{align*}
The matrix $\bIcal(\bxi)$ is also positive definite since $A_k(\rho)>0$ and $B_k(\rho)>0$ for the specified ranges of $\rho$.

\emph{Asymptotic covariance matrix of $(\hat{\rho}_{\mathrm{ML}}, \hat{\bmu}_{\mathrm{ML}})$.}

As in the proof of Theorem \ref{thm:mm1}\ref{thm:mm1:1}, we apply the delta method for $g(\bx)\defin(\|\bx\|,\bx/\|\bx\|)$, $\bx\neq\zero$, now on \eqref{eq:mlexi}. The resulting asymptotic covariance matrix is
\begin{align*}
  \mathsf{D} g(\bxi)\bIcal(\bxi)^{-1}\mathsf{D} g(\bxi)^\top%
  &= \begin{pmatrix}
    A_k(\rho)^{-1}\bmu^\top\\
    [B_k(\rho)\rho]^{-1} (\bI_{d+1} - \bmu\bmu^\top)
  \end{pmatrix} \begin{pmatrix}
    \bmu & \rho^{-1}(\bI_{d+1} - \bmu\bmu^\top)\\
  \end{pmatrix}\\
  &= \begin{pmatrix}
    A_k(\rho)^{-1} & \zero^\top\\
    \zero & [B_k(\rho)\rho^2]^{-1} (\bI_{d+1} - \bmu\bmu^\top)\\
  \end{pmatrix}.
\end{align*}

The statement in \ref{thm:mle}\ref{thm:mle:3} follows after rearranging the components:
\begin{align*}
\sqrt{n}
\begin{pmatrix}
    \hat{\rho}_n - \rho \\
    \hat{\bmu}_n - \bmu
\end{pmatrix}
\inlaw \mathcal{N}_{d+2}\lrp{\mathbf{0},\begin{pmatrix}
    \sigma^2_{\mathrm{ML}}(\rho) & \zero^\top\\
    \zero & \sigma^2_{\mathrm{ML}}(\bmu) (\bI_{d+1} - \bmu\bmu^\top)\\
  \end{pmatrix}},
\end{align*}
where $\sigma^2_{\mathrm{ML}}(\rho)=A_k(\rho)^{-1}$ and $\sigma^2_{\mathrm{ML}}(\bmu)=[B_k(\rho)\rho^2]^{-1}$.

\emph{Conditions for ML estimation asymptotics.}

We verify the ``classical conditions'' for asymptotic normality of $Z$-estimators, as given in \citet[Theorems 5.41 and 5.42]{vanderVaart1998|SM}. These conditions are straightforward to check due to the smoothness of the spherical cardioid in the $\bxi$-parametrization \eqref{eq:cardxi} and the compactness of $\Sd$. We provide them here for the sake of completeness.

Let $\bxi_0$ denote the true parameter. In the $\bxi$-parametrization, the open parameter space \eqref{eq:paramspace} is $\Xi\subset\{\bxi\in\mathbb{R}^{d+1} : 0<\|\bxi\|<\rho_\star\}$ for $0<\rho_\star<1$, and $\bxi_0\in\Xi$. Denote $\psi_{\bxi}(\bx) \defin \dot{\ell}(\bxi)=\frac{\partial}{\partial \bxi} \log f_{\mathrm{C}_k}(\bx;\bxi)$, with $\bx\in\Sd$ and $\bxi\in\Xi$. The function $\bxi\mapsto \psi_{\bxi}(\bx)$ is twice continuously differentiable for every $\bx\in\Sd$ because of the (infinite) smoothness of $f_{\mathrm{C}_k}(\bx;\bxi)=\om{d}^{-1}\{1+\|\bxi\| \tilde{C}_k^{(d-1)/2}(\bx^\top\bxi/\|\bxi\|)\}$ for $(\bx,\bxi)\in\Sd\times\Xi$. Due to the boundedness of $\bx\mapsto\psi_{\bxi}(\bx)$, the dominated convergence theorem guarantees interchangeability of derivatives and integral sign, giving
\begin{align}
  \mathrm{E}_{\bxi_0}[\psi_{\bxi_0}(\bX)]%
  =\int_{\Sd} \frac{\partial}{\partial\bxi}\Big|_{\bxi=\bxi_0} f_{\mathrm{C}_k}(\bx;\bxi) \, \sigma_d(\rd \bx)
  =\frac{\partial}{\partial\bxi}\Big|_{\bxi=\bxi_0}\int_{\Sd} f_{\mathrm{C}_k}(\bx;\bxi) \, \sigma_d(\rd \bx)=\zero.\label{eq:leibnitz1}
\end{align}
Also, trivially, $\mathrm{E}_{\bxi_0}[\|\psi_{\bxi_0}(\bX)\|^2]\leq \sup_{\bx\in\Sd}\|\psi_{\bxi_0}(\bx)\|^2 <\infty$. The Fisher information matrix $\bIcal(\bxi_0)=\mathrm{E}_{\bxi_0}\lrcbig{\dot{\ell}(\bxi_0)\dot{\ell}(\bxi_0)^\top}$ exists and is positive definite for $0<\|\bxi_0\|<1$ by previous derivations. The smoothness of $\bxi\mapsto\ell(\bxi)$,
\begin{align*}
  \frac{\partial^2}{\partial \xi_i \partial \xi_j} \ell(\bxi) %
  =-\frac{\partial}{\partial \xi_i} \ell(\bxi) \frac{\partial}{\partial \xi_j} \ell(\bxi)+
  \frac{1}{f_{\mathrm{C}_k}(\bx;\bxi)}\frac{\partial^2}{\partial \xi_i \partial \xi_j} f_{\mathrm{C}_k}(\bx;\bxi),
\end{align*}
the dominated convergence theorem, and an argument as in \eqref{eq:leibnitz1} imply that $\mathrm{E}_{\bxi_0}[\ddot{\ell}(\bxi_0)]=-\mathrm{E}_{\bxi_0}\lrcbig{\dot{\ell}(\bxi_0)\dot{\ell}(\bxi_0)^\top}$. Therefore, $\mathrm{E}_{\bxi_0}[\ddot{\ell}(\bxi_0)]^{-1}\mathrm{E}_{\bxi_0}\lrcbig{\dot{\ell}(\bxi_0) \dot{\ell}(\bxi_0)^\top} \mathrm{E}_{\bxi_0}[\ddot{\ell}(\bxi_0)]^{-1}=\bIcal(\bxi_0)^{-1}$. Finally, in a closed neighborhood $V_{\bxi_0}$ of $\bxi_0$, the second partial derivatives $\bx\mapsto\ddot{\psi}_{\bxi}(\bx)$ are dominated by $\bx\mapsto\sup_{\bxi\in V_{\bxi_0}}\|\ddot{\psi}_{\bxi}(\bx)\|$, for every $\bxi\in V_{\bxi_0}$, which is finitely integrable on $\Sd$, since $\om{d}^{-1}(1-\rho_\star)\leq f_{\mathrm{C}_k}(\bx;\bxi)\leq \om{d}^{-1}(1+\rho_\star)$ uniformly in $(\bx,\bxi)$. The conditions of Theorem 5.41 in \cite{vanderVaart1998|SM} are therefore satisfied.

The conditions of Theorem 5.42 are the same, plus an additional one stating that $\bxi_0$ is a local maximizer of $\bxi\mapsto \mathrm{E}_{\bxi_0}[\log f_{\mathrm{C}_k}(\bX;\bxi)]$. This is true by a Kullback--Leibler divergence argument: $\mathrm{E}_{\bxi_0}[\log f_{\mathrm{C}_k}(\bX;\bxi)]-\mathrm{E}_{\bxi_0}[\log f_{\mathrm{C}_k}(\bX;\bxi_0)]=-\mathrm{KL}(f_{\mathrm{C}_k}(\cdot;\bxi_0)\|f_{\mathrm{C}_k}(\cdot;\bxi))\leq 0$.

Theorems 5.41 and 5.42 in \cite{vanderVaart1998|SM} imply that:
\begin{enumerate}[label=(\emph{\roman{*}})., ref=(\emph{\roman{*}})]
  \item The probability that $\bxi\mapsto\ell_n(\bxi)$ has at least one local maximum tends to one as $n\to\infty$.
  \item There exists a sequence of local maxima $\hat{\bxi}_n$ such that $\hat{\bxi}_n\to\bxi_0$ in probability.
  \item The sequence $\hat{\bxi}_n$ is asymptotically normal:
  \begin{align}
    \sqrt{n}(\hat{\bxi}_n - \bxi_0)\inlaw \mathcal{N}_{d+1}\lrpbig{\mathbf{0}, \bIcal(\bxi_0)^{-1}} \label{eq:mlexi}
  \end{align}
  as $n\to\infty$.
\end{enumerate}
In addition, by Theorem 5.41 in ibid,
  \begin{align*}
    \sqrt{n}(\hat{\bxi}_n - \bxi_0)=\bIcal(\bxi_0)^{-1}\frac{1}{\sqrt{n}}\sum_{i=1}^n \psi_{\bxi_0}(\bX_i)+o_{\mathrm{P}}(1).
  \end{align*}
These statements conclude the proof.
\end{proof}

\subsection{Proofs of results in Section~\ref{sec:gof}}
\label{sec:proofs:gof}

\begin{proof}[Proof of Theorem \ref{thm:stat}]
We compute each of the following terms separately:
\begin{align*}
    n^{-1}P_n^{W,\lambda}%
    &=\mathrm{E}_{\bga}\left[\int_{-1}^1 \{\hat{F}_{\bga}(x)\}^2\,\rd W(\hat{F}_{\bga}(x))\right]-2\mathrm{E}_{\bga}\left[\int_{-1}^1 F_{n,\bga}(x)\hat{F}_{\bga}(x)\,\rd W(\hat{F}_{\bga}(x))\right] \\
    &\quad+\mathrm{E}_{\bga}\left[\int_{-1}^1 \{F_{n,\bga}(x)\}^2\,\rd W(\hat{F}_{\bga}(x))\right]\\
    &\indef P_n^{(1)}-2P_n^{(2)}+P_n^{(3)}.
\end{align*}

\emph{Computation of $P_n^{(1)}$.} Considering the change of variables $u=\hat{F}_{\bga}(x)$, it readily follows that
\begin{align*}
    P_n^{(1)}%
    =\mathrm{E}_{\bga}\left[\int_{0}^1 u^2\,\rd W(u)\right]=W_2(1)-W_2(0),
\end{align*}
where $W_2(x)\defin\int_{0}^x u^2\,\rd W(u)$.

\emph{Computation of $P_n^{(2)}$.} Using the same change of variables as before,
\begin{align*}
    P_n^{(2)}%
    =&\;\frac{1}{n}\sum_{i=1}^n\mathrm{E}_{\bga}\left[\int_{-1}^1 1_{\{\bga^\top\bX_i\leq x\}}\hat{F}_{\bga}(x)\,\rd W(\hat{F}_{\bga}(x))\right]\\
    =&\;\frac{1}{n}\sum_{i=1}^n\mathrm{E}_{\bga}\left[\int_{\bga^\top\bX_i}^1 \hat{F}_{\bga}(x)\,\rd W(\hat{F}_{\bga}(x))\right]\\
    =&\;\frac{1}{n}\sum_{i=1}^n\mathrm{E}_{\bga}\left[\int_{\hat{F}_{\bga}(\bga^\top\bX_i)}^{1} u\,\rd W(u)\right]\\
    =&\;W_1(1)-\frac{1}{n}\sum_{i=1}^n\mathrm{E}_{\bga}\left[W_1(\hat{F}_{\bga}(\bga^\top\bX_i))\right],
\end{align*}
where $W_1(x)\defin\int_{0}^x u\,\rd W(u)$.

\emph{Computation of $P_n^{(3)}$.} Similar as before,
\begin{align*}
    P_n^{(3)}%
    &=\frac{1}{n^2}\sum_{i=1}^n\mathrm{E}_{\bga}\left[\int_{-1}^1 1_{\{\bga^\top\bX_i\leq x\}} \,\rd W(\hat{F}_{\bga}(x))\right]\\
    &\quad+\frac{1}{n^2}\sum_{i\neq j}\mathrm{E}_{\bga}\left[\int_{-1}^1 1_{\{\bga^\top\bX_i\leq x,\bga^\top\bX_j\leq x\}} \,\rd W(\hat{F}_{\bga}(x))\right]\\
    &\indef P_n^{(3,1)}+P_n^{(3,2)}.
\end{align*}
Therefore,
\begin{align*}
    P_n^{(3,1)}%
    =&\;\frac{1}{n^2}\sum_{i=1}^n\mathrm{E}_{\bga}\left[\int_{0}^1 1_{\{\hat{F}_{\bga}(\bga^\top\bX_i)\leq u\}} \,\rd W(u)\right]\\
    =&\;\frac{W(1)}{n}-\frac{1}{n^2}\sum_{i=1}^n\mathrm{E}_{\bga}\left[W(\hat{F}_{\bga}(\bga^\top\bX_i))\right]
\end{align*}
and
\begin{align*}
    P_n^{(3,2)}%
    =&\;\frac{1}{n^2}\sum_{i\neq j}\mathrm{E}_{\bga}\left[\int_{0}^1 1_{\{\hat{F}_{\bga}(\bga^\top\bX_i)\leq u,\hat{F}_{\bga}(\bga^\top\bX_j)\leq u\}} \,\rd W(u)\right]\\
    =&\;\frac{1}{n^2}\sum_{i\neq j}\mathrm{E}_{\bga}\left[\int_{0}^1 1_{\{\hat{F}_{\bga}(\max(\bga^\top\bX_i,\bga^\top\bX_j))\leq u\}} \,\rd W(u)\right]\\
    =&\;\frac{(n-1)}{n}W(1)-\frac{1}{n^2}\sum_{i\neq j}\mathrm{E}_{\bga}\left[W(\hat{F}_{\bga}(\max(\bga^\top\bX_i,\bga^\top\bX_j)))\right].
\end{align*}

Combining all these terms:
\begin{align*}
    n^{-1}P_n^{W,\lambda}=&\; P_n^{(1)}-2P_n^{(2)}+P_n^{(3,1)}+P_n^{(3,2)}\\
    =&\;W_2(1)-W_2(0)-2W_1(1)+\frac{2}{n}\sum_{i=1}^n\mathrm{E}_{\bga}\left[W_1(\hat{F}_{\bga}(\bga^\top\bX_i))\right]\\
    &+\frac{W(1)}{n}-\frac{1}{n^2}\sum_{i=1}^n\mathrm{E}_{\bga}\left[W(\hat{F}_{\bga}(\bga^\top\bX_i))\right]\\
    &+\frac{(n-1)}{n}W(1)-\frac{1}{n^2}\sum_{i\neq j}\mathrm{E}_{\bga}\left[W(\hat{F}_{\bga}(\max(\bga^\top\bX_i,\bga^\top\bX_j)))\right],\\
    =&\;W(1)+W_2(1)-W_2(0)-2W_1(1)\\
    &+\frac{2}{n}\sum_{i=1}^n\mathrm{E}_{\bga}\left[W_1(\hat{F}_{\bga}(\bga^\top\bX_i))\right]
    -\frac{1}{n^2}\sum_{i,j=1}^n\mathrm{E}_{\bga}\left[W(\hat{F}_{\bga}(\max(\bga^\top\bX_i,\bga^\top\bX_j)))\right],%
\end{align*}
proving the result.
\end{proof}

\begin{proof}[Proof of Corollary \ref{cor:adstat}]
To derive $P_n^{\mathrm{AD},\lambda}$, we consider the truncated weight $\rd W^{\mathrm{AD}_\varepsilon}(x)\defin [x(1-x)]^{-1} 1_{\{\varepsilon<x<1-\varepsilon\}}\,\rd x$, $\varepsilon\in(0,1/2)$, obtain $P_n^{\mathrm{AD}_\varepsilon,\lambda}$ and compute $\lim_{\varepsilon\to 0}P_n^{\mathrm{AD}_\varepsilon,\lambda}=P_n^{\mathrm{AD},\lambda}$.

Let $L_\varepsilon\defin \log((1-\varepsilon)/\varepsilon)$. Direct calculations give
\begin{align*}
  W^{\mathrm{AD}_\varepsilon}(x)&=\begin{cases}
    0, & x\leq \varepsilon,\\
    L_\varepsilon+\log(x/(1-x)), & \varepsilon<x<1-\varepsilon,\\
    2L_\varepsilon, & x\geq 1-\varepsilon,
  \end{cases},\\
  W_1^{\mathrm{AD}_\varepsilon}(x)&=\begin{cases}
    0, & x\leq \varepsilon,\\
    \log((1-\varepsilon)/(1-x)), & \varepsilon<x<1-\varepsilon,\\
    L_\varepsilon, & x\geq 1-\varepsilon,
  \end{cases},\\
  W_2^{\mathrm{AD}_\varepsilon}(x)&=\begin{cases}
    0, & x\leq \varepsilon,\\
    \varepsilon-x+\log((1-\varepsilon)/(1-x)), & \varepsilon<x<1-\varepsilon,\\
    L_\varepsilon+2\varepsilon-1, & x\geq 1-\varepsilon.
  \end{cases}
\end{align*}

Denote $t_{i,\bga}\defin\bga^\top\bX_i$ and $m_{ij,\bga}=\max(\bga^\top\bX_i,\bga^\top\bX_j)$. Theorem \ref{thm:stat} gives
\begin{align}
    P_n^{\mathrm{AD}_\varepsilon,\lambda}=&\;n(W^{\mathrm{AD}_\varepsilon}(1)+W_2^{\mathrm{AD}_\varepsilon}(1)-W_2^{\mathrm{AD}_\varepsilon}(0)-2W_1^{\mathrm{AD}_\varepsilon}(1))\nonumber\\
    & +2\sum_{i=1}^n\mathrm{E}_{\bga}\left[W_1^{\mathrm{AD}_\varepsilon}(\hat{F}_{\bga}(t_{i,\bga}))\right] -\frac{1}{n}\sum_{i, j=1}^n\mathrm{E}_{\bga}\left[W^{\mathrm{AD}_\varepsilon}(\hat{F}_{\bga}(m_{ij,\bga}))\right]\nonumber\\
    =&\;n(L_\varepsilon+2\varepsilon-1)\nonumber\\
    &+2\sum_{i=1}^n\mathrm{E}_{\bga}\lrc{\log\lrp{\frac{1-\varepsilon}{1-\hat{F}_{\bga}(t_{i,\bga})}}1_{\{\hat{F}_{\bga}(t_{i,\bga})\in(\varepsilon,1-\varepsilon)\}}}\nonumber\\
    &+2 L_\varepsilon\sum_{i=1}^n \mathrm{P}_{\bga}\lrc{\hat{F}_{\bga}(t_{i,\bga})\geq 1-\varepsilon}\nonumber\\
    &-\frac{1}{n}\sum_{i, j=1}^n\mathrm{E}_{\bga}\bigg[\lrbbigg{L_\varepsilon+\log\lrpbigg{\frac{\hat{F}_{\bga}(m_{ij,\bga})}{1-\hat{F}_{\bga}(m_{ij,\bga})}}} 1_{\{\hat{F}_{\bga}(m_{ij,\bga})\in(\varepsilon,1-\varepsilon)\}}\bigg]\nonumber\\
    &-\frac{2L_\varepsilon}{n}\sum_{i, j=1}^n \mathrm{P}_{\bga}\lrc{\hat{F}_{\bga}(m_{ij,\bga})\geq 1-\varepsilon}\nonumber\\
    =&-n(1-2\varepsilon)\nonumber\\
    &+2\sum_{i=1}^n\mathrm{E}_{\bga}\lrc{\log\lrp{\frac{1-\varepsilon}{1-\hat{F}_{\bga}(t_{i,\bga})}}1_{\{\hat{F}_{\bga}(t_{i,\bga})\in(\varepsilon,1-\varepsilon)\}}}\nonumber\\
    &-\frac{1}{n}\sum_{i, j=1}^n\mathrm{E}_{\bga}\bigg[\log\lrpbigg{\frac{\hat{F}_{\bga}(m_{ij,\bga})}{1-\hat{F}_{\bga}(m_{ij,\bga})}} 1_{\{\hat{F}_{\bga}(m_{ij,\bga})\in(\varepsilon,1-\varepsilon)\}}\bigg] \nonumber \\
    &+L_\varepsilon R_\varepsilon,\label{eq:adepsilon}
\end{align}
where
\begin{align*}
   R_\varepsilon=&\;n+2\sum_{i=1}^n \mathrm{P}_{\bga}\lrc{\hat{F}_{\bga}(t_{i,\bga})\geq 1-\varepsilon}-\frac{1}{n}\sum_{i, j=1}^n \mathrm{P}_{\bga}\lrc{\hat{F}_{\bga}(m_{ij,\bga})\in(\varepsilon, 1-\varepsilon)}\\
   &-\frac{2}{n}\sum_{i, j=1}^n \mathrm{P}_{\bga}\lrc{\hat{F}_{\bga}(m_{ij,\bga})\geq 1-\varepsilon}\\
   =&\;2\sum_{i=1}^n \mathrm{P}_{\bga}\lrc{\hat{F}_{\bga}(t_{i,\bga})\geq 1-\varepsilon}+\frac{1}{n}\sum_{i, j=1}^n \mathrm{P}_{\bga}\lrc{\hat{F}_{\bga}(m_{ij,\bga})\leq \varepsilon}\\
   &-\frac{1}{n}\sum_{i, j=1}^n \mathrm{P}_{\bga}\lrc{\hat{F}_{\bga}(m_{ij,\bga})\geq 1-\varepsilon}.
\end{align*}

As $\varepsilon\to 0$, the dominated convergence theorem applied to \eqref{eq:adepsilon} readily gives the statement of the corollary, provided that $L_\varepsilon R_\varepsilon\to 0$ as $\varepsilon\to 0$. Since $L_\varepsilon\sim -\log(\varepsilon)$ as $\varepsilon\to 0$, it suffices to show that the probabilities in $R_\varepsilon$ are $O(\varepsilon)$ as $\varepsilon\to 0$.

Set $m\defin 1-|\rho|>0$. By Theorem~\ref{thm:projgamma}, $\hat{f}_{\bga}(x)\geq m f_d(x)$ for any $x\in[-1,1]$. Hence,
\begin{align*}
  \hat{F}_{\bga}(x)=\int_{-1}^x \hat{f}_{\bga}(u)\,\rd u \geq m F_d(x)\quad\text{and}\quad
  1-\hat{F}_{\bga}(x)=\int_x^1 \hat{f}_{\bga}(u)\,\rd u \geq m(1-F_d(x)).
\end{align*}
Therefore, for $x_{\bga}=t_{i,\bga}$ or $x_{\bga}=m_{ij,\bga}$,
\begin{align*}
  \{\hat{F}_{\bga}(x_{\bga})\leq \varepsilon\}\subset \{F_d(x_{\bga})\leq \varepsilon/m\}\quad\text{and}\quad\{\hat{F}_{\bga}(x_{\bga})\geq 1-\varepsilon\}\subset \{F_d(x_{\bga})\geq 1-\varepsilon/m\}.
\end{align*}
These inclusions allow us to bound the probabilities in $R_\varepsilon$ through those involving $F_d$, which are easier to handle.

Let $\lambda$ denote the bounded density of $\bga$ with respect to the surface area measure $\sigma_d$, bounded by $M>0$. Then
\begin{align*}
  \mathrm{P}_{\bga}[\bga\in A]=\int_A \lambda(\bga)\,\sigma_d(\rd\bga)\leq M\sigma_d(A)=M\om{d}\, \mathrm{P}_{\bga}[\bU\in A]
\end{align*}
where $\bU\sim\mathrm{Unif}(\Sd)$. Then,
\begin{align*}
  \mathrm{P}_{\bga}\lrcbig{\hat{F}_{\bga}(\bga^\top\bX_i)\geq 1-\varepsilon}&\leq \mathrm{P}_{\bga}\lrcbig{F_{d}(\bga^\top\bX_i)\geq 1-\varepsilon/m}\\
  &\leq M\om{d}\mathrm{P}_{\bga}\lrcbig{F_d(\bU^\top\bX_i)\geq 1-\varepsilon/m}\\
  &= M\om{d}\frac{\varepsilon}{m},
\end{align*}
with the last equality following from $F_d(\bU^\top\bX_i)\sim \mathrm{Unif}(0,1)$.

For $m_{ij,\bga}=\max(\bga^\top\bX_i,\bga^\top\bX_j)$, recall that
\begin{align*}
  \{F_d(m_{ij,\bga})\geq 1-\varepsilon/m\}\subset \{F_d(t_{i,\bga})\geq 1-\varepsilon/m\}\cup\{F_d(t_{j,\bga})\geq 1-\varepsilon/m\},
\end{align*}
implying
\begin{align*}
  \mathrm{P}_{\bga}\lrcbig{\hat{F}_{\bga}(m_{ij,\bga})\geq 1-\varepsilon}\leq 2M\om{d}\frac{\varepsilon}{m}.
\end{align*}
Finally, since $\{F_d(m_{ij,\bga})\leq \varepsilon/m\}\subset\{F_d(t_{i,\bga})\leq \varepsilon/m\}$, then
\begin{align*}
  \mathrm{P}_{\bga}\lrcbig{\hat{F}_{\bga}(m_{ij,\bga})\leq \varepsilon}\leq M\om{d} \frac{\varepsilon}{m}.
\end{align*}
This shows that $R_\varepsilon=O(\varepsilon)$ as $\varepsilon\to 0$.
\end{proof}

\begin{proof}[Proof of Theorem \ref{thm:cvmstat}]
For the CvM weight, $W(x)=x$, we have $W_1(x)=x^2/2$ and $W_2(x)=x^3/3$. Hence, the bias term is $W(1)+W_2(1)-W_2(0)-2W_1(1)=1/3$. We also define the two kernels
\begin{align*}
  \tilde{\varphi}(\bX_1)\defin 2\mathrm{E}_{\bga}\lrcbig{W_1(\hat{F}_{\bga}(\bga^\top\bX_1))}-\frac{1}{3},\quad \tilde{\psi}(\bX_1,\bX_2)\defin \mathrm{E}_{\bga}\lrcbig{W(\hat{F}_{\bga}(\max(\bga^\top\bX_1,\bga^\top\bX_2)))}.
\end{align*}
This yields the $V$-statistic expression
\begin{align}
    n^{-1}P_n^{\mathrm{CvM},\,\mathrm{Unif}}=&\;\frac{1}{3}+\frac{1}{n}\sum_{i=1}^n\lrc{\tilde{\varphi}(\bX_i)+\frac{1}{3}} -\frac{1}{n^2}\sum_{i, j=1}^n\tilde{\psi}(\bX_i,\bX_j).\label{eq:V1}
\end{align}

We obtain the forms of $\tilde{\varphi}$ and $\tilde{\psi}$ next.

\emph{Computation of $\tilde{\varphi}(\bX_1)$.}

The first kernel is
\begin{align}
  \tilde{\varphi}(\bX_1)%
  =&\;\frac{1}{\om{d}}\int_{\Sd} \lrcbig{F_{d}(\bga^\top\bX_1)}^2 \,\sigma_d(\rd \bga)-\frac{1}{3}\nonumber\\
  &-\frac{2\hat{\rho}}{\om{d}}\int_{\Sd} F_{d}(\bga^\top\bX_1)\eta_k(\bga^\top\hat{\bmu})G_k(\bga^\top\bX_1)\,\sigma_d(\rd \bga)\nonumber\\
  &+\frac{\hat{\rho}^2}{\om{d}}\int_{\Sd}\lrcbig{\eta_k(\bga^\top\hat{\bmu})G_k(\bga^\top\bX_1)}^2 \,\sigma_d(\rd \bga)\nonumber\\
  =&\;\int_{-1}^1 F_{d}(x)^2 f_d(x)\,\rd x-\frac{1}{3}\nonumber\\
  &-\frac{2\hat{\rho}}{\om{d}}\int_{\Sd} F_{d}(\bga^\top\bX_1)G_k(\bga^\top\bX_1)\eta_k(\bga^\top\hat{\bmu})\,\sigma_d(\rd \bga)\nonumber\\
  &+\frac{\hat{\rho}^2}{\om{d}}\int_{\Sd}\lrcbig{G_k(\bga^\top\bX_1)\eta_k(\bga^\top\hat{\bmu})}^2\,\sigma_d(\rd \bga)\nonumber\\
  =&-2\hat{\rho}\frac{\om{d-1}}{\om{d}^2}\frac{1}{\lrcbig{C_k^{(d-1)/2}(1)}^2}\int_{\Sd} F_{d}(\bga^\top\bX_1)G_k(\bga^\top\bX_1)C_k^{(d-1)/2}(\bga^\top\hat{\bmu})\,\sigma_d(\rd \bga)\nonumber\\
  &+\hat{\rho}^2\frac{\om{d-1}^2}{\om{d}^3}\frac{1}{\lrcbig{C_k^{(d-1)/2}(1)}^4}\int_{\Sd}\lrcbig{G_k(\bga^\top\bX_1) C_k^{(d-1)/2}(\bga^\top\hat{\bmu})}^2\,\sigma_d(\rd \bga)\nonumber\\
  &\!\!\!\!\!\!\indef-2\hat{\rho}\frac{\om{d-1}}{\om{d}^2}\frac{1}{\lrcbig{C_k^{(d-1)/2}(1)}^2}\tilde{\varphi}^{(1)}(\bX_1) +\hat{\rho}^2\frac{\om{d-1}^2}{\om{d}^3}\frac{1}{\lrcbig{C_k^{(d-1)/2}(1)}^4}\tilde{\varphi}^{(2)}(\bX_1). \label{eq:phigeneral}
\end{align}

We compute now
\begin{align*}
  \tilde{\varphi}^{(1)}(\bX_1)&=\int_{\Sd} F_{d}(\bga^\top\bX_1)G_k(\bga^\top\bX_1)C_k^{(d-1)/2}(\bga^\top\hat{\bmu})\,\sigma_d(\rd \bga),\\
  \tilde{\varphi}^{(2)}(\bX_1)&=\int_{\Sd}\lrcbig{G_k(\bga^\top\bX_1) C_k^{(d-1)/2}(\bga^\top\hat{\bmu})}^2\,\sigma_d(\rd \bga)
\end{align*}
for the cases $d=1$, $(d=2,k=1)$ and $(d=2,k=2)$. The evaluation of these integrals has been done with the help of Mathematica \citep{Mathematica|SM}.

For $d=1$,
\begin{align*}
  \tilde{\varphi}^{(1)}(\bX_1)&=\frac{1}{k}\int_{\mathbb{S}^1} \frac{\pi-\cos^{-1}(\bga^\top\bX_1)}{\pi}\sin(k\cos^{-1}(\bga^\top\bX_1))\cos(k\cos^{-1}(\bga^\top\hat{\bmu}))\,\sigma_1(\rd \bga)\\
  &=\frac{1}{k}\int_{0}^{2\pi} \frac{\pi-\cos^{-1}(\cos(\gamma-\theta_1))}{\pi}\sin(k\cos^{-1}(\cos(\gamma-\theta_1)))\cos(k\cos^{-1}(\cos(\gamma-\hat{\mu})))\,\rd \gamma\\
  &=\frac{1}{2k^2} T_k(\bX_1^\top\hat{\bmu})
\end{align*}
and
\begin{align*}
  \tilde{\varphi}^{(2)}(\bX_1)&=\frac{1}{k^2}\int_{\mathbb{S}^1} \lrcbig{\sin(k\cos^{-1}(\bga^\top\bX_1)) \cos(k\cos^{-1}(\bga^\top\hat{\bmu}))}^2\,\sigma_1(\rd \bga)\\
  &=\frac{1}{k^2}\int_{0}^{2\pi} \lrcbig{\sin(k\cos^{-1}(\cos(\gamma-\theta_1))) \cos(k\cos^{-1}(\cos(\gamma-\hat{\mu})))}^2\,\rd \gamma\\
  &=\frac{\pi}{4k^2}(2-T_{2k}(\bX_1^\top\hat{\bmu})).
\end{align*}

For $d=2$,
\begin{align*}
  \tilde{\varphi}^{(1)}(\bX_1)&=\frac{1}{2 k(k+1)}\int_{\mathbb{S}^2} (\bga^\top\bX_1+1)  C_{k-1}^{3/2}(\bga^\top\bX_1) C_k^{1/2}(\bga^\top\hat{\bmu})(1-(\bga^\top\bX_1)^2)\,\sigma_2(\rd \bga),\\
  \tilde{\varphi}^{(2)}(\bX_1)&=\frac{1}{k^2(k+1)^2}\int_{\mathbb{S}^2}\lrcbig{C_{k-1}^{3/2}(\bga^\top\bX_1) C_k^{1/2}(\bga^\top\hat{\bmu})(1-(\bga^\top\bX_1)^2)}^2\,\sigma_2(\rd \bga).
\end{align*}
For $k=1$,
\begin{align*}
  \tilde{\varphi}^{(1)}(\bX_1)=4\pi\frac{\bX_1^\top \hat{\bmu}}{30},\quad \tilde{\varphi}^{(2)}(\bX_1)&=4\pi\lrb{\frac{2}{35}-\frac{4(\bX_1^\top \hat{\bmu})^2}{105}}.
\end{align*}
For $k=2$,
\begin{align*}
  \tilde{\varphi}^{(1)}(\bX_1)=4\pi\frac{3(\bX_1^\top \hat{\bmu})^2-1}{420},\quad \tilde{\varphi}^{(2)}(\bX_1)=4\pi\lrb{\frac{1}{330}+\frac{3(\bX_1^\top \hat{\bmu})^2}{385}-\frac{(\bX_1^\top \hat{\bmu})^4}{110}}.
\end{align*}

We can now substitute these expressions into \eqref{eq:phigeneral} to obtain the final forms of $\tilde{\varphi}(\bX_1)$ in each case:
\begin{itemize}
  \item For $d=1$ and $k\geq 1$:
  \begin{align*}
    \tilde{\varphi}(\bX_1)&=-\frac{\hat{\rho}}{\pi^2}\tilde{\varphi}^{(1)}(\bX_1)+\frac{\hat{\rho}^2}{2\pi^3}\tilde{\varphi}^{(2)}(\bX_1)\\
    &=-\frac{\hat{\rho}}{2\pi^2 k^2}\lrb{T_k(\bX_1^\top\hat{\bmu})-\frac{\hat{\rho}}{4}(2-T_{2k}(\bX_1^\top\hat{\bmu}))}.
  \end{align*}
  \item For $d=2$:
  \begin{align*}
    \tilde{\varphi}(\bX_1)&=-\frac{\hat{\rho}}{4\pi \lrcbig{C_k^{1/2}(1)}^2}\tilde{\varphi}^{(1)}(\bX_1)+\frac{\hat{\rho}^2}{16\pi \lrcbig{C_k^{1/2}(1)}^4}\tilde{\varphi}^{(2)}(\bX_1)\\
    &=-\frac{\hat{\rho}}{4\pi}\tilde{\varphi}^{(1)}(\bX_1)+\frac{\hat{\rho}^2}{16\pi}\tilde{\varphi}^{(2)}(\bX_1).
  \end{align*}
  For $k=1$,
  \begin{align*}
    \tilde{\varphi}(\bX_1)&=-\frac{\hat{\rho}}{4\pi}4\pi\frac{\bX_1^\top \hat{\bmu}}{30}+\frac{\hat{\rho}^2}{16\pi}4\pi\lrb{\frac{2}{35}-\frac{4(\bX_1^\top \hat{\bmu})^2}{105}}\\
    &=-\frac{\hat{\rho}}{30}\bX_1^\top \hat{\bmu}+\frac{\hat{\rho}^2}{4}\lrb{\frac{2}{35}-\frac{4(\bX_1^\top \hat{\bmu})^2}{105}}.
  \end{align*}
  For $k=2$,
  \begin{align*}
    \tilde{\varphi}(\bX_1)&=-\frac{\hat{\rho}}{4\pi}4\pi\frac{3(\bX_1^\top \hat{\bmu})^2-1}{420}+\frac{\hat{\rho}^2}{16\pi}4\pi\lrb{\frac{1}{330}+\frac{3(\bX_1^\top \hat{\bmu})^2}{385}-\frac{(\bX_1^\top \hat{\bmu})^4}{110}}\\
    &=-\frac{\hat{\rho}}{420}(3(\bX_1^\top \hat{\bmu})^2-1)+\frac{\hat{\rho}^2}{4}\lrb{\frac{1}{330}+\frac{3(\bX_1^\top \hat{\bmu})^2}{385}-\frac{(\bX_1^\top \hat{\bmu})^4}{110}}.
  \end{align*}
\end{itemize}

\emph{Computation of $\tilde{\psi}(\bX_1,\bX_2)$.}

Denote $m_{12,\bga}\defin\max(\bga^\top\bX_1,\bga^\top\bX_2)$. The second kernel is
\begin{align}
  \tilde{\psi}(\bX_1,\bX_2)%
  =&\;\mathrm{E}_{\bga}\left[F_{\bga}(m_{12,\bga})\right]\nonumber\\
  =&\;\mathrm{E}_{\bga}\left[F_d(m_{12,\bga})\right]-\hat{\rho}\, \mathrm{E}_{\bga}\left[\eta_k(\bga^\top\hat{\bmu}) G_k(m_{12,\bga})\right]\nonumber\\
  =&\;\mathrm{E}_{\bga}\left[F_d(m_{12,\bga})\right] -\hat{\rho}\, \frac{\om{d-1}}{\om{d}} \frac{1}{\lrcbig{C_k^{(d-1)/2}(1)}^2} \mathrm{E}_{\bga}\left[C_k^{(d-1)/2}(\bga^\top\hat{\bmu}) G_k(m_{12,\bga})\right]\nonumber\\
  &\!\!\!\!\!\!\indef \tilde{\psi}^{(1)}(\bX_1,\bX_2) -\hat{\rho}\, \frac{\om{d-1}}{\om{d}} \frac{1}{\lrcbig{C_k^{(d-1)/2}(1)}^2} \tilde{\psi}^{(2)}(\bX_1,\bX_2).\label{eq:psigeneral}
\end{align}

By Propositions 2.3 and 2.4 in \cite{Garcia-Portugues2020b|SM},
\begin{align*}
  \tilde{\psi}^{(1)}(\bX_1,\bX_2)=&\;\mathrm{E}_{\bga}\left[F_d(m_{12,\bga})\right]\\
  =&\;1-\mathrm{E}_{\bga}\left[1-F_d(m_{12,\bga})\right]\\
  =&\;1-\mathrm{E}_{\bga}\left[\int_{-1}^1 1_{\{m_{12,\bga}\leq x\}}f_d(x)\,\rd x\right]\\
  =&\;1-\int_{-1}^1\mathrm{E}_{\bga}\left[ 1_{\{\bga^\top\bX_1\leq x,\bga^\top\bX_2\leq x\}}\right]f_d(x)\,\rd x
\end{align*}
equals $1-\psi^{\mathrm{CvM}}_d(\theta_{12})$, the CvM kernel for testing uniformity on $\Sd$ given in \cite{Garcia-Portugues2020b|SM}, with $\theta_{12}=\cos^{-1}(\bX_1^\top\bX_2)$. For $d=1,2$, this gives
\begin{align*}
  1-\tilde{\psi}^{(1)}(\bX_1,\bX_2)=\psi^\mathrm{CvM}_d(\theta_{12}) = \begin{cases}
    \displaystyle\frac{1}{2}+\frac{\theta_{12}}{2\pi}\left(\frac{\theta_{12}}{2\pi}-1\right), & d=1,\\
    \displaystyle\frac{1}{2}-\frac{1}{4}\sin\lrp{\frac{\theta_{12}}{2}}, & d=2.
    \end{cases}
\end{align*}
Next we consider the kernels for the cases $d=1$, $(d=2,k=1)$ and $(d=2,k=2)$. We used Mathematica \citep{Mathematica|SM} to evaluate parts of the integrals involved.

For $d=1$,
\begin{align*}
  \tilde{\psi}^{(2)}&(\bX_1,\bX_2)\\
  =&\;\frac{1}{k}\mathrm{E}_{\bga}\left[C^0_k(\bga^\top\hat{\bmu}) \sin(k\cos^{-1}(m_{12,\bga}))\right]\\
  =&\;\frac{1}{2k\pi}\int_{\mathbb{S}^1} \cos(k\cos^{-1}(\bga^\top\hat{\bmu}))\sin(k\cos^{-1}(m_{12,\bga}))\, \sigma_1(\rd\bga)\\
  =&\;\frac{1}{2k\pi}\int_0^{2\pi} \cos(k(\cos^{-1}(\cos(\gamma-\hat{\mu}))))\sin(k\cos^{-1}(\max(\cos(\gamma-\theta_1),\cos(\gamma-\theta_2))))\,\rd \gamma\\
  =&\;\frac{1}{k}\lrb{-\frac{\pi-\cos^{-1}(\bX_1^\top\bX_2)}{2\pi}T_k\lrp{\frac{(\bX_1+\bX_2)^\top\hat{\bmu}}{\sqrt{2(1+\bX_1^\top\bX_2)}}} \sin\lrp{\frac{k\cos^{-1}(\bX_1^\top\bX_2)}{2}}}.
\end{align*}

For $d\geq2$,
\begin{align}
  \tilde{\psi}^{(2)}(\bX_1,\bX_2)=\frac{d-1}{k(k+d-1)} \mathrm{E}_{\bga}\left[C_{k}^{(d-1)/2}(\bga^\top\hat{\bmu})C_{k-1}^{(d+1)/2}(m_{12,\bga})(1-m_{12,\bga}^2)^{d/2}\right]. \label{eq:psi2d2}
\end{align}

If $k=1$, \eqref{eq:psi2d2} simplifies to
\begin{align*}
  \tilde{\psi}^{(2)}(\bX_1,\bX_2)=&\;\frac{(d-1)^2}{d}\mathrm{E}_{\bga}\left[(\bga^\top\hat{\bmu})(1-m_{12,\bga}^2)^{d/2}\right]
\end{align*}
and, if $d=2$, to
\begin{align*}
  \tilde{\psi}^{(2)}(\bX_1,\bX_2)=&\;\frac{1}{2}\mathrm{E}_{\bga}\left[(\bga^\top\hat{\bmu})(1-m_{12,\bga}^2)\right]\\
  =&-\frac{1}{16} \sqrt{\frac{1-\bX_1^\top\bX_2}{2}}(\bX_1+\bX_2)^\top\hat{\bmu}.
\end{align*}

If $k=2$, \eqref{eq:psi2d2} simplifies to
\begin{align*}
  \tilde{\psi}^{(2)}(\bX_1,\bX_2)=&\;\frac{(d-1)(d+1)}{2(d+1)}\frac{d-1}{2} \mathrm{E}_{\bga}\left[[(d+1)(\bga^\top\hat{\bmu})^2-1]m_{12,\bga}(1-m_{12,\bga}^2)^{d/2}\right]\\
  =&\;\frac{(d-1)^2}{4}\mathrm{E}_{\bga}\left[[(d+1)(\bga^\top\hat{\bmu})^2-1]m_{12,\bga}(1-m_{12,\bga}^2)^{d/2}\right]
\end{align*}
and, if $d=2$, to
\begin{align*}
  \tilde{\psi}^{(2)}(\bX_1,\bX_2)=&\;\frac{1}{4}\mathrm{E}_{\bga}\left[[3(\bga^\top\hat{\bmu})^2-1]m_{12,\bga}(1-m_{12,\bga}^2)\right]\\
  =&\;\frac{\sqrt{(1-\bX_1^\top\bX_2)/2}}{64}\Bigg\{\frac{1+\bX_1^\top\bX_2}{2}+\frac{3(3\bX_1^\top\bX_2-1)}{4(1-\bX_1^\top\bX_2)}\lrcbig{(\bX_1^\top\hat{\bmu})^2+(\bX_2^\top\hat{\bmu})^2}\\
  &+\frac{3(\bX_1^\top\bX_2-3)}{2(1-\bX_1^\top\bX_2)}(\bX_1^\top\hat{\bmu}) (\bX_2^\top\hat{\bmu})\Bigg\}.
\end{align*}

We can now substitute these expressions into \eqref{eq:psigeneral} to obtain the final forms of $\psi(\bX_1,\bX_2)$ in each case:
\begin{itemize}
  \item For $d=1$ and $k\geq 1$:
  \begin{align*}
    \tilde{\psi}(\bX_1,\bX_2)=&\;\tilde{\psi}^{(1)}(\bX_1,\bX_2)-\frac{\hat{\rho}}{\pi} \tilde{\psi}^{(2)}(\bX_1,\bX_2)\\
    =&\;1-\psi^\mathrm{CvM}_d(\cos^{-1}(\bX_1^\top\bX_2))\\
    &+\hat{\rho}\lrb{\frac{\pi-\cos^{-1}(\bX_1^\top\bX_2)}{2\pi^2 k}T_k\lrp{\frac{(\bX_1+\bX_2)^\top\hat{\bmu}}{\|\bX_1+\bX_2\|}} \sin\lrp{\frac{k\cos^{-1}(\bX_1^\top\bX_2)}{2}}}.
  \end{align*}
  \item For $d=2$ and $k=1$:
  \begin{align*}
    \tilde{\psi}(\bX_1,\bX_2)=&\;\tilde{\psi}^{(1)}(\bX_1,\bX_2)-\frac{\hat{\rho}}{2} \tilde{\psi}^{(2)}(\bX_1,\bX_2)\\
    =&\;1-\psi^\mathrm{CvM}_2(\cos^{-1}(\bX_1^\top\bX_2)) +\frac{\hat{\rho}}{32} \sqrt{\frac{1-\bX_1^\top\bX_2}{2}}(\bX_1+\bX_2)^\top\hat{\bmu}.
  \end{align*}
  \item For $d=2$ and $k=2$:
  \begin{align*}
    \tilde{\psi}(\bX_1,\bX_2)=&\;\tilde{\psi}^{(1)}(\bX_1,\bX_2)-\frac{\hat{\rho}}{2} \tilde{\psi}^{(2)}(\bX_1,\bX_2)\\
    =&\;1-\psi^\mathrm{CvM}_2(\cos^{-1}(\bX_1^\top\bX_2))\\
    &-\frac{\hat{\rho}}{128}\sqrt{\frac{1-\bX_1^\top\bX_2}{2}}\Bigg\{\frac{1+\bX_1^\top\bX_2}{2}+\frac{3(3\bX_1^\top\bX_2-1)}{4(1-\bX_1^\top\bX_2)}\lrcbig{(\bX_1^\top\hat{\bmu})^2+(\bX_2^\top\hat{\bmu})^2}\\
    &+\frac{3(\bX_1^\top\bX_2-3)}{2(1-\bX_1^\top\bX_2)}(\bX_1^\top\hat{\bmu}) (\bX_2^\top\hat{\bmu})\Bigg\}.
  \end{align*}
\end{itemize}
Since $\psi^\mathrm{CvM}_1(0)=\psi^\mathrm{CvM}_2(0)=1/2$, it is easy to check that $\tilde{\psi}(\bX_1,\bX_1)=1/2$.

\emph{Final expression.}

Returning to expression \eqref{eq:V1} for the $V$-statistic, it can be re-expressed as:
\begin{align*}
    n^{-1}P_n^{\mathrm{CvM},\,\mathrm{Unif}}=&\;\frac{1}{3}+\frac{1}{n}\sum_{i=1}^n\lrc{\tilde{\varphi}(\bX_i)+\frac{1}{3}} -\frac{1}{n^2}\sum_{i, j=1}^n\tilde{\psi}(\bX_i,\bX_j)\\
    =&\;\frac{2}{3}-1-\frac{1}{n}\sum_{i=1}^n [-\tilde{\varphi}(\bX_i)]+\frac{1}{n^2}\sum_{i, j=1}^n[1-\tilde{\psi}(\bX_i,\bX_j)]\\
    =&-\frac{1}{3}-\frac{1}{n}\sum_{i=1}^n [-\tilde{\varphi}(\bX_i)]+\frac{2}{n^2}\sum_{i<j}[1-\tilde{\psi}(\bX_i,\bX_j)]+\frac{1}{2n}\\
    =&\;\frac{3-2n}{6n}-\frac{1}{n}\sum_{i=1}^n [-\tilde{\varphi}(\bX_i)]+\frac{2}{n^2}\sum_{i<j}[1-\tilde{\psi}(\bX_i,\bX_j)].
\end{align*}
We set $\varphi(\bX_i)=-\tilde{\varphi}(\bX_i)$ and $\psi(\bX_i,\bX_j)=1-\tilde{\psi}(\bX_i,\bX_j)$ to obtain \eqref{eq:Ustatcvm}.
\end{proof}

\subsection{Auxiliary result}
\label{sec:aux}

\begin{lemma} \label{lem:Bmuotimes}
Let $d\ge 1$ and let $\bB_{\bmu}\in\mathcal{M}_{d+1,d}$ be a semi-orthogonal matrix such that
\begin{align*}
  \bB_{\bmu}\bB_{\bmu}^\top =\bI_{d+1}-\bmu\bmu^\top\indef \mathbf{P}_{\boldsymbol{\mu}}\quad\text{and}\quad \bB_{\bmu}^\top \bB_{\bmu} = \bI_d.
\end{align*}
Then, for any $p\ge 0$,
\begin{align*}
    \bB_{\bmu}^{\otimes 2p} (\operatorname{vec}\bI_d)^{\otimes p} = (\operatorname{vec}\mathbf{P}_{\boldsymbol{\mu}})^{\otimes p}.%
\end{align*}
\end{lemma}

\begin{proof}[Proof of Lemma \ref{lem:Bmuotimes}]
We use two standard equalities for the Kronecker product: the vec-operator identity
\begin{align}
  \operatorname{vec}(\bA \bX \bB^\top) = (\bB\otimes \bA)\,\operatorname{vec}\bX \label{eq:Kvec}
\end{align}
and the mixed-product property $(\bA\otimes \bB)(\bC\otimes \bD)=(\bA \bC)\otimes(\bB \bD)$, both for conformable matrices. The latter can be inductively extended to
\begin{align}
   \bA^{\otimes p}\bB^{\otimes p}=(\bA \bB)^{\otimes p}, \quad p\geq 0. \label{eq:mixedprod}
\end{align}

Applying \eqref{eq:Kvec},
\begin{align*}
  (\bB_{\bmu}\otimes \bB_{\bmu})\,\operatorname{vec} \bI_d = \operatorname{vec}\lrpbig{\bB_{\bmu} \bI_d \bB_{\bmu}^\top} = \operatorname{vec}\lrpbig{\bB_{\bmu}\bB_{\bmu}^\top}= \operatorname{vec}\mathbf{P}_{\boldsymbol{\mu}}
\end{align*}
while, due to the associativity of the Kronecker product,
\begin{align*}
  \bB_{\bmu}^{\otimes 2p} = (\bB_{\bmu}\otimes \bB_{\bmu})^{\otimes p}
\end{align*}
Using these results and \eqref{eq:mixedprod}, we obtain
\begin{align*}
  \bB_{\bmu}^{\otimes 2p}(\operatorname{vec}\bI_d)^{\otimes p}=(\bB_{\bmu}\otimes \bB_{\bmu})^{\otimes p}(\operatorname{vec}\bI_d)^{\otimes p}=[(\bB_{\bmu}\otimes \bB_{\bmu})\,\operatorname{vec} \bI_d]^{\otimes p}=(\operatorname{vec}\mathbf{P}_{\boldsymbol{\mu}})^{\otimes p},
\end{align*}
proving the claimed result.%
\end{proof}

%
%
%
%
%
%
%
%
%
%
%
%
%

\section{Asymptotic relative efficiencies of estimators}
\label{sec:ares}

%
In this section, we explore the Asymptotic Relative Efficiencies (AREs) of the moment estimators with respect to the maximum likelihood estimators. These AREs are defined as
\begin{align*}
\mathrm{ARE}_{\mathrm{MM}}(\rho)\defin\frac{\sigma^2_{\mathrm{ML}}(\rho)}{\sigma^2_{\mathrm{MM}}(\rho)},\quad \mathrm{ARE}_{\mathrm{MM}}(\bmu)\defin\frac{\sigma^2_{\mathrm{ML}}(\bmu)}{\sigma^2_{\mathrm{MM}}(\bmu)},\quad \mathrm{ARE}_{\mathrm{GM}}(\rho)\defin\frac{\sigma^2_{\mathrm{ML}}(\rho)}{\sigma^2_{\mathrm{GM},k}(\rho)}.
\end{align*}

Figure~\ref{fig:are1} shows the ARE curves $\rho\mapsto\mathrm{ARE}_{\mathrm{MM}}(\bmu)$ and $\rho\mapsto\mathrm{ARE}_{\mathrm{MM}}(\rho)$ for $\rho\in(0,1)$, $k=1,2$, and $d=1,\ldots,10$. As expected, the AREs are smaller than one. The efficiency of both moment estimators is maximal for $\rho=0$ and decreases as $\rho$ increases, but differently for $\hat{\bmu}_\mathrm{MM}$ and $\hat{\rho}_\mathrm{MM}$. Indeed, $\mathrm{ARE}_{\mathrm{MM}}(\bmu)$ is almost always above $0.8$, while $\mathrm{ARE}_{\mathrm{MM}}(\rho)$ can attain minima below $0.25$ for $\rho\approx 1$. The AREs are larger for $k=2$ than for $k=1$ when $d>1$, except for $d=1$ where $\rho\mapsto\mathrm{ARE}_{\mathrm{MM}}(\bmu)$ and $\rho\mapsto\mathrm{ARE}_{\mathrm{MM}}(\rho)$ coincide for $k=1,2$. Recall that in the displayed case $\rho>0$ (assumption of Theorem \ref{thm:mle}), $(\hat{\bmu}_{\mathrm{MM},2},\hat{\rho}_{\mathrm{MM},2})$ are based on the largest eigenpair of $\bS$, i.e., the most stable version of the estimator. The efficiency of the moment estimators increases with the dimension $d$.

\begin{figure}[h!]
  \centering
  \includegraphics[width=0.8\textwidth]{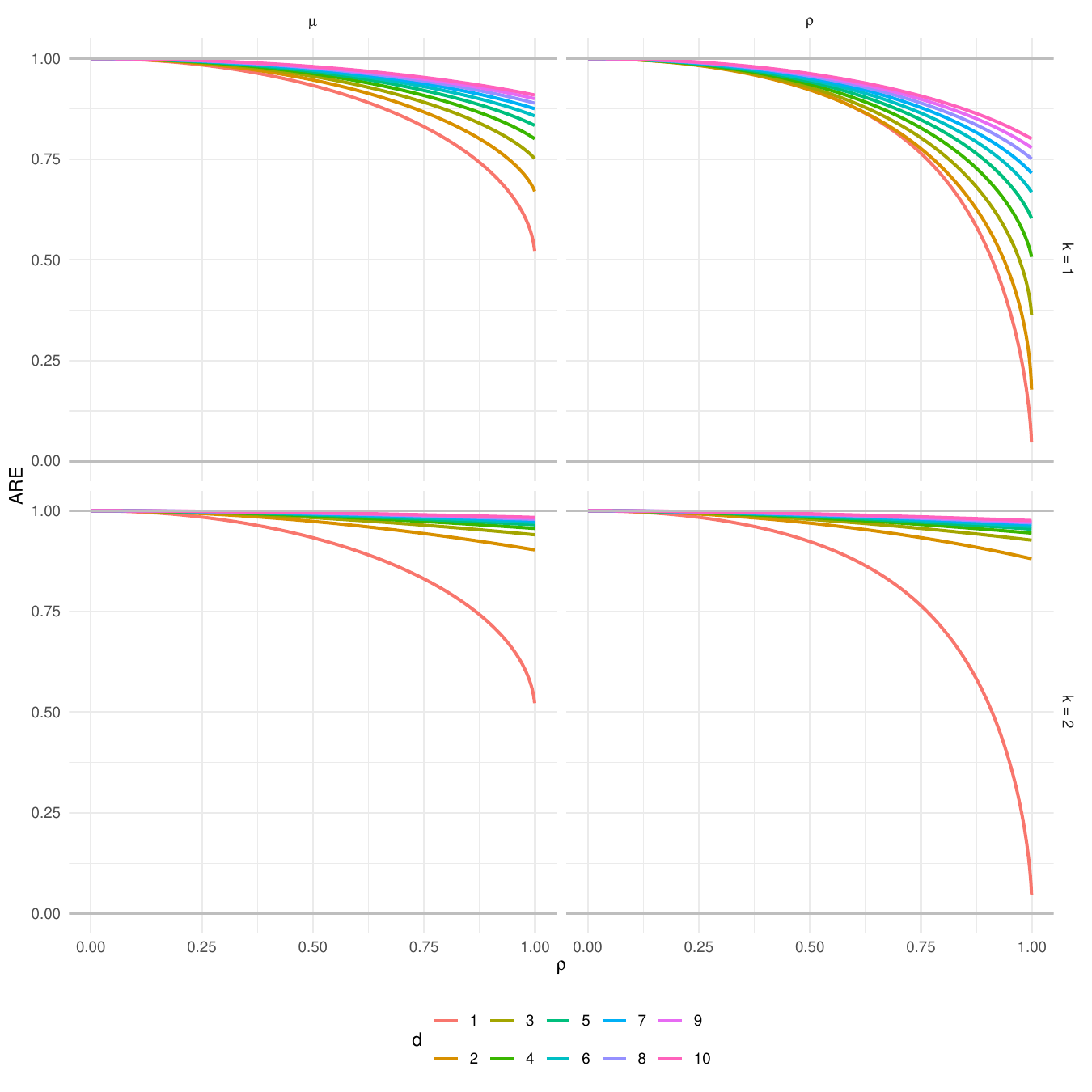}
  \caption{\small Asymptotic relative efficiencies $\rho\mapsto\mathrm{ARE}_{\mathrm{MM}}(\bmu)$ and $\rho\mapsto\mathrm{ARE}_{\mathrm{MM}}(\rho)$ for $k=1,2$ and $d=1,\ldots,10$.}
  \label{fig:are1}
\end{figure}

Figure~\ref{fig:are2} shows the ARE curves $\rho\mapsto\mathrm{ARE}_{\mathrm{GM}}(\rho)$ for several choices of $d$ and $k$. For $k=1,2$, the ARE curves coincide with those of $\mathrm{ARE}_{\mathrm{MM}}(\rho)$ shown in Figure~\ref{fig:are1}. For $k=3$ fixed, the efficiency pattern is similar to that of $\mathrm{ARE}_{\mathrm{MM}}(\rho)$: maximum ARE at $\rho=0$, decay as $\rho$ increases, and increase as $d$ grows. Despite knowing $\bmu$, efficiency of $\hat{\rho}_\mathrm{GM}$ is still smaller than that of $\hat{\rho}_\mathrm{ML}$. With $d=2$ fixed it is seen that the AREs increase with $k$, in an uneven fashion depending on the parity of $k$: even $k$ yield larger AREs, presumably because estimating $\bmu$ becomes harder owing to axial patterns, by which $\hat{\rho}_\mathrm{GM}$ is unaffected.

\begin{figure}[h!]
  \centering
  \includegraphics[width=0.4\textwidth]{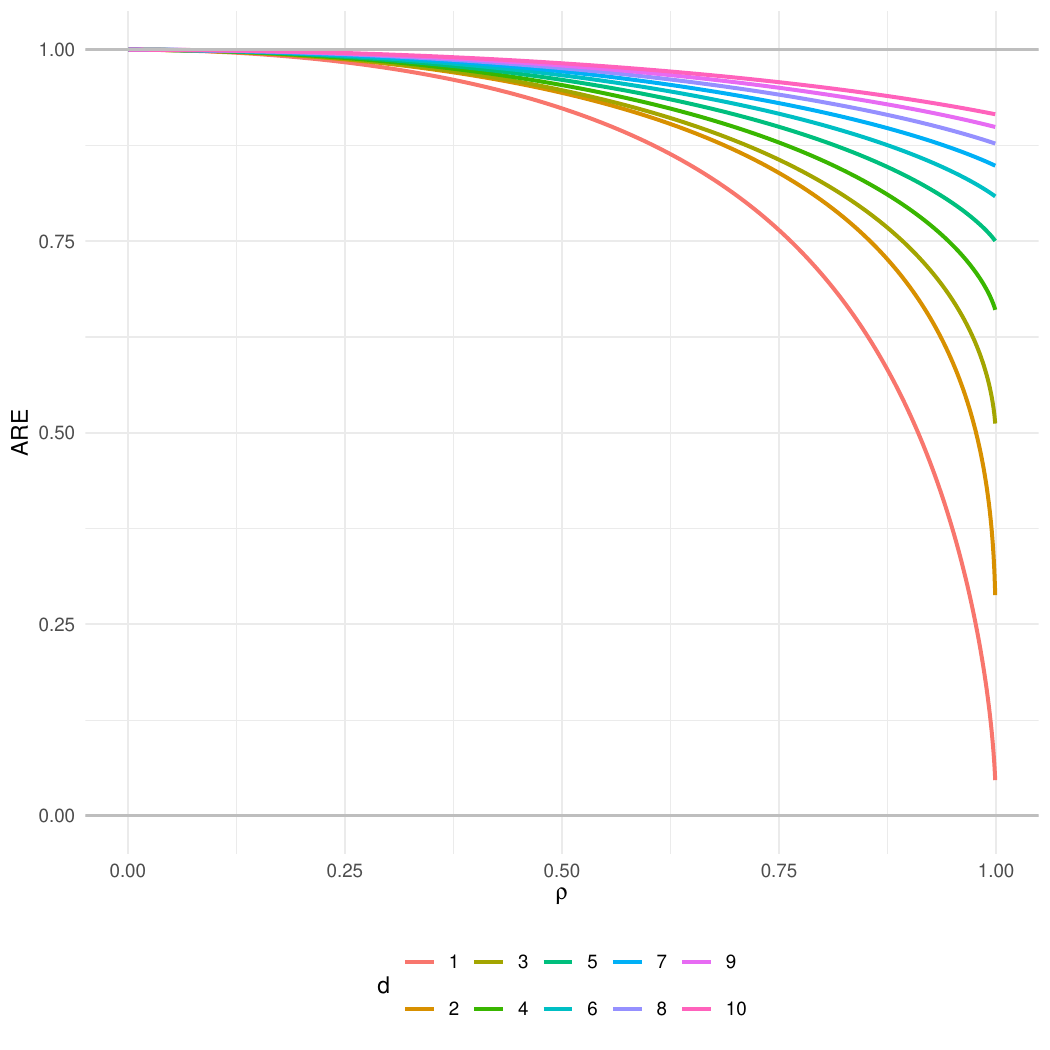}\includegraphics[width=0.4\textwidth]{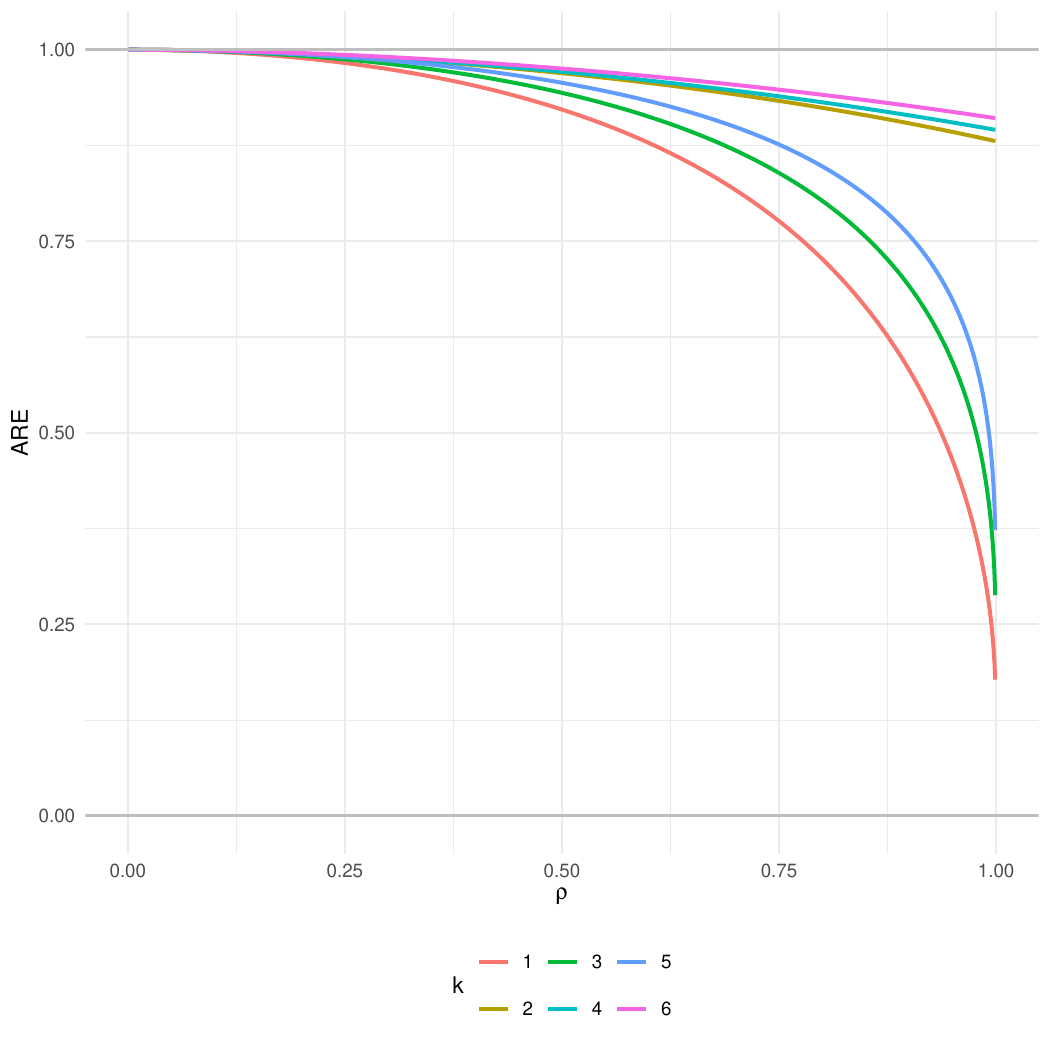}
  \caption{\small Asymptotic relative efficiencies $\rho\mapsto\mathrm{ARE}_{\mathrm{GM}}(\rho)$ for $k=3$ and $d=1,\ldots,10$ (left panel) and $k=1,\ldots,6$ and $d=2$ (right panel).}
  \label{fig:are2}
\end{figure}

\section{Finite-sample comparison of estimators}
\label{sec:rmse}

We compare in this section the finite-sample performance of the three estimators introduced in Sec.~\ref{sec:estimation}: the maximum likelihood estimator $(\hat{\bmu}_{\mathrm{ML}},\hat{\rho}_{\mathrm{ML}})$ (with $\mathrm{sign}(\rho)=1$), the method of moments estimator $(\hat{\bmu}_{\mathrm{MM},k},\hat{\rho}_{\mathrm{MM},k})$ (with truncation and $\mathrm{sign}(\rho)=1$), and the Gegenbauer moment estimator $\hat{\rho}_{\mathrm{GM}}$ (with known $\bmu$ and truncation). We set $\bmu_0=(0,\ldots,0,1)^\top$ and considered $k\in\{1,2\}$, $d\in\{1,2\}$, $\rho\in\{0.1,0.25,0.5,0.75,1\}$, and sample sizes $n=2^3,\ldots,2^{12}$. We computed the root mean squared errors $\mathrm{RMSE}(\hat{\bmu})=\{M^{-1}\sum_{j=1}^M\|\hat{\bmu}^{(j)}-\bmu_0\|^2\}^{1/2}$ and $\mathrm{RMSE}(\hat{\rho})=\{M^{-1}\sum_{j=1}^M(\hat{\rho}^{(j)}-\rho)^2\}^{1/2}$ for $M=10^4$ Monte Carlo replications in each configuration.

Figures~\ref{fig:rmse_k1}--\ref{fig:rmse_k2} display the RMSE curves on $\log_2$--$\log_2$ scales. All estimators exhibit the expected $n^{-1/2}$ convergence rate. For $\mathrm{RMSE}(\hat{\bmu})$, the ML and MM estimators perform similarly at small $\rho$, with ML gaining a slight advantage as $\rho$ increases, and having a clear edge for $k=2$. For $\mathrm{RMSE}(\hat{\rho})$, the MM estimator can outperform ML for small sample sizes and small $\rho$, with ML improving as $\rho$ increases. GM shows high RMSE values for low sample sizes. At the boundary $\rho=1$, the MM and GM estimators exhibit slower convergence in some configurations, consistent with the truncation effects noted by \citet{Pewsey2026|SM}.

\begin{figure}[H]
  \centering
  \includegraphics[width=\textwidth]{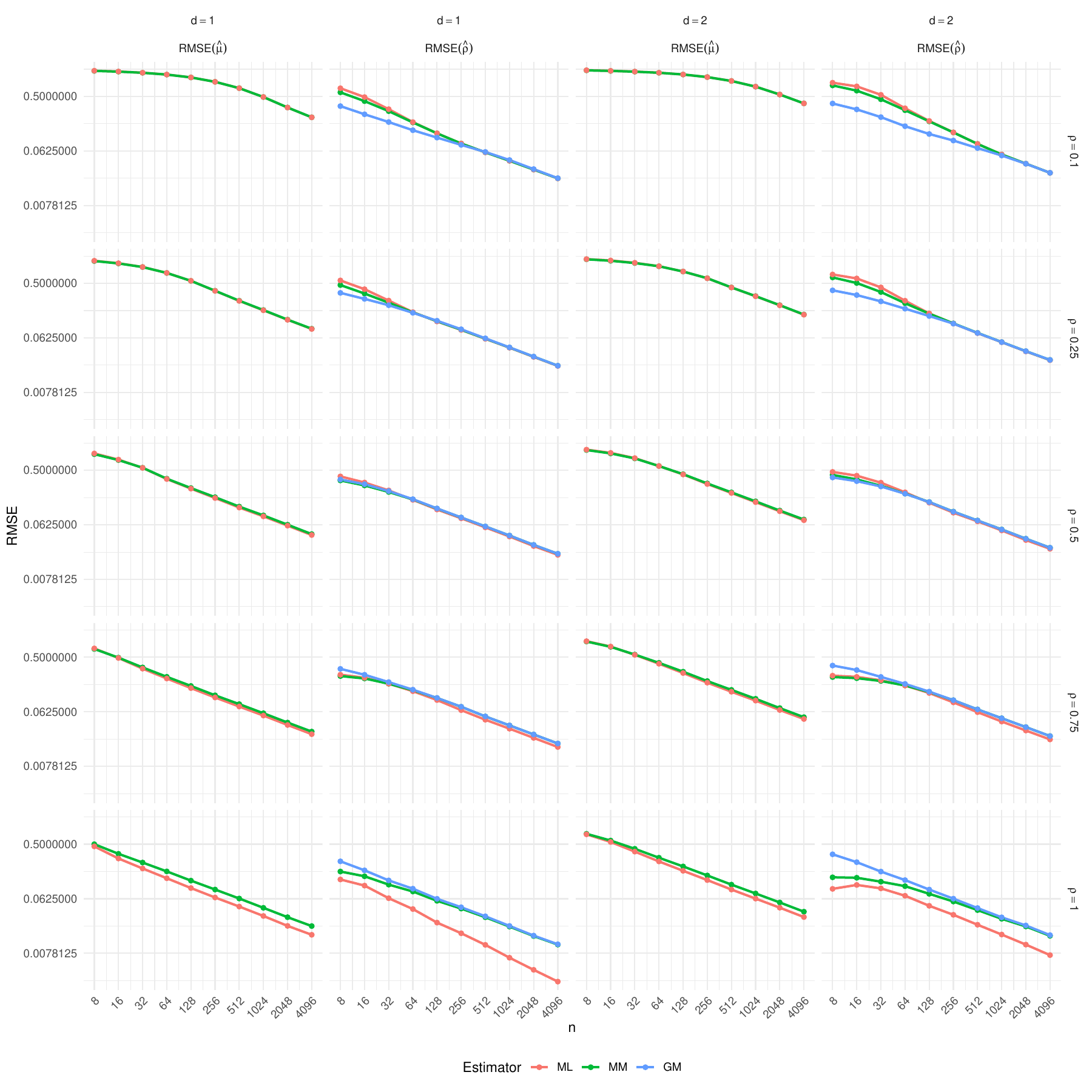}
  \caption{\small RMSEs of the ML, MM, and GM estimators as a function of $n$ for $k=1$, with $d=1$ (two leftmost columns) and $d=2$ (two rightmost columns), across $\rho\in\{0.1,0.25,0.5,0.75,1\}$ (rows). Within each $d$, the left column shows $\mathrm{RMSE}(\hat{\bmu})$ and the right column $\mathrm{RMSE}(\hat{\rho})$.}
  \label{fig:rmse_k1}
\end{figure}

\begin{figure}[H]
  \centering
  \includegraphics[width=\textwidth]{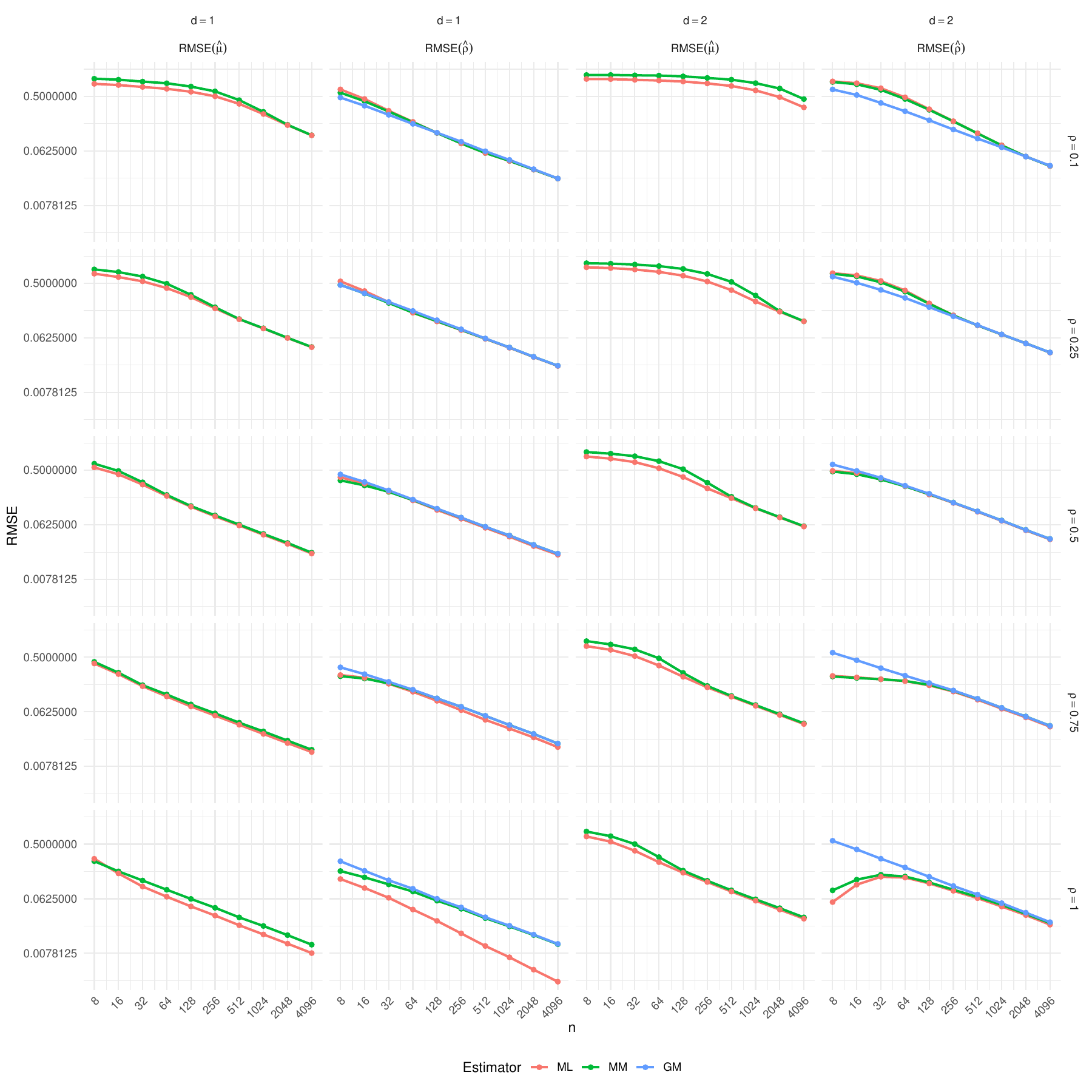}
  \caption{\small Same description as Figure~\ref{fig:rmse_k1}, but for $k=2$.}
  \label{fig:rmse_k2}
\end{figure}

\section{Additional goodness-of-fit experiments}
\label{sec:gofextra}

This appendix complements Sec.~\ref{sec:experiments:gof} by reporting the goodness-of-fit size and power experiments for the sample sizes $n\in\{50,100,200,400\}$. The simulation design is identical to that of Sec.~\ref{sec:experiments:gof}: MM estimation, $M=B=K=1000$, and the six test statistics $P_n^{W,\lambda}$. Table~\ref{tab:h0all} reports the empirical sizes and Table~\ref{tab:h1all} reports the powers against the three families of alternatives. Both contain the results for Tables~\ref{tab:h0}--\ref{tab:h1} for completeness. As anticipated, the empirical size approaches the nominal level as $n$ grows, and the power increases with $n$.

The goodness-of-fit experiments of Sec.~\ref{sec:experiments:gof} use the MM estimators. The whole size and power study was also run with ML, and a summary of the comparative performance of both estimators for the goodness-of-fit test is given in Tables~\ref{tab:mmmlsize}--\ref{tab:mmmlpow}. Table~\ref{tab:mmmlsize} reports the null calibration: MM places more empirical sizes inside the equal-tail $95\pct$ binomial band and keeps the mean size nearer the nominal $5\pct$, most clearly on the sphere ($d=2$) at small $n$, where the ML test is undersized. Table~\ref{tab:mmmlpow} reports the power against two departures, the vMF vs. $\mathrm{C}_1$ ($\kappa=1.5$) and the Watson vs. $\mathrm{C}_2$ ($\kappa=2.5$): the estimators agree to within about a percentage point for the former, except for the Watson departure on the circle, where ML clearly outperforms MM.

\begin{table}[h!]
\centering
\scriptsize
\setlength{\tabcolsep}{3pt}
\begin{tabular}{lll|rr|rr|rr|rr|rr|rr}
\toprule
 &  &  & \multicolumn{6}{c|}{$d=1$} & \multicolumn{6}{c}{$d=2$}\\
\cmidrule(lr){4-9}\cmidrule(lr){10-15}
 &  &  & \multicolumn{2}{c|}{$\mathrm{Unif}$} & \multicolumn{2}{c|}{$\mathrm{P}_n$} & \multicolumn{2}{c|}{$\mathrm{C}_k$} & \multicolumn{2}{c|}{$\mathrm{Unif}$} & \multicolumn{2}{c|}{$\mathrm{P}_n$} & \multicolumn{2}{c}{$\mathrm{C}_k$}\\
$n$ & $k$ & $\rho$ & CvM & AD & CvM & AD & CvM & AD & CvM & AD & CvM & AD & CvM & AD\\
\midrule
50 & 1 & 0.25 & 4.5 & 4.5 & 4.7 & 3.8 & 4.8 & 4.5 & 4.1 & 4.7 & 4.2 & 3.7 & 4.3 & 4.1 \\
 &  & 0.50 & 4.9 & 5.0 & 5.0 & 3.7 & 4.7 & 4.8 & \underline{3.6} & 4.0 & \underline{3.0} & 3.7 & 3.8 & 3.7 \\
 &  & 0.75 & \underline{3.5} & \underline{3.5} & \underline{3.1} & \underline{2.9} & 3.8 & \underline{3.4} & \underline{2.4} & \underline{3.3} & \underline{2.7} & \underline{2.7} & \underline{2.4} & \underline{2.9} \\
 & 2 & 0.25 & 4.5 & 4.6 & 4.7 & 4.8 & 4.6 & 4.8 & 5.8 & 5.6 & 5.4 & 4.1 & 5.1 & 5.4 \\
 &  & 0.50 & 6.1 & 6.3 & 5.6 & 6.4 & 5.7 & 5.4 & 5.6 & 5.7 & 5.5 & 4.1 & 6.0 & 5.4 \\
 &  & 0.75 & 4.7 & 4.6 & 4.5 & 4.8 & 4.5 & 4.8 & 5.8 & 6.0 & 5.7 & 4.2 & 5.6 & 5.7 \\
\midrule
100 & 1 & 0.25 & 4.1 & 4.4 & 4.0 & 3.9 & 3.7 & 4.4 & 4.4 & 4.4 & 4.4 & 4.0 & 4.0 & 4.3 \\
 &  & 0.50 & 4.2 & 4.2 & 4.0 & 3.8 & 4.3 & 3.9 & 4.4 & 4.5 & 4.7 & 4.4 & 3.9 & 4.3 \\
 &  & 0.75 & \underline{3.1} & 3.8 & \underline{3.4} & \underline{3.3} & \underline{3.6} & \underline{3.3} & \underline{2.5} & \underline{2.9} & \underline{2.9} & \underline{3.1} & \underline{2.5} & \underline{2.7} \\
 & 2 & 0.25 & 5.3 & 5.2 & 5.6 & 5.7 & 5.2 & 5.2 & 5.0 & 5.1 & 5.6 & 4.4 & 5.5 & 4.6 \\
 &  & 0.50 & 4.7 & 4.8 & 4.4 & 5.0 & 4.7 & 4.8 & 4.8 & 5.0 & 5.6 & 4.8 & 5.5 & 5.3 \\
 &  & 0.75 & 5.2 & 4.9 & 5.2 & 5.1 & 5.2 & 5.4 & 5.0 & 4.9 & 5.5 & 4.7 & 5.6 & 5.1 \\
\midrule
200 & 1 & 0.25 & 4.9 & 5.0 & 4.7 & 4.9 & 5.0 & 5.5 & 4.4 & 4.2 & 4.1 & 3.9 & 4.2 & 4.0 \\
 &  & 0.50 & 4.5 & 4.3 & 4.4 & 4.1 & 4.2 & 4.8 & 3.8 & 3.8 & 4.1 & \underline{3.4} & 4.1 & 3.9 \\
 &  & 0.75 & 4.6 & 4.4 & 4.6 & 4.3 & 4.3 & 4.1 & \underline{3.6} & \underline{3.6} & \underline{3.4} & \underline{3.1} & \underline{3.5} & 3.8 \\
 & 2 & 0.25 & 4.4 & 4.6 & 4.5 & 4.5 & 4.5 & 4.7 & 4.2 & 4.0 & 4.1 & 4.0 & 4.0 & 4.8 \\
 &  & 0.50 & 4.0 & 4.1 & 4.1 & 4.5 & 4.2 & 4.4 & 4.2 & 4.4 & 3.9 & 4.2 & 3.8 & 4.6 \\
 &  & 0.75 & 3.8 & 4.0 & 4.1 & 4.3 & 3.9 & 4.3 & 4.4 & 4.4 & 4.2 & 4.1 & 4.2 & 4.4 \\
\midrule
400 & 1 & 0.25 & 4.8 & 4.6 & 4.9 & 4.8 & 4.9 & 4.8 & 5.4 & 5.4 & 5.3 & 5.0 & 5.2 & 5.5 \\
 &  & 0.50 & 5.7 & 6.0 & 5.9 & 5.5 & 5.9 & 5.9 & 5.4 & 5.6 & 5.2 & 5.0 & 5.6 & 5.4 \\
 &  & 0.75 & 5.1 & 5.2 & 5.1 & 5.0 & 5.4 & 5.7 & 5.0 & 4.9 & 4.9 & 5.2 & 4.8 & 4.9 \\
 & 2 & 0.25 & 5.0 & 4.7 & 4.5 & 4.7 & 4.2 & 5.0 & 4.8 & 4.7 & 5.3 & 5.1 & 5.2 & 5.1 \\
 &  & 0.50 & \underline{3.2} & \underline{3.1} & \underline{3.3} & \underline{3.5} & \underline{3.1} & \underline{3.0} & 5.0 & 4.9 & 5.6 & 5.0 & 5.8 & 5.4 \\
 &  & 0.75 & 4.5 & 4.2 & 4.3 & 4.3 & 4.2 & 4.3 & 5.1 & 4.8 & 5.6 & 5.2 & 5.2 & 5.5 \\
\bottomrule
\end{tabular}
\caption{\small Rejection percentages of the goodness-of-fit test for the null hypothesis of spherical cardioidness of order $k$ at the significance level $\alpha=5\pct$, for $n\in\{50,100,200,400\}$. The null hypothesis holds: the data-generating process is $\mathrm{C}_k(\bmu,\rho)$. For each dimension $d\in\{1,2\}$, the columns give the statistics $P_n^{W,\lambda}$ with $\lambda\in\{\mathrm{Unif},\mathrm{P}_n,\mathrm{C}_k\}$ and $W\in\{\mathrm{CvM},\mathrm{AD}\}$. Underlined values indicate empirical sizes outside the equal-tail $95\pct$ prediction interval for $\mathrm{Bin}(M, \alpha)\times 100/M$.}
\label{tab:h0all}
\end{table}

\begin{table}[h!]
\centering
\scriptsize
\setlength{\tabcolsep}{3pt}
\begin{tabular}{lll|rr|rr|rr|rr|rr|rr}
\toprule
 &  &  & \multicolumn{6}{c|}{$d=1$} & \multicolumn{6}{c}{$d=2$}\\
\cmidrule(lr){4-9}\cmidrule(lr){10-15}
 &  &  & \multicolumn{2}{c|}{$\mathrm{Unif}$} & \multicolumn{2}{c|}{$\mathrm{P}_n$} & \multicolumn{2}{c|}{$\mathrm{C}_{k_0}$} & \multicolumn{2}{c|}{$\mathrm{Unif}$} & \multicolumn{2}{c|}{$\mathrm{P}_n$} & \multicolumn{2}{c}{$\mathrm{C}_{k_0}$}\\
$n$ & $H_1$ & $\rho/\kappa$ & CvM & AD & CvM & AD & CvM & AD & CvM & AD & CvM & AD & CvM & AD\\
\midrule
50 & $\mathrm{C}_2$ & 0.10 & 4.9 & 5.3 & 5.1 & 6.4 & 5.1 & 5.4 & 4.0 & 4.8 & 5.1 & 4.1 & 4.2 & 5.2 \\
 &  & 0.25 & 15.9 & 16.6 & 15.3 & 16.8 & 14.9 & 16.2 & 6.4 & 6.5 & 7.2 & 6.1 & 6.2 & 7.1 \\
 &  & 0.50 & 56.1 & 55.7 & 53.4 & 50.4 & 54.5 & 54.1 & 16.6 & 17.2 & 16.8 & 14.3 & 15.2 & 15.5 \\
 & $\mathrm{C}_1$ & 0.10 & 6.2 & 6.2 & 6.0 & 5.2 & 5.8 & 6.2 & 6.7 & 6.2 & 5.9 & 3.8 & 5.2 & 5.6 \\
 &  & 0.25 & 18.4 & 17.6 & 18.1 & 16.5 & 18.1 & 17.4 & 11.9 & 11.7 & 11.3 & 8.8 & 10.4 & 11.3 \\
 &  & 0.50 & 62.5 & 62.3 & 61.0 & 58.9 & 59.8 & 60.2 & 36.6 & 36.7 & 33.9 & 28.4 & 34.0 & 33.7 \\
 & $\mathrm{vMF}$ & 0.50 & 4.3 & 4.6 & 4.7 & 4.1 & 4.8 & 5.2 & 2.7 & 3.8 & 2.5 & 3.1 & 3.3 & 4.3 \\
 &  & 1.50 & 49.5 & 49.1 & 49.7 & 44.1 & 48.9 & 48.8 & 51.6 & 50.6 & 53.6 & 48.4 & 50.7 & 49.2 \\
 &  & 2.00 & 92.0 & 91.1 & 91.9 & 88.5 & 91.4 & 90.8 & 93.7 & 93.2 & 93.7 & 91.1 & 93.0 & 92.2 \\
 & $\mathrm{W}$ & 0.50 & 6.1 & 6.0 & 5.6 & 5.2 & 5.8 & 6.2 & 5.7 & 5.7 & 5.5 & 4.4 & 4.6 & 5.2 \\
 &  & 2.50 & 7.3 & 7.3 & 7.5 & 7.5 & 7.5 & 7.5 & 12.3 & 13.5 & 17.6 & 21.2 & 13.1 & 16.1 \\
 &  & 5.00 & 34.1 & 51.9 & 49.9 & 61.2 & 28.1 & 43.4 & 86.3 & 98.0 & 99.9 & 99.9 & 90.1 & 98.4 \\
\midrule
100 & $\mathrm{C}_2$ & 0.10 & 7.5 & 7.7 & 7.3 & 7.0 & 7.6 & 8.1 & 5.6 & 6.0 & 5.4 & 4.3 & 5.7 & 5.2 \\
 &  & 0.25 & 30.8 & 31.0 & 29.8 & 29.7 & 30.8 & 30.2 & 10.0 & 10.2 & 10.4 & 8.7 & 10.1 & 9.4 \\
 &  & 0.50 & 89.4 & 88.7 & 87.9 & 86.3 & 89.6 & 88.2 & 33.4 & 33.6 & 35.3 & 30.5 & 32.5 & 32.9 \\
 & $\mathrm{C}_1$ & 0.10 & 8.5 & 8.5 & 8.1 & 7.9 & 8.8 & 8.0 & 5.6 & 5.8 & 5.8 & 4.8 & 5.6 & 5.8 \\
 &  & 0.25 & 33.2 & 32.9 & 33.1 & 31.9 & 32.6 & 32.9 & 17.6 & 17.8 & 16.8 & 14.8 & 15.9 & 16.6 \\
 &  & 0.50 & 91.4 & 91.4 & 89.9 & 89.7 & 90.0 & 90.5 & 68.0 & 68.0 & 65.1 & 62.0 & 66.1 & 65.5 \\
 & $\mathrm{vMF}$ & 0.50 & 5.2 & 5.9 & 5.4 & 5.2 & 5.2 & 6.4 & 4.7 & 4.7 & 4.9 & 4.2 & 4.1 & 4.8 \\
 &  & 1.50 & 77.0 & 76.5 & 77.1 & 74.1 & 76.9 & 76.5 & 77.0 & 75.5 & 79.7 & 76.2 & 76.9 & 76.1 \\
 &  & 2.00 & 99.9 & 99.8 & 99.9 & 99.7 & 99.8 & 99.8 & 100.0 & 99.9 & 99.9 & 99.8 & 99.9 & 100.0 \\
 & $\mathrm{W}$ & 0.50 & 5.8 & 5.8 & 5.4 & 5.2 & 5.5 & 6.1 & 6.7 & 6.3 & 7.0 & 6.5 & 6.5 & 6.9 \\
 &  & 2.50 & 7.3 & 8.0 & 8.0 & 8.2 & 7.8 & 8.7 & 18.9 & 26.3 & 31.7 & 41.1 & 21.7 & 32.0 \\
 &  & 5.00 & 85.0 & 96.9 & 92.3 & 96.7 & 75.5 & 91.1 & 100.0 & 100.0 & 100.0 & 100.0 & 100.0 & 100.0 \\
\midrule
200 & $\mathrm{C}_2$ & 0.10 & 11.7 & 11.8 & 11.6 & 11.7 & 11.5 & 11.4 & 6.6 & 6.9 & 6.5 & 5.9 & 6.1 & 7.5 \\
 &  & 0.25 & 55.7 & 54.6 & 55.7 & 54.6 & 55.2 & 54.3 & 20.0 & 19.5 & 20.3 & 18.4 & 20.1 & 19.3 \\
 &  & 0.50 & 99.5 & 99.4 & 99.5 & 99.4 & 99.4 & 99.3 & 62.6 & 62.0 & 64.3 & 60.6 & 62.6 & 62.2 \\
 & $\mathrm{C}_1$ & 0.10 & 14.1 & 14.2 & 13.5 & 13.7 & 13.9 & 13.7 & 8.6 & 8.3 & 8.0 & 7.1 & 8.5 & 8.5 \\
 &  & 0.25 & 59.9 & 59.1 & 59.3 & 59.0 & 59.8 & 59.2 & 37.1 & 35.9 & 36.4 & 35.3 & 35.9 & 36.2 \\
 &  & 0.50 & 99.8 & 99.8 & 99.7 & 99.8 & 99.7 & 99.7 & 93.5 & 93.6 & 93.0 & 92.5 & 93.5 & 93.2 \\
 & $\mathrm{vMF}$ & 0.50 & 7.0 & 7.8 & 6.9 & 6.5 & 7.1 & 7.7 & 5.1 & 5.3 & 5.6 & 5.1 & 5.0 & 5.2 \\
 &  & 1.50 & 95.5 & 95.9 & 95.4 & 95.2 & 95.6 & 95.9 & 97.0 & 96.7 & 97.2 & 96.6 & 97.2 & 96.5 \\
 &  & 2.00 & 100.0 & 100.0 & 100.0 & 100.0 & 100.0 & 100.0 & 100.0 & 100.0 & 100.0 & 100.0 & 100.0 & 100.0 \\
 & $\mathrm{W}$ & 0.50 & 5.1 & 5.2 & 5.2 & 5.3 & 5.3 & 5.2 & 5.1 & 5.5 & 4.8 & 4.8 & 5.1 & 5.2 \\
 &  & 2.50 & 7.9 & 10.8 & 7.7 & 10.6 & 7.4 & 9.8 & 41.0 & 61.7 & 66.8 & 79.8 & 48.5 & 69.7 \\
 &  & 5.00 & 100.0 & 100.0 & 100.0 & 100.0 & 100.0 & 100.0 & 100.0 & 100.0 & 100.0 & 100.0 & 100.0 & 100.0 \\
\midrule
400 & $\mathrm{C}_2$ & 0.10 & 20.6 & 20.2 & 20.0 & 19.6 & 20.8 & 19.2 & 8.6 & 8.3 & 9.4 & 8.4 & 8.7 & 8.8 \\
 &  & 0.25 & 87.0 & 86.1 & 86.5 & 85.4 & 86.7 & 85.9 & 32.3 & 31.9 & 34.6 & 31.6 & 32.3 & 31.9 \\
 &  & 0.50 & 100.0 & 100.0 & 100.0 & 100.0 & 100.0 & 100.0 & 92.7 & 92.4 & 93.6 & 92.9 & 92.2 & 92.1 \\
 & $\mathrm{C}_1$ & 0.10 & 22.6 & 22.8 & 22.8 & 22.6 & 22.9 & 22.6 & 12.2 & 11.5 & 11.9 & 11.3 & 11.4 & 12.3 \\
 &  & 0.25 & 89.8 & 89.3 & 89.6 & 89.4 & 89.8 & 89.6 & 66.3 & 65.2 & 65.6 & 65.2 & 65.8 & 65.0 \\
 &  & 0.50 & 100.0 & 100.0 & 100.0 & 100.0 & 100.0 & 100.0 & 99.9 & 100.0 & 99.9 & 99.9 & 99.9 & 100.0 \\
 & $\mathrm{vMF}$ & 0.50 & 10.5 & 10.5 & 10.1 & 9.7 & 10.6 & 10.6 & 6.2 & 6.5 & 6.8 & 6.4 & 6.1 & 6.8 \\
 &  & 1.50 & 99.9 & 99.9 & 99.9 & 99.9 & 99.9 & 99.9 & 99.7 & 99.7 & 99.7 & 99.7 & 99.7 & 99.7 \\
 &  & 2.00 & 100.0 & 100.0 & 100.0 & 100.0 & 100.0 & 100.0 & 100.0 & 100.0 & 100.0 & 100.0 & 100.0 & 100.0 \\
 & $\mathrm{W}$ & 0.50 & 6.2 & 6.2 & 6.0 & 5.2 & 6.1 & 5.5 & 4.6 & 4.3 & 4.9 & 4.8 & 5.2 & 4.8 \\
 &  & 2.50 & 13.4 & 22.3 & 13.0 & 19.2 & 11.7 & 16.5 & 91.2 & 98.0 & 98.7 & 99.5 & 94.6 & 99.0 \\
 &  & 5.00 & 100.0 & 100.0 & 100.0 & 100.0 & 100.0 & 100.0 & 100.0 & 100.0 & 100.0 & 100.0 & 100.0 & 100.0 \\
\bottomrule
\end{tabular}
\caption{\small Rejection percentages of the goodness-of-fit test for the null hypothesis of spherical cardioidness of order $k_0$ at the significance level $\alpha=5\pct$, for $n\in\{50,100,200,400\}$. The second column gives the data-generating alternative $H_1$: spherical cardioids $\mathrm{C}_k(\bmu,\rho)$ of orders $k=2,1$ for $k_0=1,2$, $\mathrm{vMF}(\bmu,\kappa)$ for $k_0=1$, and $\mathrm{W}(\bmu,\kappa)$ for $k_0=2$. The column layout is that of Table~\ref{tab:h0all}.}
\label{tab:h1all}
\end{table}

\clearpage

\begin{table}[h!]
\centering
\small
\setlength{\tabcolsep}{4pt}
\begin{tabular}{c|cccc|cccc}
\toprule
 & \multicolumn{4}{c|}{$d=1$} & \multicolumn{4}{c}{$d=2$}\\
\cmidrule(lr){2-5}\cmidrule(lr){6-9}
 & \multicolumn{2}{c}{\pct Band} & \multicolumn{2}{c|}{$\overline{\text{Size}}$} & \multicolumn{2}{c}{\pct Band} & \multicolumn{2}{c}{$\overline{\text{Size}}$}\\
$n$ & MM & ML & MM & ML & MM & ML & MM & ML\\ \midrule
50  & 86.1 & 86.1 & 4.62 & 4.45 & 77.8 & 50.0 & 4.44 & 3.96\\
100 & 86.1 & 55.6 & 4.47 & 4.00 & 83.3 & 50.0 & 4.45 & 4.03\\
200 & 100.0 & 91.7 & 4.43 & 4.17 & 83.3 & 63.9 & 4.02 & 3.69\\
400 & 83.3 & 83.3 & 4.67 & 4.48 & 100.0 & 100.0 & 5.19 & 5.18\\
\bottomrule
\end{tabular}
\caption{\small Calibration under the null at $\alpha=5\pct$: percentage of the $36$ cells ($k$, $\rho$, and six statistics) inside the equal-tail $95\pct$ binomial prediction band (\pct Band), and mean empirical size ($\overline{\text{Size}}$), by dimension $d$, estimator, and sample size $n$.}
\label{tab:mmmlsize}
\end{table}

\begin{table}[h!]
\centering
\small
\setlength{\tabcolsep}{4pt}
\begin{tabular}{c|cccc|cccc}
\toprule
 & \multicolumn{4}{c|}{$d=1$} & \multicolumn{4}{c}{$d=2$}\\
\cmidrule(lr){2-5}\cmidrule(lr){6-9}
 & \multicolumn{2}{c}{vMF} & \multicolumn{2}{c|}{Watson} & \multicolumn{2}{c}{vMF} & \multicolumn{2}{c}{Watson}\\
$n$ & MM & ML & MM & ML & MM & ML & MM & ML\\ \midrule
50  & 48.4 & 49.5 & 7.4 & 8.4 & 50.7 & 51.0 & 15.6 & 15.5\\
100 & 76.4 & 75.3 & 8.0 & 9.8 & 76.9 & 76.4 & 28.6 & 28.7\\
200 & 95.6 & 95.4 & 9.0 & 14.0 & 96.9 & 95.6 & 61.3 & 60.9\\
400 & 99.9 & 99.9 & 16.0 & 31.3 & 99.7 & 99.8 & 96.8 & 96.8\\
\bottomrule
\end{tabular}
\caption{\small Power (mean over the six statistics, \pct) against the von Mises--Fisher ($\kappa=1.5$) and Watson ($\kappa=2.5$) alternatives, by dimension $d$ and sample size $n$, for the MM and ML estimators.}
\label{tab:mmmlpow}
\end{table}


\begin{thebibliography}{}

\bibitem[Azzalini and Capitanio, 2014]{Azzalini2014}
Azzalini, A. and Capitanio, A. (2014).
\newblock {\em The Skew-Normal and Related Families}, volume~3 of {\em
  Institute of Mathematical Statistics Monographs}.
\newblock Cambridge University Press, Cambridge.

\bibitem[Baringhaus and Gr{\"u}bel, 2024]{Baringhaus2024}
Baringhaus, L. and Gr{\"u}bel, R. (2024).
\newblock Discrete mixture representations of spherical distributions.
\newblock {\em Stat. Pap.}, 65(2):557--596.

\bibitem[Borodavka and Ebner, 2026]{Borodavka2026}
Borodavka, J. and Ebner, B. (2026).
\newblock A general maximal projection approach to uniformity testing on the
  hypersphere.
\newblock {\em Bernoulli}, 32(2):996--1019.

\bibitem[Chac\'{o}n and Duong, 2018]{Chacon2018}
Chac\'{o}n, J.~E. and Duong, T. (2018).
\newblock {\em Multivariate Kernel Smoothing and its Applications}, volume 160
  of {\em Monographs on Statistics and Applied Probability}.
\newblock CRC Press, Boca Raton.

\bibitem[Chac\'on et~al., 2026]{Chacon2026}
Chac\'on, J.~E., Garc\'ia-Portugu\'es, E., and Meil\'an-Vila, A. (2026).
\newblock Blessing of dimensionality in cross-validated bandwidth selection on
  the sphere.
\newblock {\em arXiv:2601.20442}.

\bibitem[{DLMF}, 2025]{NIST}
{DLMF} (2025).
\newblock {\em NIST Digital Library of Mathematical Functions}.
\newblock F. W. J. Olver, A. B. {Olde Daalhuis}, D. W. Lozier, B. I. Schneider,
  R. F. Boisvert, C. W. Clark, B. R. Miller and B. V. Saunders, eds.

\bibitem[Dones et~al., 2015]{Dones2015}
Dones, L., Brasser, R., Kaib, N., and Rickman, H. (2015).
\newblock Origin and evolution of the cometary reservoirs.
\newblock {\em Space Sci. Rev.}, 197(1):191--269.

\bibitem[Ebner et~al., 2024]{Ebner2024}
Ebner, B., Henze, N., and Meintanis, S. (2024).
\newblock A unified approach to goodness-of-fit testing for spherical and
  hyperspherical data.
\newblock {\em Stat. Pap.}, 65(6):3447--3475.

\bibitem[Fern\'andez-Dur\'an, 2004]{Fernandez-Duran2004}
Fern\'andez-Dur\'an, J.~J. (2004).
\newblock Circular distributions based on nonnegative trigonometric sums.
\newblock {\em Biometrics}, 60(2):499--503.

\bibitem[Fern\'andez-Dur\'an and Gregorio-Dom\'inguez,
  2014]{Fernandez-Duran2014a}
Fern\'andez-Dur\'an, J.~J. and Gregorio-Dom\'inguez, M.~M. (2014).
\newblock Distributions for spherical data based on nonnegative trigonometric
  sums.
\newblock {\em Stat. Pap.}, 55(4):983--1000.

\bibitem[Garc\'ia-Portugu\'es et~al., 2023]{Garcia-Portugues2020b}
Garc\'ia-Portugu\'es, E., Navarro-Esteban, P., and Cuesta-Albertos, J.~A.
  (2023).
\newblock On a projection-based class of uniformity tests on the hypersphere.
\newblock {\em Bernoulli}, 29(1):181--204.

\bibitem[Garc\'ia-Portugu\'es et~al., 2025]{Garcia-Portugues2025}
Garc\'ia-Portugu\'es, E., Paindaveine, D., and Verdebout, T. (2025).
\newblock On a class of {S}obolev tests for symmetry, their detection
  thresholds, and asymptotic powers.
\newblock {\em J. Am. Stat. Assoc.}

\bibitem[Garc\'ia-Portugu\'es and Verdebout, 2025]{Garcia-Portugues:sphunif}
Garc\'ia-Portugu\'es, E. and Verdebout, T. (2025).
\newblock {\em {sphunif}: Uniformity Tests on the Circle, Sphere, and
  Hypersphere}.
\newblock {R} package version 1.4.4.

\bibitem[Gin\'e, 1975]{Gine1975}
Gin\'e, E. (1975).
\newblock Invariant tests for uniformity on compact {R}iemannian manifolds
  based on {S}obolev norms.
\newblock {\em Ann. Stat.}, 3(6):1243--1266.

\bibitem[Jammalamadaka and SenGupta, 2001]{Jammalamadaka2001}
Jammalamadaka, S.~R. and SenGupta, A. (2001).
\newblock {\em Topics in Circular Statistics}, volume~5 of {\em Series on
  Multivariate Analysis}.
\newblock World Scientific, Singapore.

\bibitem[Jeffreys, 2003]{Jeffreys2003}
Jeffreys, H. (2003).
\newblock {\em Theory of Probability}.
\newblock Oxford University Press, Oxford, third edition.

\bibitem[Jones and Pewsey, 2005]{Jones2005}
Jones, M.~C. and Pewsey, A. (2005).
\newblock A family of symmetric distributions on the circle.
\newblock {\em J. Am. Stat. Assoc.}, 100(472):1422--1428.

\bibitem[Jupp et~al., 2003]{Jupp2003}
Jupp, P.~E., Kim, P.~T., Koo, J.-Y., and Wiegert, P. (2003).
\newblock The intrinsic distribution and selection bias of long-period cometary
  orbits.
\newblock {\em J. Am. Stat. Assoc.}, 98(463):515--521.

\bibitem[Kalf, 1995]{Kalf1995}
Kalf, H. (1995).
\newblock On the expansion of a function in terms of spherical harmonics in
  arbitrary dimensions.
\newblock {\em Bull. Belg. Math. Soc. Simon Stevin}, 2(4):361--380.

\bibitem[Kato and Jones, 2015]{Kato2015}
Kato, S. and Jones, M.~C. (2015).
\newblock A tractable and interpretable four-parameter family of unimodal
  distributions on the circle.
\newblock {\em Biometrika}, 102(1):181--190.

\bibitem[Klemel\"a, 2000]{Klemela2000}
Klemel\"a, J. (2000).
\newblock Estimation of densities and derivatives of densities with directional
  data.
\newblock {\em J. Multivar. Anal.}, 73(1):18--40.

\bibitem[Mardia and Jupp, 1999]{Mardia1999a}
Mardia, K.~V. and Jupp, P.~E. (1999).
\newblock {\em Directional Statistics}.
\newblock Wiley Series in Probability and Statistics. Wiley, Chichester.

\bibitem[Pewsey, 2026]{Pewsey2026}
Pewsey, A. (2026).
\newblock On {J}effreys's cardioid distribution.
\newblock {\em Comput. Stat. Data Anal.}, 213:108248.

\bibitem[Pewsey et~al., 2013]{Pewsey2013}
Pewsey, A., Neuh\"auser, M., and Ruxton, G.~D. (2013).
\newblock {\em Circular Statistics in {R}}.
\newblock Oxford University Press, Oxford.

\bibitem[Saw, 1984]{Saw1984}
Saw, J.~G. (1984).
\newblock Ultraspherical polynomials and statistics on the {$m$}-sphere.
\newblock {\em J. Multivar. Anal.}, 14(1):105--113.

\bibitem[Self and Liang, 1987]{Self1987}
Self, S.~G. and Liang, K.-Y. (1987).
\newblock Asymptotic properties of maximum likelihood estimators and likelihood
  ratio tests under nonstandard conditions.
\newblock {\em J. Am. Stat. Assoc.}, 82(398):605--610.

\end{thebibliography}

\begin{thebibliography}{}

\bibitem[Chac\'on et~al., 2026]{Chacon2026|SM}
Chac\'on, J.~E., Garc\'ia-Portugu\'es, E., and Meil\'an-Vila, A. (2026).
\newblock Blessing of dimensionality in cross-validated bandwidth selection on
  the sphere.
\newblock {\em arXiv:2601.20442}.

\bibitem[Dai and Xu, 2013]{Dai2013|SM}
Dai, F. and Xu, Y. (2013).
\newblock {\em Approximation Theory and Harmonic Analysis on Spheres and
  Balls}.
\newblock Springer Monographs in Mathematics. Springer, New York.

\bibitem[{DLMF}, 2025]{NIST|SM}
{DLMF} (2025).
\newblock {\em NIST Digital Library of Mathematical Functions}.
\newblock F. W. J. Olver, A. B. {Olde Daalhuis}, D. W. Lozier, B. I. Schneider,
  R. F. Boisvert, C. W. Clark, B. R. Miller and B. V. Saunders, eds.

\bibitem[{Fern\'andez-de-Marcos} and Garc\'ia-Portugu\'es,
  2023]{Fernandez-de-Marcos2023|SM}
{Fern\'andez-de-Marcos}, A. and Garc\'ia-Portugu\'es, E. (2023).
\newblock On new omnibus tests of uniformity on the hypersphere.
\newblock {\em Test}, 32(4):1508--1529.

\bibitem[Garc\'ia-Portugu\'es et~al., 2023]{Garcia-Portugues2020b|SM}
Garc\'ia-Portugu\'es, E., Navarro-Esteban, P., and Cuesta-Albertos, J.~A.
  (2023).
\newblock On a projection-based class of uniformity tests on the hypersphere.
\newblock {\em Bernoulli}, 29(1):181--204.

\bibitem[Gradshteyn and Ryzhik, 2014]{Gradshteyn2014|SM}
Gradshteyn, I.~S. and Ryzhik, I.~M. (2014).
\newblock {\em Table of Integrals, Series, and Products}.
\newblock Academic Press, Amsterdam, eighth edition.

\bibitem[Magnus and Neudecker, 1999]{Magnus1999|SM}
Magnus, J.~R. and Neudecker, H. (1999).
\newblock {\em Matrix Differential Calculus with Applications in Statistics and
  Econometrics}.
\newblock Wiley Series in Probability and Statistics. Wiley, Chichester.

\bibitem[Magnus et~al., 1966]{Magnus1966|SM}
Magnus, W., Oberhettinger, F., and Soni, R.~P. (1966).
\newblock {\em Formulas and Theorems for the Special Functions of Mathematical
  Physics}, volume~52 of {\em Die Grundlehren der mathematischen
  Wissenschaften}.
\newblock Springer, Berlin.

\bibitem[Pewsey, 2026]{Pewsey2026|SM}
Pewsey, A. (2026).
\newblock On {J}effreys's cardioid distribution.
\newblock {\em Comput. Stat. Data Anal.}, 213:108248.

\bibitem[van~der Vaart, 1998]{vanderVaart1998|SM}
van~der Vaart, A.~W. (1998).
\newblock {\em Asymptotic Statistics}, volume~3 of {\em Cambridge Series in
  Statistical and Probabilistic Mathematics}.
\newblock Cambridge University Press, Cambridge.

\bibitem[{Wolfram Research, Inc.}, 2021]{Mathematica|SM}
{Wolfram Research, Inc.} (2021).
\newblock Mathematica.
\newblock Computer software, Version 13.0.

\end{thebibliography}


\fi

\end{document}